\documentclass[fleqn,usenatbib]{mnras}

\usepackage{newtxtext,newtxmath}
\usepackage[T1]{fontenc}

\DeclareRobustCommand{\VAN}[3]{#2}
\let\VANthebibliography\thebibliography
\def\thebibliography{\DeclareRobustCommand{\VAN}[3]{##3}\VANthebibliography}

\usepackage{graphicx}	% Including figure files
\usepackage{amsmath}	% Advanced maths commands
\usepackage[labelfont=bf]{caption}
\usepackage{float}
\restylefloat{table}
\restylefloat{figure}
\usepackage{longtable}
\usepackage{xcolor}
\usepackage{soul}
\usepackage[title]{appendix}
\usepackage[switch,columnwise]{lineno}
\usepackage{changepage}
\usepackage{enumitem}
\usepackage{hyperref}
\usepackage{tikz}

\usepackage{academicons}
\usepackage{xcolor}
\usepackage{orcidlink}
\newcommand{\orcid}[1]{\href{https://orcid.org/#1}{\textcolor[HTML]{A6CE39}{\aiOrcid}}}
\usepackage{eso-pic}% http://ctan.org/pkg/eso-pic
\AddToShipoutPictureBG*{%
  \AtPageUpperLeft{%
    \hspace{0.75\paperwidth}%
    \raisebox{-4.5\baselineskip}{%
      \makebox[0pt][l]{\textnormal{DES-2022-0741}}
}}}%

\AddToShipoutPictureBG*{%
  \AtPageUpperLeft{%
    \hspace{0.75\paperwidth}%
    \raisebox{-5.5\baselineskip}{%
      \makebox[0pt][l]{\textnormal{FERMILAB-PUB-23-020-PPD}}
}}}%

\title[XCS tests of the DES Y3 redMaPPer Catalogue]{
The XMM Cluster Survey: Exploring scaling relations and completeness of the Dark Energy Survey Year 3 redMaPPer cluster catalogue}

\author[E.W. Upsdell et al.]
{E. W. Upsdell$^{1}$\thanks{E-mail: e.upsdell@sussex.ac.uk}\orcidlink{0000-0003-0628-7201},
P. A. Giles$^{1}$\orcidlink{0000-0003-4937-8453}, 
A. K. Romer$^{1}$\orcidlink{0000-0002-9328-879X},
R. Wilkinson$^{1}$\orcidlink{0000-0002-3908-7313}, 
D. J. Turner$^{1}$\orcidlink{0000-0001-9658-1396},
\newauthor
M. Hilton$^{2,3}$\orcidlink{0000-0002-8490-8117},
E. Rykoff$^{4}$,
A. Farahi$^{5}$\orcidlink{0000-0003-0777-4618},
S. Bhargava$^{6}$,
T. Jeltema$^{7}$\orcidlink{0000-0001-6089-0365},
M. Klein$^{8}$,
A. Bermeo$^{1}$,
\newauthor
C. A. Collins$^{9}$,
L. Ebrahimpour$^{10,11}$\orcidlink{0000-0001-7243-8296},
D. Hollowood$^{7}$,
R. G. Mann$^{12}$,
M. Manolopoulou$^{12}$,
\newauthor
C. J. Miller$^{13}$,
P. J. Rooney$^{1}$,
Martin Sahl\'en,$^{14}$
J. P. Stott$^{15}$\orcidlink{0000-0002-1679-9983},
P. T. P. Viana$^{10,11}$\orcidlink{0000-0003-1572-8531},
\newauthor
S.~Allam$^{31}$,
O.~Alves$^{16}$,
D.~Bacon$^{17}$,
E.~Bertin$^{18}$,
S.~Bocquet$^{19}$\orcidlink{0000-0002-4900-805X},
D.~Brooks$^{20}$\orcidlink{0000-0002-8458-5047},
D.~L.~Burke$^{4,52}$,
\newauthor
M.~Carrasco~Kind$^{21,22}$\orcidlink{0000-0002-4802-3194},
J.~Carretero$^{23}$\orcidlink{0000-0002-3130-0204},
M.~Costanzi$^{24,25,26}$,
L.~N.~da Costa$^{27}$,
M.~E.~S.~Pereira$^{28}$,
\newauthor
J.~De~Vicente$^{29}$\orcidlink{0000-0001-8318-6813},
S.~Desai$^{30}$\orcidlink{0000-0002-0466-3288},
H.~T.~Diehl$^{31}$\orcidlink{0000-0002-8357-7467},
J.~P.~Dietrich$^{19}$\orcidlink{0000-0002-8134-9591},
S.~Everett$^{32}$,
I.~Ferrero$^{33}$,
J.~Frieman$^{31,34}$\orcidlink{0000-0003-4079-3263},
\newauthor
J.~Garc\'ia-Bellido$^{53}$\orcidlink{0000-0002-9370-8360},
D.~W.~Gerdes$^{35}$\orcidlink{0000-0001-6942-2736},
G.~Gutierrez$^{31}$\orcidlink{0000-0003-0825-0517},
S.~R.~Hinton$^{36}$,
K.~Honscheid$^{37,38}$\orcidlink{0000-0002-6550-2023},
D.~J.~James$^{39}$\orcidlink{0000-0001-5160-4486},
\newauthor
K.~Kuehn$^{40,41}$\orcidlink{0000-0003-0120-0808},
N.~Kuropatkin$^{31}$\orcidlink{0000-0003-2511-0946},
M.~Lima$^{42,27}$,
J.~L.~Marshall$^{43}$\orcidlink{0000-0003-0710-9474},
J. Mena-Fern{\'a}ndez$^{29}$\orcidlink{0000-0001-9497-7266},
\newauthor
F.~Menanteau$^{21,22}$\orcidlink{0000-0002-1372-2534},
R.~Miquel$^{44,23}$\orcidlink{0000-0002-6610-4836},
J.~J.~Mohr$^{19,50}$,
R.~L.~C.~Ogando$^{45}$\orcidlink{0000-0003-2120-1154},
A.~Pieres$^{27,45}$\orcidlink{0000-0001-9186-6042},
M.~Raveri$^{46}$,
\newauthor
M.~Rodriguez-Monroy$^{29}$,
E.~Sanchez$^{29}$\orcidlink{0000-0002-9646-8198},
V.~Scarpine$^{31}$,
I.~Sevilla-Noarbe$^{29}$\orcidlink{0000-0002-1831-1953},
M.~Smith$^{47}$\orcidlink{0000-0002-3321-1432},
E.~Suchyta$^{48}$\orcidlink{0000-0002-7047-9358},
\newauthor
M.~E.~C.~Swanson$^{}$,
G.~Tarle$^{16}$\orcidlink{0000-0003-1704-0781},
C.~To$^{37}$\orcidlink{0000-0001-7836-2261},
N.~Weaverdyck$^{16,49}$,
J.~Weller$^{50,51}$\orcidlink{0000-0002-8282-2010}
and P.~Wiseman$^{47}$
\\
\\
Author affiliations are listed at the end of this paper.
}

\date{Accepted 2023 April 20. Received 2023 April 20; in original form 2023 January 24}

\pubyear{2023}

\begin{document}
%\linenumbers
%\runningpagewiselinenumbers
\label{firstpage}
\pagerange{\pageref{firstpage}--\pageref{lastpage}}
\maketitle

\begin{abstract}
We cross-match and compare characteristics of galaxy clusters identified in
observations from two sky surveys using two completely different techniques.  One sample is optically selected from the analysis of three years of Dark Energy Survey observations using the redMaPPer cluster detection algorithm.  The second is X-ray selected from {\em XMM} observations analysed by the {\em XMM} Cluster Survey.  The samples comprise a total area of 57.4 deg$^2$, bounded by the area of 4 contiguous XMM survey regions that overlap the DES footprint.  We find that the X-ray selected sample is fully matched with entries in the redMaPPer catalogue, above $\lambda>$20 and within  0.1$< z <$0.9.  Conversely, only 38\% of the redMaPPer catalogue is matched to an X-ray extended source. Next, using 120 optically clusters and 184 X-ray selected clusters, we investigate the form of the X-ray luminosity-temperature ($L_{X}-T_{X}$), luminosity-richness ($L_{X}-\lambda$) and temperature-richness ($T_{X}-\lambda$) scaling relations.  We find that the fitted forms of the $L_{X}-T_{X}$ relations are consistent between the two selection methods and also with other studies in the literature.  However, we find tentative evidence for a steepening of the slope of the relation for low richness systems in the X-ray selected sample.  When considering the scaling of richness with X-ray properties, we again find consistency in the relations (i.e., $L_{X}-\lambda$ and $T_{X}-\lambda$) between the optical and X-ray selected samples.  This is contrary to previous similar works that find a significant increase in the scatter of the luminosity scaling relation for X-ray selected samples compared to optically selected samples.  
\end{abstract}


%%%%%%%%%%%%%%%%%%%%%%%%%%%%%%%%%%%%%%%%%%%%%%%%%%

%%%%%%%%%%%%%%%%% BODY OF PAPER %%%%%%%%%%%%%%%%%%
\newpage
\clearpage

\section{Introduction}
\begin{figure*}
\centering
 	\includegraphics[width=18cm,clip=true,trim={4.5cm 0cm 4.5cm 0cm}]{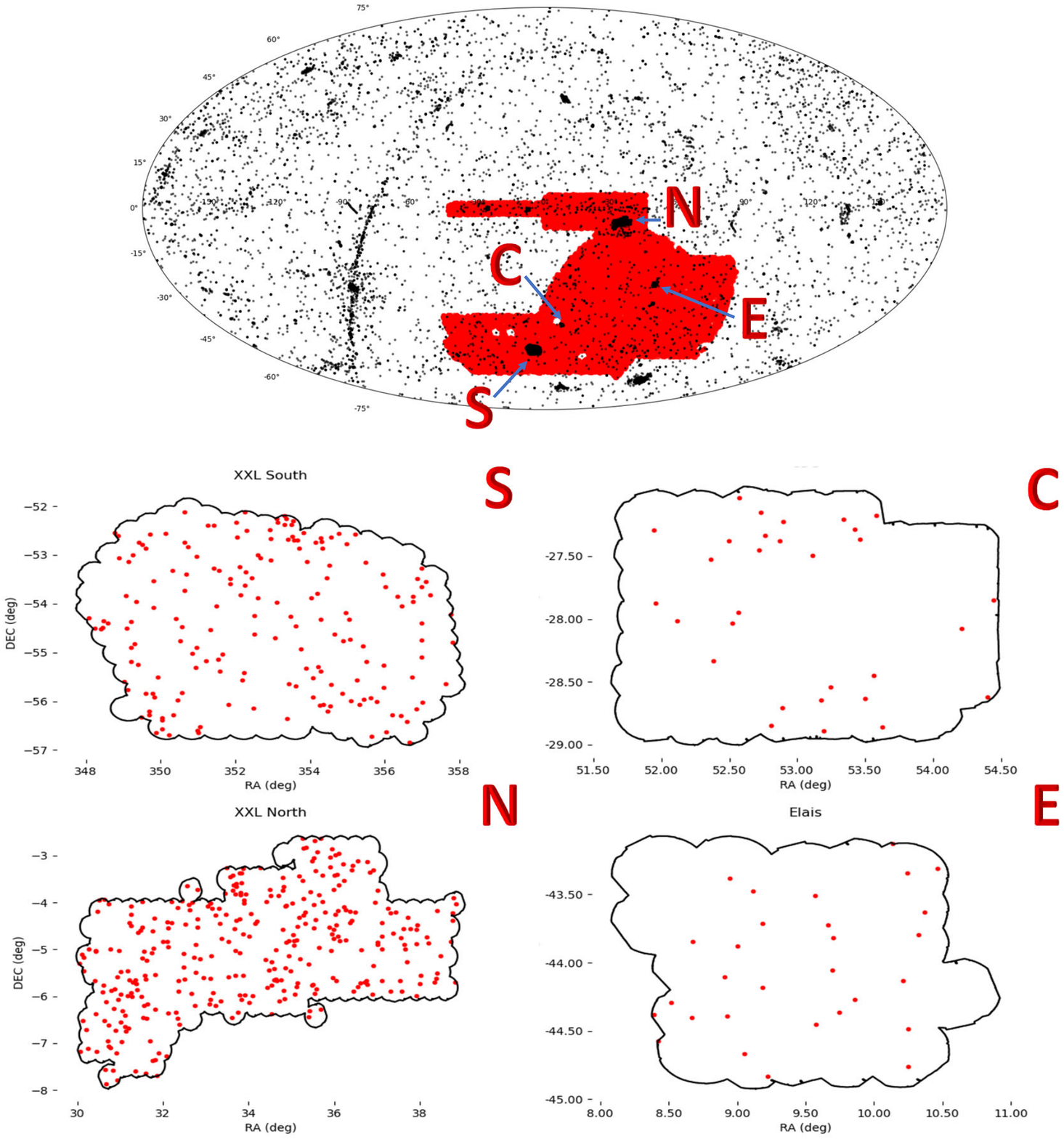}
 	\vspace{1pt}
	\caption{(Top) XMM observations over the whole sky (black points) with the DES footprint highlighted by the red shaded region. We highlight the survey regions used in this work, corresponding to XXL-North(N), XXL-South (S), ELAIS (E) and CDS (C). (Bottom) Outlines of the footprints covered for each of the four corresponding contiguous regions, with the red points indicating the location of redMaPPer clusters in the DES catalogue (with $\lambda>20$ and $0.1<z<0.9$, see Sect.~\ref{sec:optical_sample}). Note: the 4 subplots are not to equal scale: N:\,27\,deg$^2$; S:\,23.7\,deg$^2$; C:\,4.6\,deg$^2$; E:\,2.1\,deg$^2$}.
	\label{fig:footprint}
\end{figure*}

\begin{table*}
	\setlength{\tabcolsep}{0.4em}
\centering
	\begin{tabular}{|l|l|c|c|c|c|c}
	\hline
		Region &   & Area & RM Clusters & XMM & RM Density  & Median X-Ray\\
		 &   & (deg$^{2}$) & ($\lambda>$20) & Observations & (clusters deg$^{-2}$) & Exposure Time (ks)  \\
		\hline
		N-XXL\textsubscript{North} &   & 27 & 337 & 375 & 11.5 & 10.8\\
		S-XXL\textsubscript{South} &   & 23.7 & 180 & 276 & 7.6 & 8.7\\
		C-CDS &   & 4.6 & 30 & 107 & 6.5 & 30\\
		E-ELAIS &   & 2.1 & 29 & 37 & 9.5 & 28\\
		\hline
	\end{tabular}

	\caption{The four contiguous regions comprised of {\em XMM} observations falling within the DES footprint (see Fig.~\ref{fig:footprint}).  Median exposure time is given for all clusters at the redMaPPer location.}
    \label{tab:regions}
\end{table*}

Galaxy clusters are the most massive collapsed objects in the Universe, forming at the intersections of large scale structure filaments and provide an ideal laboratory for cosmological studies.  The formation of large scale structure is predicated on the gravitational collapse of primordial density fluctuations and, therefore, the halo mass function (as measured by the number of clusters of mass M per unit comoving volume) is sensitive to cosmological models \citep[e.g.,][]{2009ApJ...692.1060V}.  Probing number counts and mass can therefore place constraints on cosmology and this is particularly powerful when used in complement with other cosmological markers, such as the angular power spectrum of the Cosmic Microwave Background \citep[e.g.,][]{spt_cosmo,2016A&A...594A..24P} or Baryonic Acoustic Oscillations \citep[e.g.,][]{2016PhRvD..93b3530C}.

In \citet[][hereafter A20]{2020PhRvD.102b3509A}, the Dark Energy Survey \citep[hereafter DES,][]{2016MNRAS.460.1270D} collaboration released cluster cosmology results, estimated using the number density of clusters and a stacked weak lensing mass calibration.  These results highlighted a significant tension between the calculated values of cosmological parameters (namely $\Omega_{\rm m}$ and ${\sigma_8}$) versus those produced by other surveys, including other DES analyses.  The tensions in the $\sigma_{8}$-$\Omega_m$ plane were 1.1$\sigma$ vs SPT-2500 \citep{Bocquet_2019}, 1.7$\sigma$ vs Weighing the Giants \citep{10.1093/mnras/stt2129}, 2.4$\sigma$ vs DES 3x2pt \citep{PhysRevD.98.043526} and 5.6$\sigma$ vs Planck \citep{2016A&A...594A..24P}.  A20 suggests that the tension is most likely explained by a failure in the understanding of the optical selection on the weak-lensing signal such as biases due to cluster orientation and projection effects \citep[e.g.][]{2022MNRAS.515.4471W}. A20 notes that the tension is reduced if lower mass clusters with a richness $(\lambda)<30$ are removed from the sample.  Richness is redMaPPer's (see section \ref{subsec:surveys_catalogues} for an introduction to redMaPPer) probabilistic measure of the number of galaxies in a cluster and is the main optical observable in the DES cluster analysis.  The A20 authors point out that had the analysis been performed only on the higher mass systems, this systematic tension would not have been discovered.

It is therefore of critical importance for inferring cosmology to understand the differences at these lower richnesses/masses, either their physical properties or unknown systematics in the modelling of these systems.
We therefore have two primary considerations:\\ 
\begin{adjustwidth}{0.25cm}{}
\begin{enumerate}[label=\arabic*.,listparindent=1.5em]
    \item the completeness and purity of the catalogue used for number count  analysis, particularly at lower masses
    \begin{itemize}
        \item[] The completeness and purity of the RM sample has been confirmed at $\lambda>40$ using the SPT galaxy cluster sample  \citep{2021MNRAS.504.1253G}. In this work we are able to use X-ray surveys and cross-matching to the DESY3 redMaPPer optical catalogue (see Section \ref{subsec:surveys_catalogues}) to probe the lower lambda redMaPPer systems.  
    \end{itemize}
    \item the mass and scatter of lower mass halos
    \begin{itemize}
        \item[] Measuring masses for individual galaxy clusters directly is inherently difficult and expensive (especially over relatively shallow survey regions) and so A20 used stacked weak-lensing data \cite[see][]{10.1093/mnras/sty2711}.  However, in using stacked data, information about the scatter of the observable versus mass is lost.  The bottom up hierarchical formation model allows us to relate mass to more readily observable properties such as X-ray temperature and luminosity via simple power law relations. Further, cosmological hydrodynamical simulations using first principles suggest the power-law slope and scatter may be scale dependant {\citep[e.g.,][]{2017MNRAS.466.4442L,2018MNRAS.478.2618F,2020MNRAS.495..686A}}. By considering the scatter and evolution of these scaling relations, we can re-introduce the lost scatter into the mass-calibrations and cosmological models can be constrained. 
    \end{itemize}
\end{enumerate}
\end{adjustwidth}
Currently, there are no all-sky X-ray surveys to significant and consistent depth, making cross-correlation analysis between X-ray and optical studies difficult, particularly at higher redshifts and lower richnesses. We note that, in the future, eROSITA \citep{Predehl_2021}, which took first light in 2019, will create and release the deepest, most detailed X-ray all-sky survey ever made having 30–50 times the sensitivity of the previous all-sky X-ray survey by ROSAT. The few studies that consider the cross-correlation between optical and X-ray surveys, either  suffer from small overlapping contiguous areas \citep[e.g.,][$<1$~deg\textsuperscript{2}]{Connelly_2012}, use cross-matches from catalogues in non-contiguous regions \citep[e.g.,][]{2019MNRAS.490.3341F,2022MNRAS.516.3878G} or use targeted X-ray follow-up of optically selected samples \citep[e.g.,][]{2016A&A...585A.147A}, although again, across a non-contiguous region. The most comparable study to this one in terms of using serendipitous detections overlapping survey areas is \cite{2022MNRAS.511.1227G} which used XXL and GAMA.  However, the sample in \cite{2022MNRAS.511.1227G} is limited in size by the spectroscopic selection to $\sim$30 clusters and the sky coverage is only 14.6\,deg\textsuperscript{2}. 

In this paper we overcome these shortcomings by using four contiguous fields of XMM observations, totalling 57.4 deg$^{2}$, within the DES footprint.  X-ray clusters are found using analysis performed by the XMM Cluster Survey and for the optical dataset, we use the redMaPPer cluster catalogue, derived from three years of DES observations .  Compared to G22a who only used clusters designated as C1 in the XXL catalogue, our sample area is $\approx$4 times larger (57.4$^{\circ}$ vs 14$^{\circ}$) and our cluster sample size is $\approx$11 times larger (341 vs 30).

The outline of the paper is as follows: In \S\ref{sec:samples}, the construction of the samples used is detailed; \S\ref{sec:completeness} outlines the overlaps of the optically selected and X-ray selected samples and explains the differences; \S\ref{sec:data_extract} describes the methods used for recovering the X-ray observables and the resultant sub-samples used for the scaling relations. \S\ref{sec:Scaling_Relations} presents the scaling relations and the fitted results. In \S\ref{sec:discussion}, we briefly discuss low signal-to-noise clusters and binning on our scaling relation results, compare our results to analogous literature works and consider any implications for samples derived from the Legacy Survey of Space and Time.  Finally, we summarise our conclusions in \S\ref{sec:summary}.  Throughout this paper we assume a cosmology of $\Omega_{M}$=0.3, $\Omega_{\Lambda}$=0.7 and $H_{0}$=70 km s$^{-1}$ Mpc$^{-1}$. 

\section{Samples}
\label{sec:samples}
In this section we describe how the optical and X-ray data were selected and combined to form the basis of the samples used throughout this work.

\subsection{The Surveys and Catalogues}
\label{subsec:surveys_catalogues}

The optical data were taken from the DES, an optical survey covering approximately 5,000 square degrees of the Southern sky (the DES footprint is highlighted in Figure \ref{fig:footprint} (top), given by the red shaded region).  Observations were made using a 570 megapixel camera, DECam \citep{2015AJ....150..150F}, made up of 62 2048$\times$4096 CCDs and 12 2048$\times$2048 CCDs, mounted on the 4m \textit{Blanco} telescope at the Cerro Tololo Inter-American Observatory in Chile.  Specifically, we make use of the cluster catalogue generated by the red-sequence Matched-filter Probabilistic Percolation cluster finder algorithm\footnote{version 6.4.22+2\_lgt5\_vl50} \citep[hereafter redMaPPer,][]{Rykoff_2014,2016ApJS..224....1R}, run on the DES Year 1 - Year 3 data \citep[hereafter DESY3,][]{2021ApJS..254...24S}.  redMaPPer iteratively calculates photometric redshifts for probable clusters by self-training the red sequence model and assigning a characteristic richness ($\lambda$) based on the sum of the probabilities of membership for all galaxies within a scale radius, R$_{\lambda}$, where R$_{\lambda}=1.0h^{-1}{\rm Mpc}(\lambda / 100)^{0.2}$.

The XMM Cluster Survey \citep[hereafter XCS,][]{1999astro.ph.11499R} is a serendipitous survey of {\em XMM-Newton} observations (see Figure~\ref{fig:footprint}, top-plot, black points) with the primary aim of detecting galaxy clusters.  XCS pipelines process and clean all publicly available observations from the XMM Science Archive
\citep{https://doi.org/10.48550/arxiv.astro-ph/0206412}, with the ultimate aim of finding galaxy clusters.  

\subsection{The Sky Regions Used}

In this work, we make use of four large contiguous fields that have complete X-ray coverage from {\em XMM}, within the DES footprint. The four regions are shown in Figure \ref{fig:footprint}.  The two larger contiguous regions are the XXL\textsubscript{North} and XXL\textsubscript{South} regions (denoted as N and S in Figure~\ref{fig:footprint} respectively) that form the basis of the XXL Survey \citep{2016A&A...592A...1P}. The two smaller regions, CDS and ELAIS (denoted C and E in Figure~\ref{fig:footprint} respectively), are part of the extended SERVS survey \citep{Mauduit_2012}.  Within the outline of the footprints in Figure~\ref{fig:footprint}, the location of redMaPPer cluster detections with $\lambda> $20 are given by the red points (see Sect.~\ref{sec:optical_sample}).  Details of the regions are outlined in Table \ref{tab:regions}.  In total, the regions constitute 57.4~deg$^{2}$ of contiguous overlap between the DES and XCS observations although it is noted that the median X-ray exposure times at the redMaPPer cluster locations within the survey regions is 3 times greater in CDS and ELAIS than in XXL. 

\subsection{The Optical Sample}
\label{sec:optical_sample}

Within the 57.4 deg$^{2}$ of the four contiguous regions, the full redMaPPer catalogue ($\lambda>5$) contains 9,792 entries. We designate this sample set as RM\textsubscript{all} and use this sample when considering the completeness of the X-ray selected sample (see Section \ref{sec:completeness}). For the purpose of creating the optical sample used to derive scaling relations (see Section~\ref{sec:data_extract}), we cut this catalogue by both richness and redshift.  First, we set a minimum $\lambda$ limit of 20 for two reasons:
\begin{enumerate}
\item to be consistent with the DES cluster cosmology analysis \citep{PhysRevD.98.043526};
\item it is likely that only a small fraction of $\lambda <$20 clusters would be detected in our current X-ray data, leading to a large amount of incompleteness (especially at high redshifts, see Sect.~\ref{sec:completeness}).
\end{enumerate}

This results in a sub-sample of 576 cluster entries. We then make a further cut to confine the redshift range between 0.1 < $z$ < 0.9. This leaves a final redMaPPer optical candidate list of 469 potential clusters, which we designate as RM\textsubscript{cut}. 

We then use XCS's image processing suite OCTAVIUS (\textbf{O}bject \textbf{C}lassification \textbf{T}ools for \textbf{A}stronomy Images and \textbf{VI}s\textbf{U}ali\textbf{S}ation) to confirm, or otherwise, the presence of an XCS extended source in the corresponding XMM observation.  The process is similar to that as undertaken in \citet[][hereafter G22b]{2022MNRAS.516.3878G}.  Briefly, we matched each redMaPPer cluster to its nearest XCS X-ray counterpart within 2 $h$\textsuperscript{-1} Mpc (based on the redMaPPer redshift). This was chosen to encapsulate the entire range of mis-centering between redMaPPer and {\sc xapa} centroids \citep[see][]{2019MNRAS.487.2578Z}.  Each potential match is visually inspected to confirm whether the XCS extended source is likely physically associated with the redMaPPer cluster in question (see Appendix \ref{app:Appendix_A} for examples). After visual inspection, 178 redMaPPer clusters are retained for having a viable counterpart in the XCS catalogue.  We designate the sample of 178 confirmed redMaPPer clusters as RM\textsubscript{XCS}.  The remaining 291 redMaPPer clusters are unmatched to an X-ray extended source.  These are discussed further in Section \ref{sec:contamination}.

\subsection{The X-ray Sample}
\label{sec:xray_data}

The original X-ray data reduction process is fully described in \citet[][hereafter LD11]{2011MNRAS.418...14L}, with  updates described by G22b.  Briefly, the data were processed using XMM-SAS version 14.0.0, and events lists generated using the {\sc EPCHAIN} and {\sc EMCHAIN} tools.  Periods of high background levels and particle contamination were excluded using an iterative 3$\sigma$ clipping process performed on the light curves with time bins falling outside this range excluded.  Single camera (i.e. PN, MOS1 and MOS2) images and exposure maps were then generated from the cleaned events files, spatially binned with a pixel size of 4.35$^{\prime\prime}$.  The images and exposure maps were extracted in the 0.5 -- 2.0 keV band, with individual camera images and exposure maps merged to create a single image per XMM observation.  

Following reduction, the resultant images are  run through the XCS source detection routine, the XCS Automated Pipeline Algorithm ({\sc xapa}, see LD11), based upon a bespoke {\sc wavdetect} \citep{2002ApJS..138..185F} analysis, to detect both point-like and extended sources. {\sc xapa} collates unique entries into a Master Source List (MSL) which is our starting X-ray detection catalogue.  Over the 57.4 deg$^{2}$ of the four contiguous regions, there are 25,213 sources in the MSL of which 1,987 are classified as extended sources.  When a large extended source is found across multiple XMM observations, {\sc xapa} can accidentally identify the same cluster twice as distinct objects.  We remove these duplicates leaving 1,972 extended sources, which we designate as the XCS\textsubscript{ext} sample.  Using OCTAVIUS, we visually inspect each X-ray extended source against its corresponding contrast enhanced DES image to confirm the presence of an overabundance of red galaxies (see Appendix \ref{app:Appendix_A} for image examples). Following this visual inspection process, we produce a list of 341 clusters that are X-ray selected and optically confirmed.  We designate this sample XCS\textsubscript{opt}.  Although XCS\textsubscript{opt} represents only 17\% of the 1,972 extended candidates, {\sc xapa} classifies many point spread function sized detections as extended sources when they are often AGN. Hence why all classifications are visually inspected.

\section{Crossmatching the \lowercase{red}M\lowercase{a}PP\lowercase{er} and XCS Samples}
\label{sec:completeness}
In this section, we investigate the overlap of the optical and X-ray selected samples described in Sections~\ref{sec:optical_sample} and \ref{sec:xray_data}. In \ref{sec:xray_completeness} and \ref{sec:opt_comp}, we consider how much of the redMaPPer catalogue is covered by the XCS catalogue and vice-versa. We discuss potential unmatched clusters in Section \ref{sec:contamination}. 

\subsection{X-ray to Optical Matching}
\label{sec:xray_completeness}

\begin{figure}

\centering

 	\includegraphics[width=\linewidth]{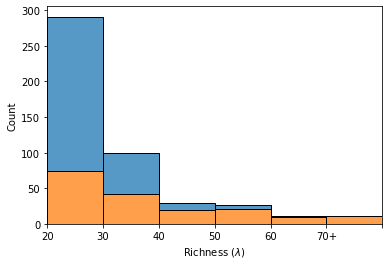}

 	\includegraphics[width=\linewidth]{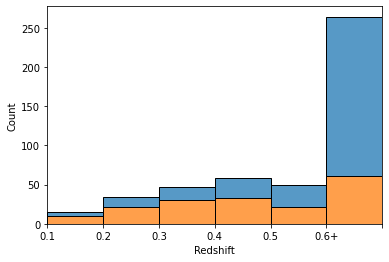}

 	\includegraphics[width=\linewidth]{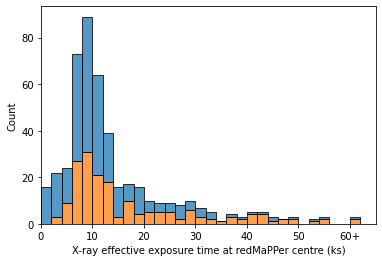}

  \vspace{2pt}
	\caption{Stacked histograms of redMaPPer clusters with (orange) and without (blue) an X-ray counterpart as a function of Richness (Top), Redshift (Middle) and {\em XMM} effective exposure time (Bottom). 
	}
    \label{fig:completeness}
 
\end{figure}

\begin{figure}
\centering
\includegraphics[width=\linewidth]{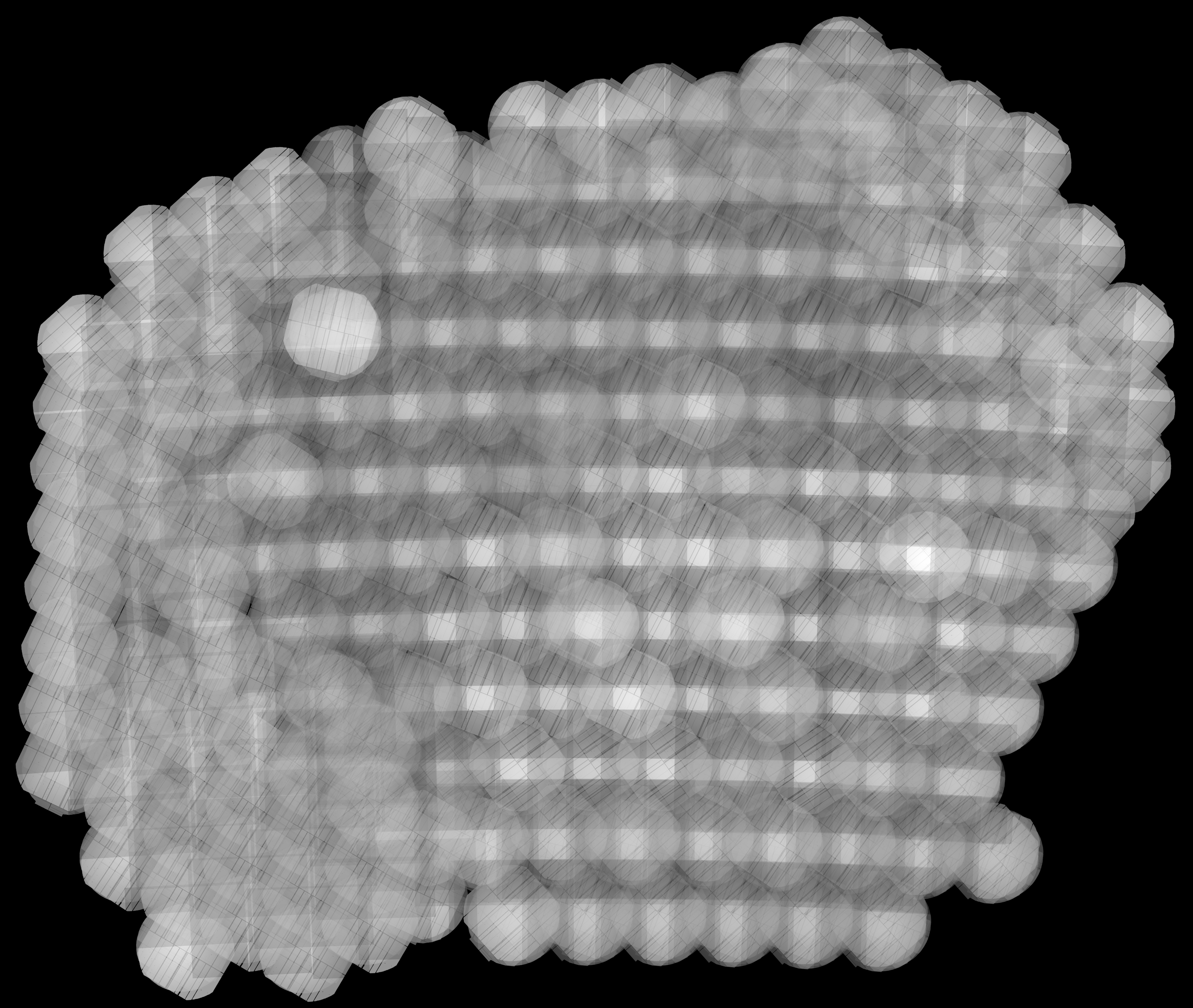}
\caption{Composite mosaic of the exposure maps of  {\em XMM} observations in the XXL\textsubscript{south} region.  Brighter areas show regions with multiple exposures}
\label{fig:xxl_south_exp}
\end{figure}

As shown in Section~\ref{sec:optical_sample}, there are 469 clusters in the RM\textsubscript{cut} sample, but only 178 in the RM\textsubscript{XCS} sub-sample.  We find 38\% of the RM\textsubscript{cut} sample (see Section \ref{sec:optical_sample}) are matched to an X-ray counterpart.  The main driver in finding an X-ray cluster detection is the cluster's X-ray flux (and additionally on the distribution of that flux).  We therefore investigate two properties of a cluster that have a bearing on the cluster flux, namely the richness (as a proxy for mass) and redshift.  Additionally, we investigate the effective exposure time of the observation. Effective exposure time is calculated by adjusting raw exposure time with correction factors including telescope's effective area, field of view and background radiation to accurately measure the amount of X-ray photons collected and decreases as a function of off-axis position on the {\em XMM} detector.  Since the clusters can fall anywhere on the detector due to the survey nature of the observations, the effective exposure time is a property of interest to explore.  One can assume that in most cases higher richnesses, lower redshifts and greater exposure times will all increase the likelihood of detection.  Figure~\ref{fig:completeness} highlights these distributions through histograms of richness (top plot), redshift (middle plot) and exposure time (bottom plot).  We note the wide range of exposure times despite these being survey regions which is due to 3 reasons: 1/ although smaller than the XXL regions, the CDS and ELAIS regions are about 3 times deeper (based on median exposures, see Table \ref{tab:regions}), 2/ the off-axis location of the serendipitously detected clusters and 3/ there have been a number of specific deep observations outside of the original surveys as illustrated by the example mosaic exposure map of XXL\textsubscript{south} shown in Figure \ref{fig:xxl_south_exp}.


As expected, the trends clearly show that as a function of decreasing richness, increasing redshift and decreasing exposure time, the crossmatch success of the sample is reduced.  95\% of redMaPPer clusters with a $\lambda$ above 60 are matched to an X-ray source (22 clusters) and all redMaPPer clusters with $\lambda>70$ are recovered, although we note this complete sample is small (only 11 clusters). In contrast, for $\lambda<30$, only 25\% of redMaPPer clusters have a corresponding X-ray detection.

\begin{figure}
    \centering
    \includegraphics[width=\linewidth]{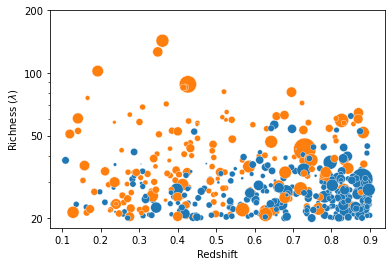}
    \caption{(Redshift vs Richness scatter distribution for confirmed (orange) and unconfirmed (blue) clusters. Size of bubble relates to observation exposure time - larger=longer}
    \label{fig:bubble}
\end{figure}

We explore further whether these properties are intrinsically linked in terms of the likelihood of detection.  As shown in Figure \ref{fig:bubble}, there is a large number of high redshift clusters with short exposure observations at these lower richnesses.  This observation correlates with the expectation from the cluster mass function that we should see a greater number of lower mass clusters but our current X-ray surveys are not sensitive enough to detect them across the full DES redMaPPer redshift range.

\subsection{Optical to X-ray Matching}
\label{sec:opt_comp}

While we have shown that the X-ray selected sample does indeed suffer from a low level of matches with the optical sample (especially at low richnesses and high redshifts), it is also necessary to confirm whether the redMaPPer catalogue detects all known X-ray clusters.  To do this we consider the X-ray sample XCS\textsubscript{opt} (see Section~\ref{sec:xray_data}), which is the sample of X-ray extended sources with a visually confirmed overabundance of red galaxies in the DES imagery. As there are no redshift or richness values associated with this X-ray data set, we use redMaPPer in ``scanning mode'' to probe them.  In ``scanning mode'', redMaPPer takes the position of the X-ray centroid as a prior and determines the likelihood of there being a cluster at a grid of redshifts within a projected distance of 500$h^{-1}\,$kpc. It then considers the maximum likelihoods and returns a redshift and richness property for each cluster, if one is found.  Of the 341 scanned X-ray clusters, 31 lie in sky regions that have been masked out in DES. Regions are masked out of DES images for reasons such as a bright star or a CCD artefact that renders the area unusable for scientific analysis. redMaPPer returns a mask fraction (MASKFRAC) for each sky location that shows how much of the 500kpc region is affected.  We ignore any entry that has a MASKFRAC greater than 20\% and and these are therefore removed from further analysis.   Of the remaining 310 clusters:
\begin{itemize}
\item Using the default parameters of $\lambda>20$ and $0.1<z<0.9$, the 177 clusters from the RM\textsubscript{XCS} sample are directly matched. (The reason this is not the full 178 clusters from the RM\textsubscript{XCS} sample is because redMaPPer optically detects 2 clusters along the line of sight at 2 different redshifts whereas XCS only catalogues 1 X-ray extended source at this location).    
\item Expanding the parameter space to be unconstrained, we use the \texttt{kdtree} algorithm from Python's \texttt{scikitlearn} module to find closest neighbours between the redMaPPer scanned X-ray centres and redMaPPer catalogue entries that fulfill the following criteria: (1) the distance between the centres is <3 arcminutes; (2) the richness difference is <30\% and (3) the redshift difference is <10\%. Thus, a further 94 clusters are directly matched.
\end{itemize}

This leaves 39 X-ray cluster candidates without a direct match in the redMaPPer catalogue based on the conditions above.  Of these remaining candidates:
\begin{itemize}
\item Likely matches

\begin{itemize}
    \item7 pairs agree on location and redshift but have a $\lambda$ difference greater than 30\% between the scan and the catalogue
    \item3 pairs agree on location and $\lambda$ but have a redshift difference greater than 10\% between the scan and the catalogue
\end{itemize}

    \item Do not match
\begin{itemize}

    \item13 high redshift systems ($z>$0.92) found in the scan are not in the redMaPPer catalogue (see Appendix \ref{app:Appendix_A} for images)
    \item11 sources were not found by the redMaPPer scan, i.e. redMaPPer cannot determine a cluster's presence and therefore returns null values for richness and redshift.  Visual inspection suggests these sources are likely high redshift and thus outside the functioning limits of redMaPPer. (See Appendix \ref{app:Appendix_A} for image examples.)
    \item4 clusters found in the scan are not in the redMaPPer catalogue. They are low richness ($\lambda=5.95, 9.23, 15.26, 13.93$) systems across a wide range of redshifts ($z=0.09$ to $z=0.34$). 

\end{itemize}

\end{itemize}

\begin{figure*}
\centering
 	\includegraphics[width=\linewidth]{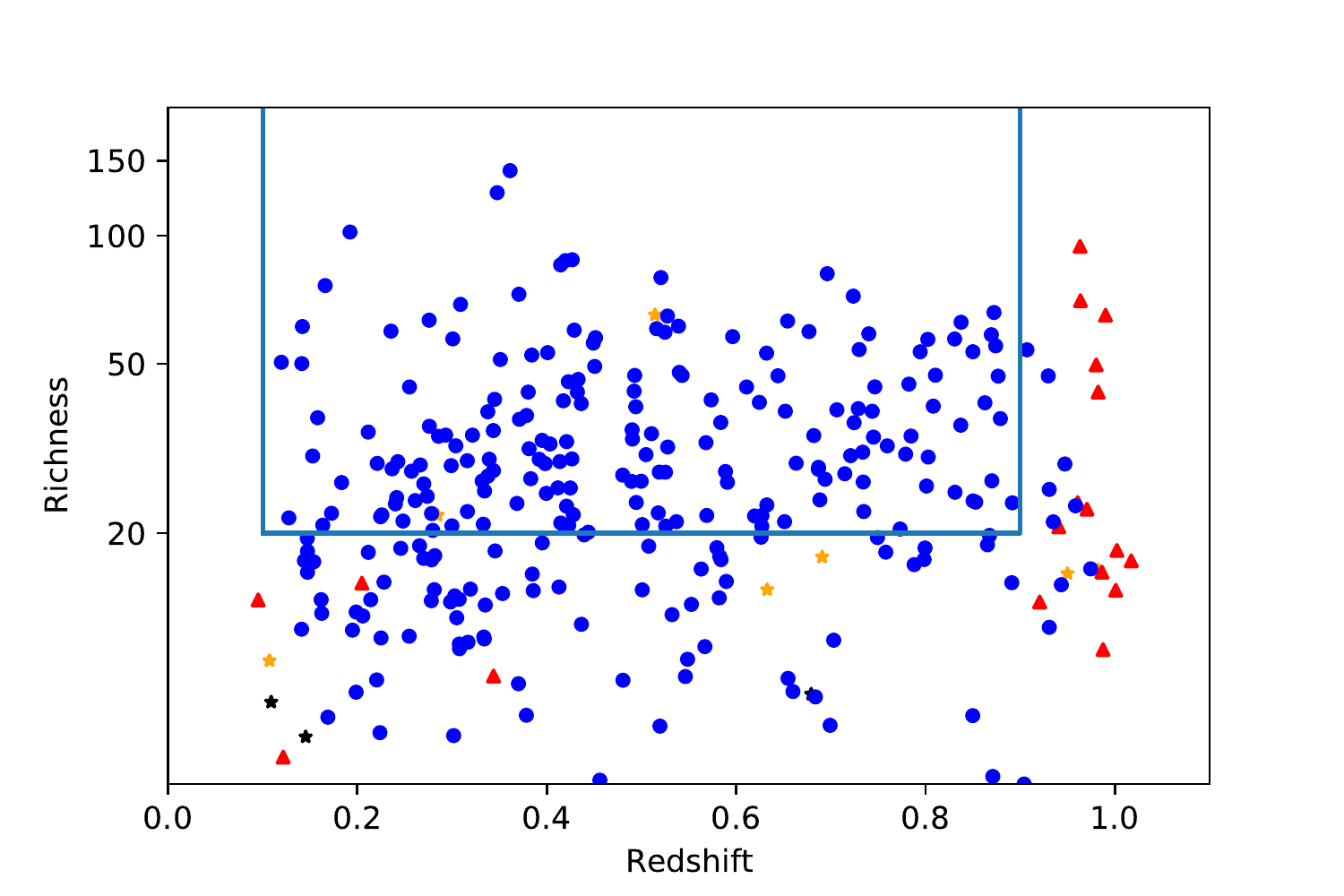} 	
 	\vspace{1pt}
	\caption{The distribution of redshift and richness of the 310 clusters in the X-ray selected cluster sample (see section \ref{sec:opt_comp}). The values were determined using redMaPPer in scanning mode at the position of the {\sc Xapa} centroid. Blue dots also have a counterpart  in the optically selected cluster sample (see section \ref{sec:optical_sample}) with consistent redshift and richness values (i.e. (1): the distance modulus of the centres is <3 arcminutes, (2): the richness modulus is <30\% and (3): the redshift modulus is <10\% ). Orange (Black) stars indicate clusters that also have a counterpart, but the richness (redshift) values  are inconsistent. Red triangles indicate clusters that only appear in the X-ray selected sample. The blue box bounds the parameter constraints placed on the RM\textsubscript{XCS} sample ($\lambda>20$ and 0.1<z<0.9)}
    \label{fig:completeness_scatter}
\end{figure*}
Figure \ref{fig:completeness_scatter} shows these results as scatter points within the redshift/richness parameter space.  The blue box bounds the parameter constraints placed on the RM\textsubscript{XCS} sample and all matched clusters are highlighted in blue.  The unmatched clusters are shown in red and clusters that match on location but with a difference in redshift or richness are shown as black and orange dots respectively. 

We can therefore state that, based on the matching process outlined above, the redMaPPer Y3 catalogue is fully matched to the XCS catalogue above $\lambda>$20 and within $0.1<z<0.9$ (i.e. the redMaPPer catalogue recovers all X-ray clusters within these limits).

\subsection{redMaPPer Clusters Undetected in X-ray Observations}
\label{sec:contamination}

In this section, we investigate further why there are 291 redMaPPer clusters, with $\lambda>20$, undetected by the current X-ray observations (see Section \ref{sec:optical_sample}).  We do this by comparing the required {\em XMM} exposure time needed to achieve a minimum of 20 counts and a signal-to-noise (SNR) ratio of at least 3, versus the actual exposure times of the observations used.  We note that, for this exploratory test, we are only assuming the use of the PN camera (to match estimates of the SNR of the detected cluster, see Section~\ref{sec:snr_cut}).  These SNR and count values were chosen as a cut-off because 85\% of the detected redMaPPer clusters (i.e., the RM$_{\rm XCS}$ sample) have an SNR$>3$ (see Sect.~\ref{sec:snr_cut}) and 85\% had at least 20 PN counts.  However, in running the analysis below, it became clear that, for this sample, the count criteria is dominant as the SNR was always greater than 3. (Hereafter, ``SNR$>3$ and counts$>20$'' are referred to as ``the detection criteria'').  To estimate the required exposure time for each cluster, we use the following process: we estimated the X-ray luminosity based upon the redMaPPer measured $\lambda$, using the best-fit luminosity-$\lambda$ relation for the RM\textsubscript{scaling} sample including Upper Limits presented in Section~\ref{sec:x-ray_lambda_relations} (note for this test we ignore the uncertainties on the relation).  Using the redMaPPer determined redshift, this luminosity is converted into an expected flux.  Then, using the {\sc HEASARC} {\sc PIMMS}\footnote{\url{https://heasarc.gsfc.nasa.gov/docs/software/tools/pimms.html}} software and assuming an {\sc apec} model, with the same parameters we use in Section \ref{sec:x-ray_analysis}, we convert the flux into an {\em XMM} PN count rate assuming the redMaPPer redshift and a temperature estimated from the best fit temperature-richness relation given in Section~\ref{sec:x-ray_lambda_relations} (again using the input $\lambda$ values). To account for the background, we use the existing {\em XMM} observations and determined the background rate within an annulus 1.05-1.5$\,\times\,$r$_{500}$ centred on the redMaPPer centroid in order to be consistent with the X-ray analysis methods used throughout this paper.  Values of $r_{500}$ were estimated assuming Equation~\ref{equ:rdelta}, with the temperature estimated again from the best-fit temperature-$\lambda$ relation. (We were unable to estimate a reliable background for 15 clusters due to their close proximity to the edge of the field-of-view.)  Finally, we estimated the required minimum exposure time (Exp$_{\rm req}$) needed to ``detect'' the cluster whilst fulfilling the above detection criteria.

\begin{figure}
    \centering
    \includegraphics[width=\linewidth]{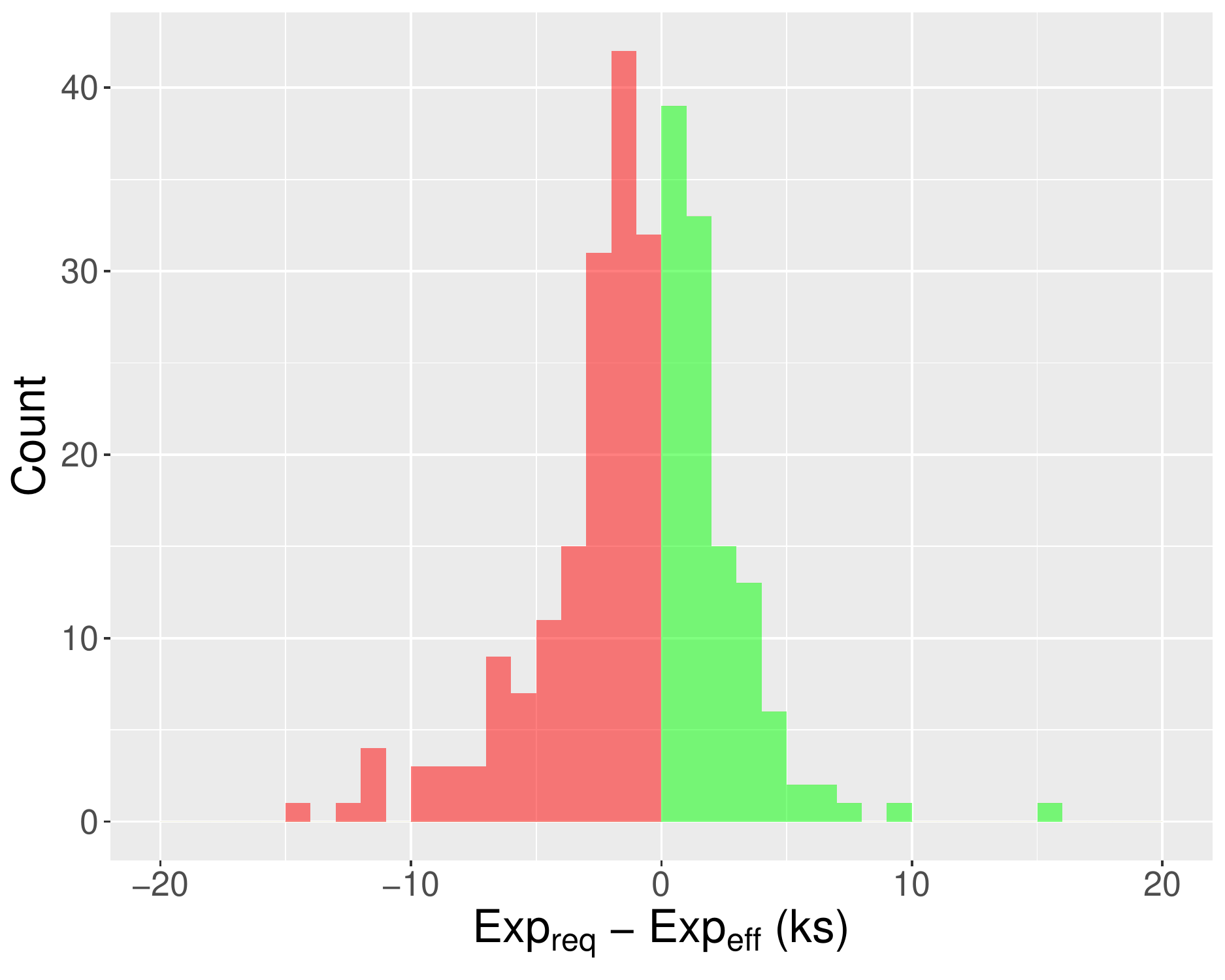}
    \caption{Histogram showing the distribution of how many extra seconds of exposure time would be required to detect a minimum of 20 PN counts with a SNR$>3$ for the 276 redMaPPer clusters with a retrieved background estimate that are not confirmed by X-ray (see Sect.~\ref{sec:contamination}). Red bars to the left of zero indicate the 163 clusters with a current effective exposure time long enough to meet the same detection criteria assuming our best-fit Luminosity-$\lambda$ (including Upper Limits) scaling relation and background estimates.  Green bars to the right show the extra time in ks that would be required to fulfil the same criteria for the remaining 113 clusters based on our best-fit scaling relations and with background estimates taken from the respective {\em XMM} observation.}
    \label{fig:completeness_comparison.png}
\end{figure}

In order to determine whether a cluster is considered ``detectable'', we subtract the estimated required exposure times from the effective exposure time of the {\em XMM} observation, Exp$_{\rm eff}$ (estimated at the location of the redMaPPer cluster and assuming an average exposure within the r$_{500}$ region).  Figure~\ref{fig:completeness_comparison.png} shows the distribution of these time differences, with red bars indicating clusters with observations that should be long enough to detect them (i.e., Exp$_{\rm req}$<Exp$_{\rm eff}$); green bars indicate those clusters with observations that were not long enough to detect the respective clusters. For example, if an estimated exposure time of 15~ks is required to meet the detection criteria but the existing exposure time of the XMM observation at that location was only 10~ks, it would appear in the +5ks bar (as a green bar in the distribution). Based upon this analysis, there are 113 clusters ($\approx$43\%) in Figure~\ref{fig:completeness_comparison.png} where the actual observation exposure times are not sufficient to fulfil the detection criteria, and we would not expect to detect them.  Conversely, there are 163 clusters where the current exposure times should be sufficient to detect a cluster. Therefore, we need to explore whether not detecting these clusters is a concern. 

\begin{figure}

\centering

 	\includegraphics[width=\linewidth]{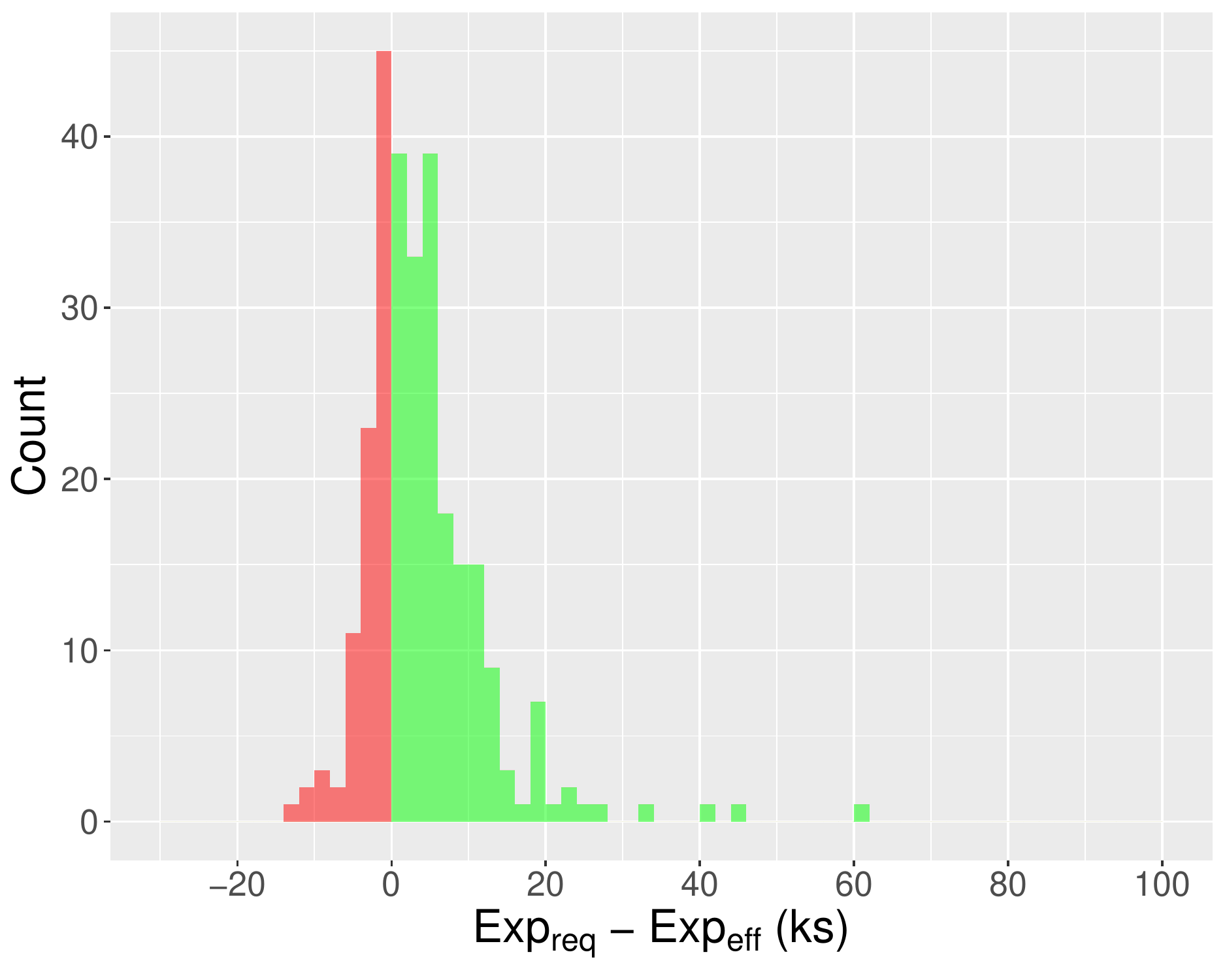}

 	\includegraphics[width=\linewidth]{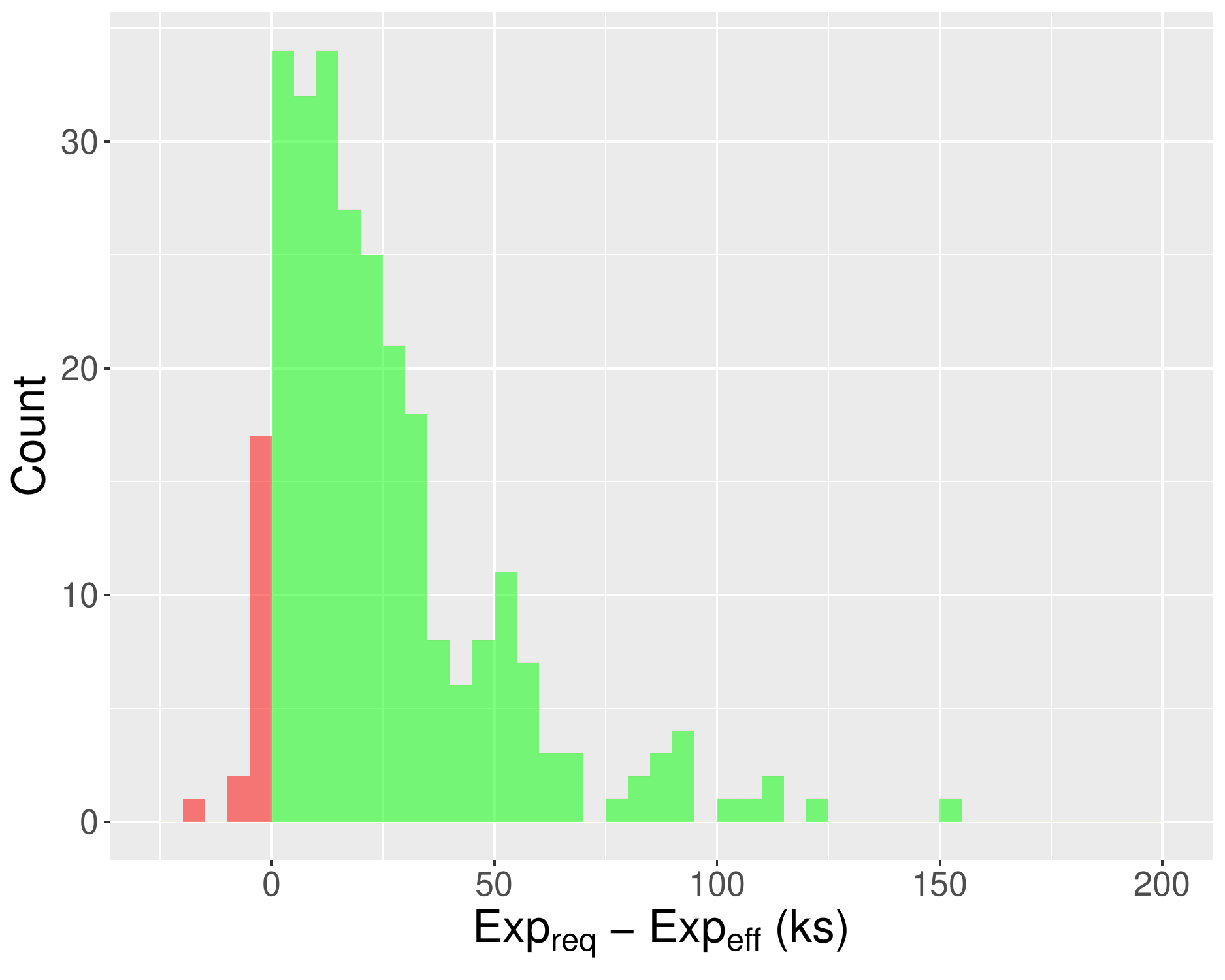}

 	\includegraphics[width=\linewidth]{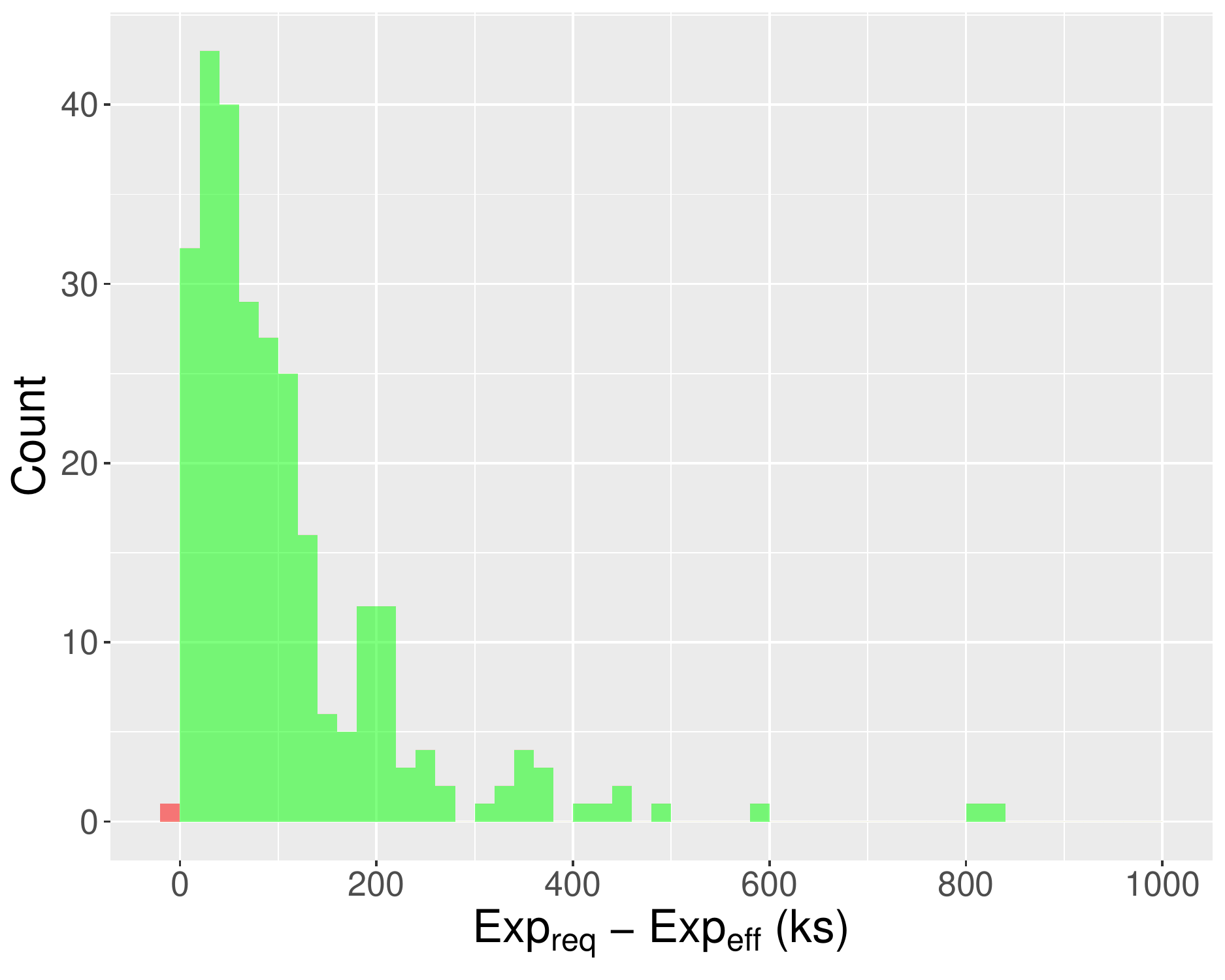}

  \vspace{2pt}
	\caption{As per Figure \ref{fig:completeness_comparison.png} but with the scaling relation best-fit line moved to the 1$\sigma$ (top), 2$\sigma$ (middle) and 3$\sigma$ (bottom) channel boundaries. Using the 1$\sigma$ channel, 88 clusters remain ``detectable'', for the 2$\sigma$ channel, 20 clusters remain ``detectable'' and within the 3$\sigma$ channel only 1 cluster should be ``detectable'' and this is explained in Section \ref{sec:contamination}.
	}
    \label{fig:comp_sig_bands}
 
\end{figure}

One reasonable explanation is that these undetected clusters are less luminous for their given richness than the luminosities estimated from the best-fit scaling relation. This is plausible given the detected clusters used to generate the best-fit luminosity-$\lambda$ relation shows significant scatter around the mean.  Therefore, we shifted the scaling relation best-fit line to the 1$\sigma$, 2$\sigma$ and 3$\sigma$ scatter boundaries and re-ran the above analysis for each to show how many clusters would become undetectable if their luminosity was at each scatter band.  The distributions are illustrated in Figure \ref{fig:comp_sig_bands}.  As shown, moving through the sigma channels, more clusters become undetectable and at the 3$\sigma$ limit all but one cluster are undetectable.  On visual inspection, although the X-ray emission appears to be extended, this cluster is classified as a point source by {\sc XAPA}.  The cluster falls very close to a PN chip gap and visually has a peaked emission profile, possibly leading to the point source classification by {\sc XAPA}.  

Thus, using the method outlined in this section, we have shown it is plausible that all 158 redMaPPer clusters that are deemed ``detectable'' (when assuming the best-fit scaling relation and available X-ray observations) but were not detected in the {\em XMM} X-ray observations become ``undetectable'' within 3$\sigma$ of this best-fit scaling relation. Therefore we should not be concerned at failing to detect these clusters given the observation exposure times available to us.  It should be noted that we cannot rule out the possibility that the redMaPPer richness values may be overestimated which would, via the scaling relation,  infer an higher than actual X-ray luminosity for a cluster.  It is clear that deeper X-ray data is needed for a complete sample of clusters (including the current non-detections), along with a well understood selection function, to truly understand both the completeness and properties of redMaPPer selected clusters.

\section{X-ray analysis}
\label{sec:data_extract}
\subsection{Recovering X-ray temperature and luminosities}
\label{sec:x-ray_analysis}

We use the XCS Post Processing Pipeline (XCS3P) to extract X-ray temperatures and luminosities from XMM observations.  A detailed description of XCS3P can be found in LD11, with recent improvements described in G22b.  A brief overview of the process is detailed below.

Cluster spectra are extracted using the SAS tool {\sc evselect}.  Spectra are extracted within an iteratively determined radius of r$_{500}$\footnote{r$_{500}$ is the radius at which the density of the cluster is 500 times the critical density of the Universe} (see Section 3.1.1 of G22b). Using the relation given in \cite{2005A&A...441..893A}, r$_{500}$ is estimated from the X-ray temperature ($T_{X}$), using the equation:  
\begin{align}
\label{equ:rdelta}
E(z)r_{\Delta} &= B_{\Delta}\left(\frac{T_{\rm X}}{5~{\rm keV}} \right)^{\beta},
\end{align}\\
where $E(z)$=$\sqrt{\Omega_{M}(1 + z)^3 + \Omega_{\Lambda}}$.   For $r_{500}$, $B_{500}$=1104~kpc and $\beta$=0.57.  Iterations are performed until r$_{500}$ converges to within 10\% of the previous iteration, with a minimum requirement of 3 iterations. If no convergence is achieved after 10 iterations, the process is stopped and no X-ray properties are obtained.  

{\sc xspec} \citep{1996ASPC..101...17A} is used to fit the spectrum with an absorbed {\tt APEC} \citep{2001ApJ...556L..91S} model, accounting for the elemental lines in the hot diffuse gas.  The absorption is taken into account with a multiplicative {\tt Tbabs} model \citep{2000ApJ...542..914W}, with the absorption ($nH$) fixed at a value estimated from \cite{2016A&A...594A.116H}, determined at the coordinates of the cluster.  {\sc xspec} fits are performed in the 0.3-7.9~keV band using a fixed abundance of 0.3\,$Z_{\odot}$ \citep[as the typical value for the intracluster medium used in the relevant literature e.g.,][]{2012ARA&A..50..353K} and the redshifts are as per the RM scan.  Note, we do not assign any uncertainties to the redshift as these are insignificant in the fit (the typical error on the redMaPPer photometric redshift is $\approx$1\%).  The {\tt APEC} temperature ($T_{\rm X}$) and normalisation are then free to vary to find the best fit. Finally, luminosities ($L_{\rm X}$) are estimated using the {\sc xspec} LUMIN command in both the bolometric ($L_{X,bol}$) and soft (0.5-2.0~keV) bands ($L_{X,52}$).  While we include the enhancements as detailed in G22b, we note here a further change used in this analysis.  The binning of spectra for use in the {\sc xspec} fits performed above uses the {\tt ftgrouppha} command (as opposed to {\tt grppha} used in XCS3P), with the "optimised binning" parameter\footnote{Following private communication with K.Arnaud, author of {\sc xspec}}.

\subsection{Samples used for recovering temperatures and luminosities}
\subsubsection{The optical sample}

Starting with the RM\textsubscript{XCS} sample (see Sect.~\ref{sec:optical_sample}) and using the process outlined above, XCS3P recovered 135 temperatures and luminosities.  Of the 43 clusters that failed to return $T_{X}$ or $L_{X}$ values, 27 failed during the iteration process before the required minimum 3 iterations completed; a further 16 failed to converge after 10 iterations.  Additionally, we removed clusters where the average $T_{X}$ errors bars were greater than 50\% of the central value (removing a further 12). We also remove 3 clusters that had a coefficient of variation>0.5 where the coefficient of variation is defined as the ratio of the standard deviation to the mean. The final sample used for fitting the scaling relations is 120 clusters, designated as the RM$_{\rm scaling}$ sample.

\subsubsection{The X-ray sample}
\label{sec:LT_xray_sample}

We use the XCS\textsubscript{opt} sample of 341 clusters (Section \ref{sec:xray_data}) but remove the 27 clusters for which redMaPPer was unable to assign a redshift. We therefore pass 314 clusters to the XCS3P pipeline.  Using the process outlined above, XCS3P recovers 239 temperatures and luminosities. For the 75 clusters that failed to return $T_{X}$ or $L_{X}$ values, 59 failed during the iteration process before the required minimum 3 iterations completed; 16 failed to converge after 10 iterations.  Furthermore, 29 clusters had $T_{X}$ errors bars greater than 50\% of the central value and are therefore removed. Another 3 clusters are removed due to extensive variation in the temperature fit as measured by the coefficient of variance.  Furthermore, during the initial eyeballing process, 21 of these clusters were highlighted as being potentially affected by chips gaps, low counts or dominant point sources affecting the {\sc XAPA} region.  Although these clusters ran through the XCS3P process, we are not confident in the temperature or luminosity outputs.  Therefore, we remove these clusters entirely from the scaling relation fit; however, they are retained on the scaling plots for reference circled in red (e.g. Figure \ref{fig:LT_scaling_relations}) and it should be noted that many are not outliers suggesting we have been overly cautious. Finally, we remove the 3 clusters that returned a redshift in scanning mode $>10\%$ than the catalogue value (see section \ref{sec:opt_comp}) but again leave these on the plot highlighted in red.  We designate this final sample of 184 clusters as XCS\textsubscript{scaling}.

For the scaling relations involving richness, we exclude a further 10 clusters from the fit, but again, retain these on the plot for reference highlighted by black circles.  6 of these 10 clusters are because they have a $MASKFRAC>0.20$ and the other 4 are removed from the fit because the richness measured for the X-ray cluster in scanning mode differs from the matched catalogue entry by $>30\%$ (see section \ref{sec:opt_comp}).  We therefore use 174 clusters from the XCS\textsubscript{scaling} sample for scaling relations involving richness.

\subsubsection{Luminosity Upper Limits}
\label{sec:UL}

For the 291 optically detected clusters in RM\textsubscript{cut} with no matched XCS source, we estimate upper limit luminosities using the same methodology as G22b (see Section 3.3 of that paper). Briefly, we estimate an initial $r_{500}$ using a fixed temperature of 3~keV in Equation~\ref{equ:rdelta}.  This is chosen to avoid introducing any bias from estimating temperature from the $T_{\rm X}-\lambda$ relation, given temperature is correlated to luminosity.  We  measure a $3\sigma$ count rate upper limit using the SAS tool {\sc eregionanalyse} and convert the count rate upper limit to a flux.  This is done using an energy conversion factor again assuming a fixed temperature of 3~keV and the redMaPPer estimated redshift.  This flux is then converted to an upper limit. While these upper limits are a simplistic estimate (the initial count rate estimate from {\sc eregionanalyse} does not assume cluster emission), they qualitatively follow a similar distribution to the analysis presented in Section~\ref{sec:contamination} (which presents a more detailed analysis of undetected redMaPPer clusters in our X-ray data).
\section{Scaling relations}
\label{sec:Scaling_Relations}
Here, we present the measured scaling relation between X-ray luminosity and X-ray temperature ($L_{X}$-$T_{X}$) and between the X-ray properties and the optical observable, $\lambda$ ($T_{X}$-$\lambda$ and $L_{X}$-$\lambda$). We assume self-similarity \citep{1986MNRAS.222..323K} and note that the analysis presented herein does not account for selection biases (namely, Eddington and Malmquist) but these will be explored in a future paper when a simulation based XCS selection function is well established. The scaling relations presented should be considered with this in mind.  

\subsection{Fitting the data}
\label{sec:fitting}

We fit the data using the LIRA (\textbf{LI}near \textbf{R}egression in \textbf{A}stronomy, see \cite{2016ascl.soft02006S} for further details on the LIRA) package (in R).  Each scaling relation is fitted with a power-law of the form
\begin{align}
Y = A + B\cdot Z \pm \epsilon
\end{align}
where ${\rm var}(\epsilon)=\sigma^{2}_{Y|Z}$ and $Z$ is the intrinsic cluster property.  For simplicity, the scaling relations are denoted by the cluster properties in question and the scatter given by $\sigma$ (for example, see Equation~\ref{equ:lt}).

For the $L_{X}$-$T_{X}$ relation, we fit the data using a power law relation between $L^{r500}_{X,bol}$ and $T_{X}$, expressed as:
\begin{align}
{\rm log}\left(\frac{L^{r500}_{X,bol}}{E(z)^{\gamma_{LT}}L_{0}}\right) &=
{\rm log}(A_{LT}) + B_{LT}{\rm
log}\left(\frac{T_{X}}{T_{0}}\right) \pm \sigma_{L|T},
\label{equ:lt}
\end{align}
where $A_{LT}$ denotes the normalisation, $B_{LT}$ the slope and
$\sigma_{L|T}$ the intrinsic scatter. $\gamma_{LT}$ is the evolution with redshift and is set equal to 1 as per the self-similar expectation.  Note that the intrinsic scatter is given in natural log space and can be interpreted as the fractional scatter. Normalisation values were set to $L_{0}$=0.7$\times$10$^{44}$~erg~s$^{-1}$ and $T_{0}$=2.5~keV, roughly median values of the samples.
\begin{figure*}
\centering
  \includegraphics[width=\textwidth]{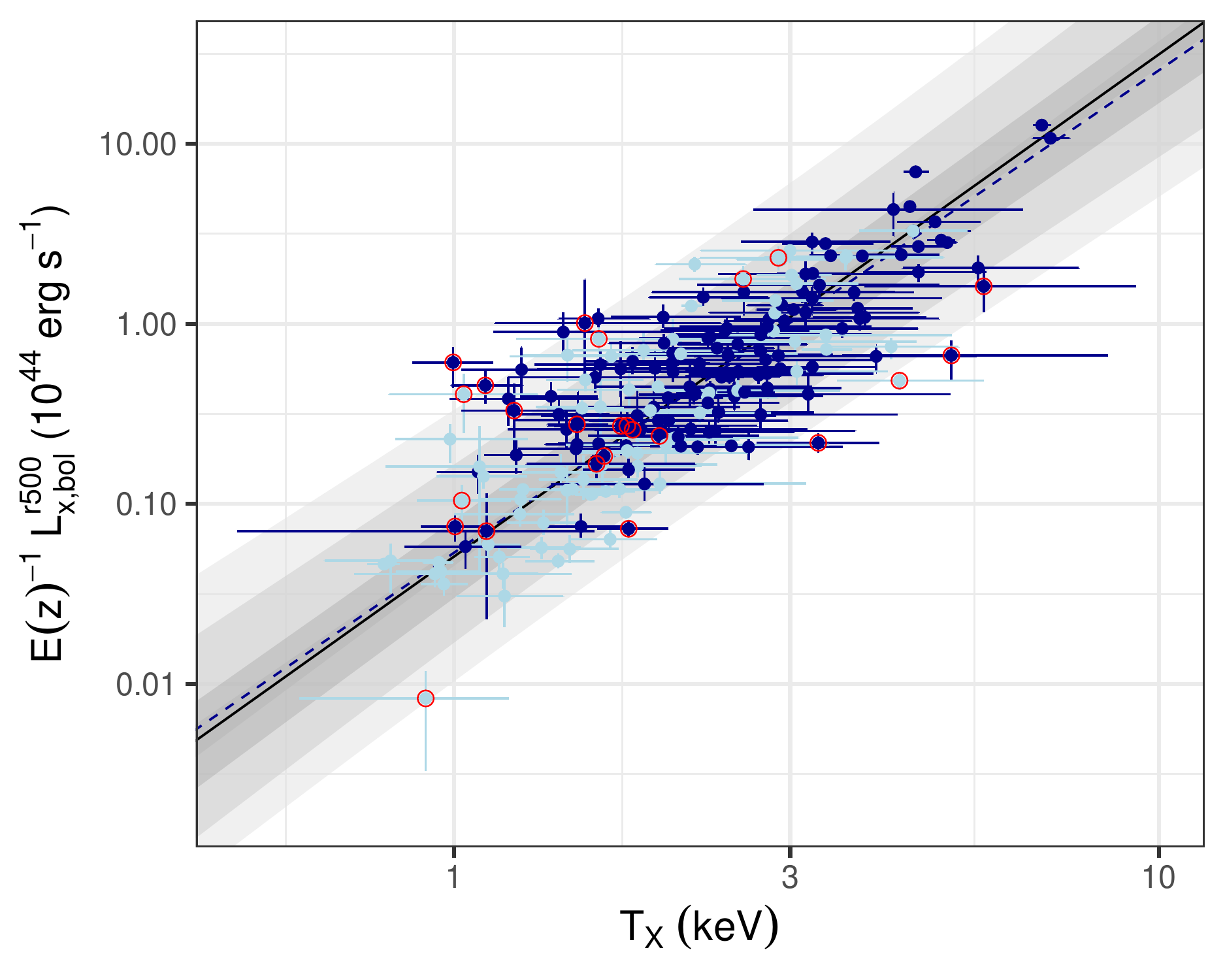}
\begin{tabular}{ccc}

     \includegraphics[width=0.33\textwidth]{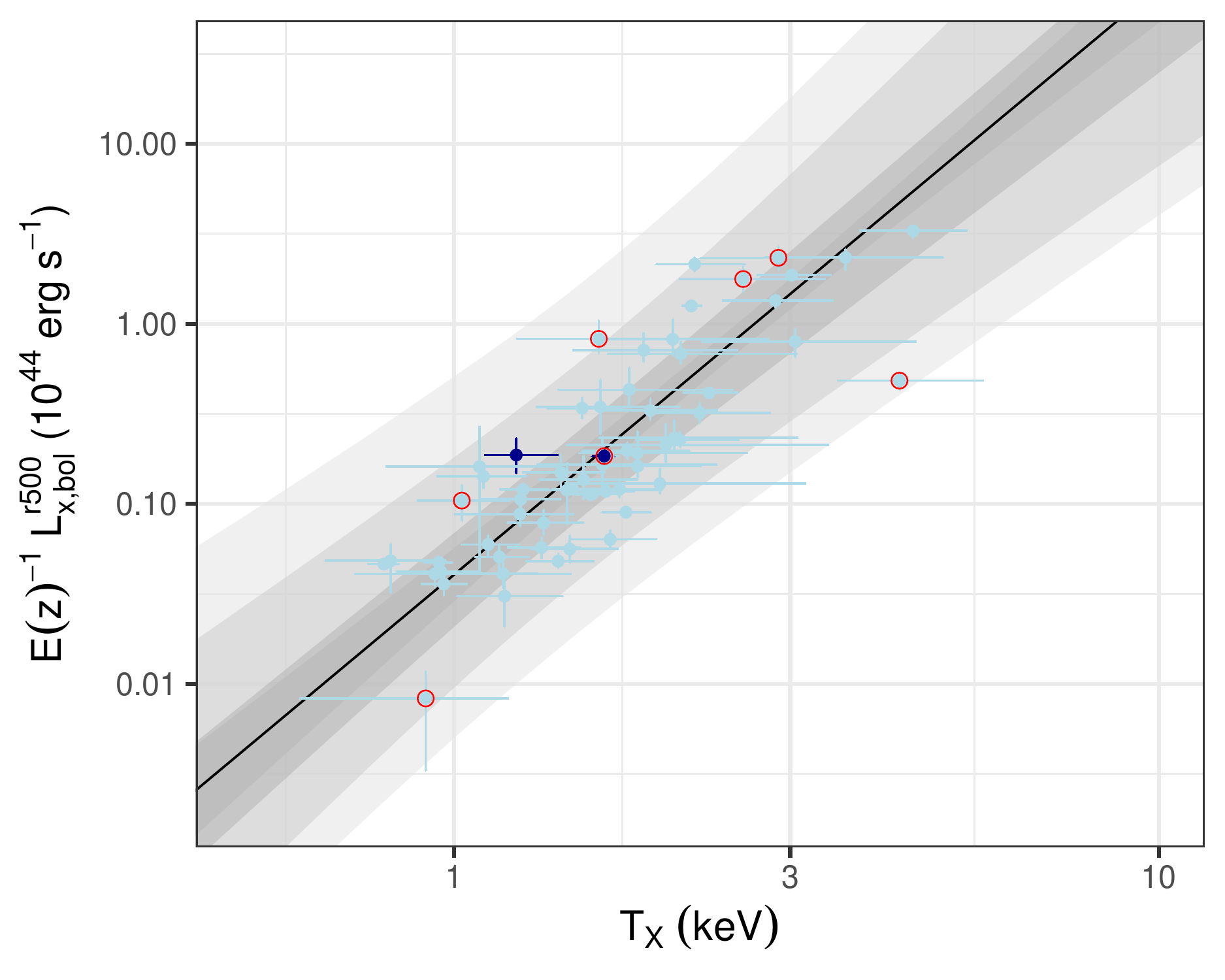} &
     
     \includegraphics[width=0.33\textwidth]{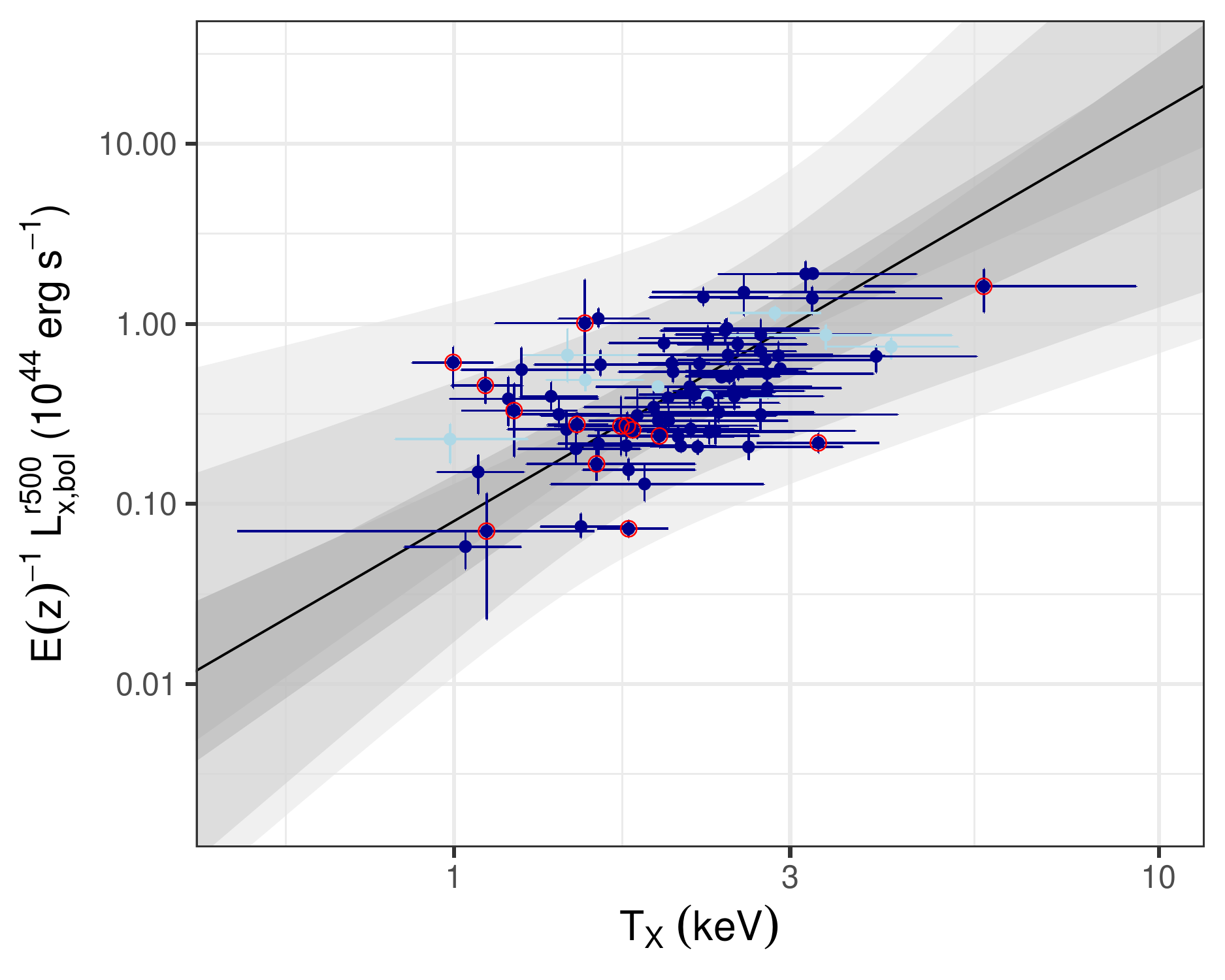} &

  \includegraphics[width=0.33\textwidth]{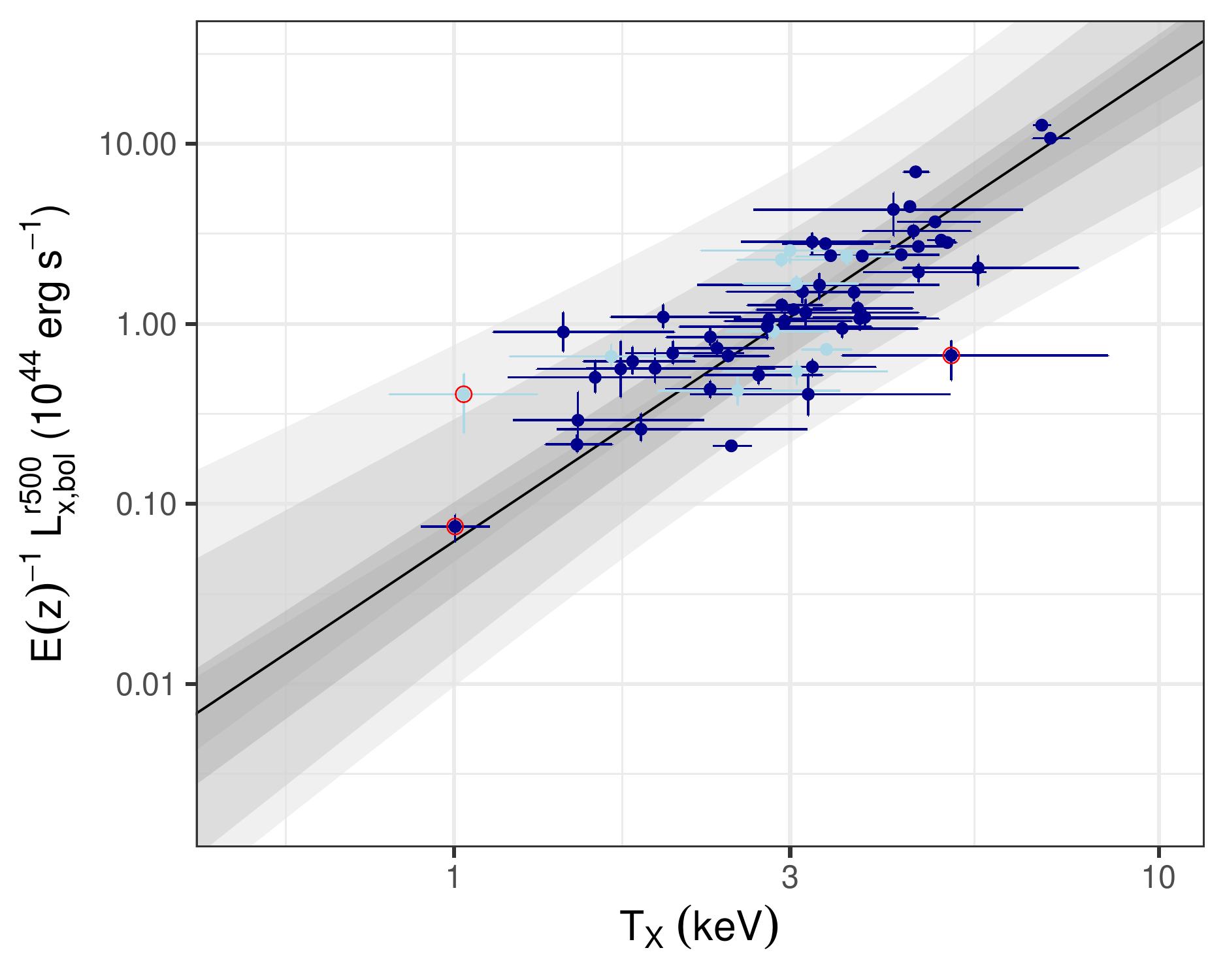}
    \\
    (a): XCS-DES$_{5-20}$ &
    (b): XCS-DES$_{20-40}$ &
    (c): XCS-DES$_{40+}$ \\

\end{tabular}
 	\caption{Luminosity-Temperature ($L_{\rm X}-T_{\rm X}$) relations in different richness bins. These were derived from the XCS\textsubscript{scaling} sample (see section \ref{sec:LT_xray_sample}). Dots in dark (light) blue represent clusters from XCS\textsubscript{scaling} with (without) a match in the optical catalogue (with $\lambda>20$ and $0.1<z<0.9$). The best fit and the corresponding  1, 2, 3 $\sigma$ regions for each relation are shown by the black line and the dark, medium, and light grey regions respectively. Dots circled in red are included for completeness, but were not included in the $L_{\rm X}-T_{\rm X}$ (see section \ref{sec:LT_xray_sample} for details of the fitting procedure).  The dark blue dashed line is the best fit for the RM\textsubscript{XCS} sample. Top: XCS-DES$_{All}$.  Bottom (a):XCS-DES$_{5-20}$. (b) :XCS-DES$_{20-40}$. (c) XCS-DES$_{40+}$.}
    \label{fig:LT_scaling_relations}  

\end{figure*}

\begin{figure*}
 \centering
  \includegraphics[width=0.75\textwidth]{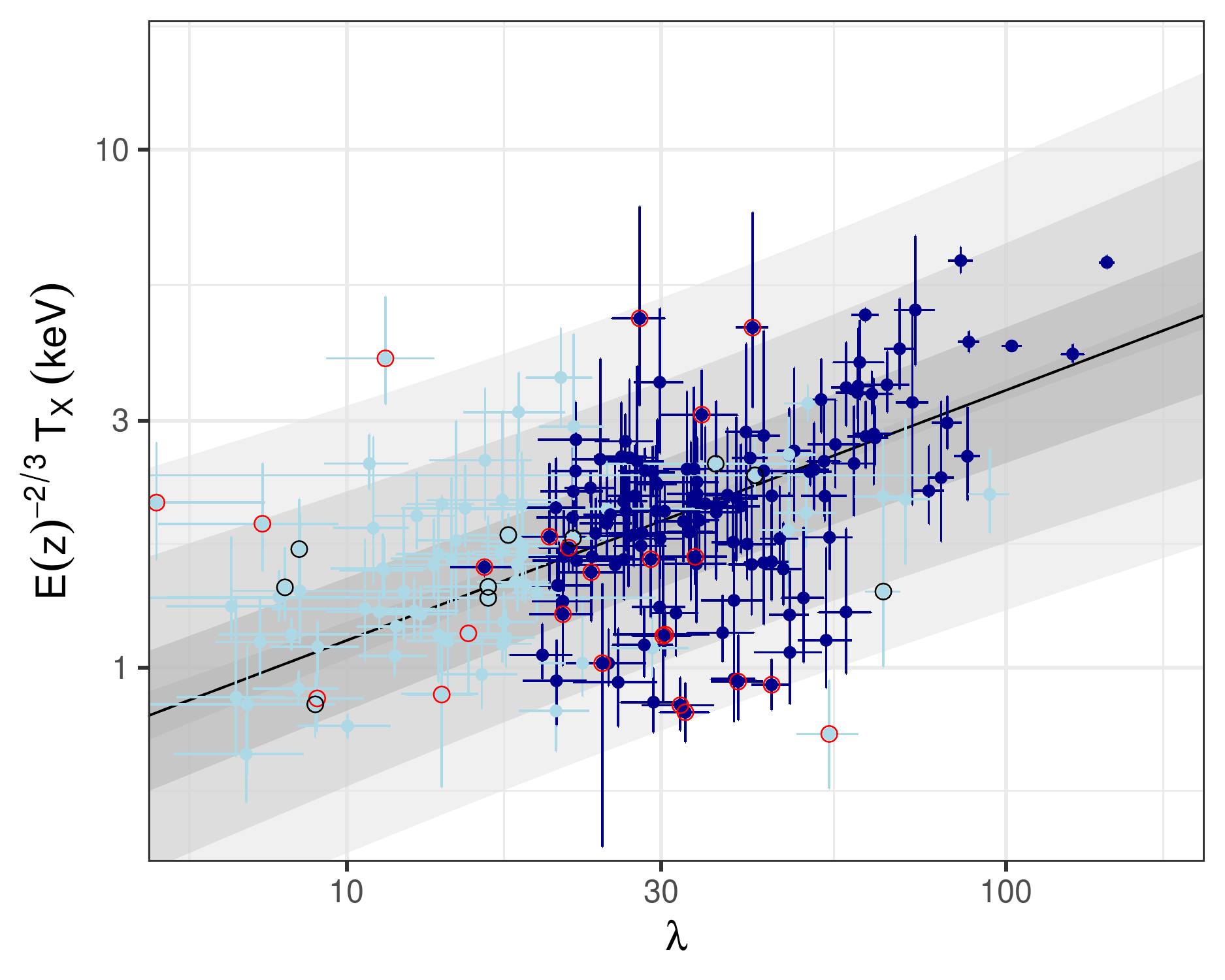}
  	\caption{The Temperature-Richness ($T_{\rm X}-\lambda$) relation of the XCS$_{\rm scaling}$ sample.  Points in dark (light) blue represent clusters from the XCS$_{\rm scaling}$ sample with (without) a match in the optical catalogue (with $\lambda>20$ and $0.1<z<0.9$). The best fit line is given by the black solid line and the corresponding  1, 2, 3 $\sigma$ residual scatters are shown by the dark, medium, and light grey regions respectively.  Points highlighted in red (indicating an unreliable X-ray value) and black (indicating an unreliable richness value), are excluded from the fit (see section \ref{sec:LT_xray_sample} for details of the fitting procedure).}
  	\label{fig:TR_scaling_relation}
\end{figure*}

\begin{figure*}
\centering
\begin{tabular}{cc}
     \includegraphics[width=0.47\textwidth]{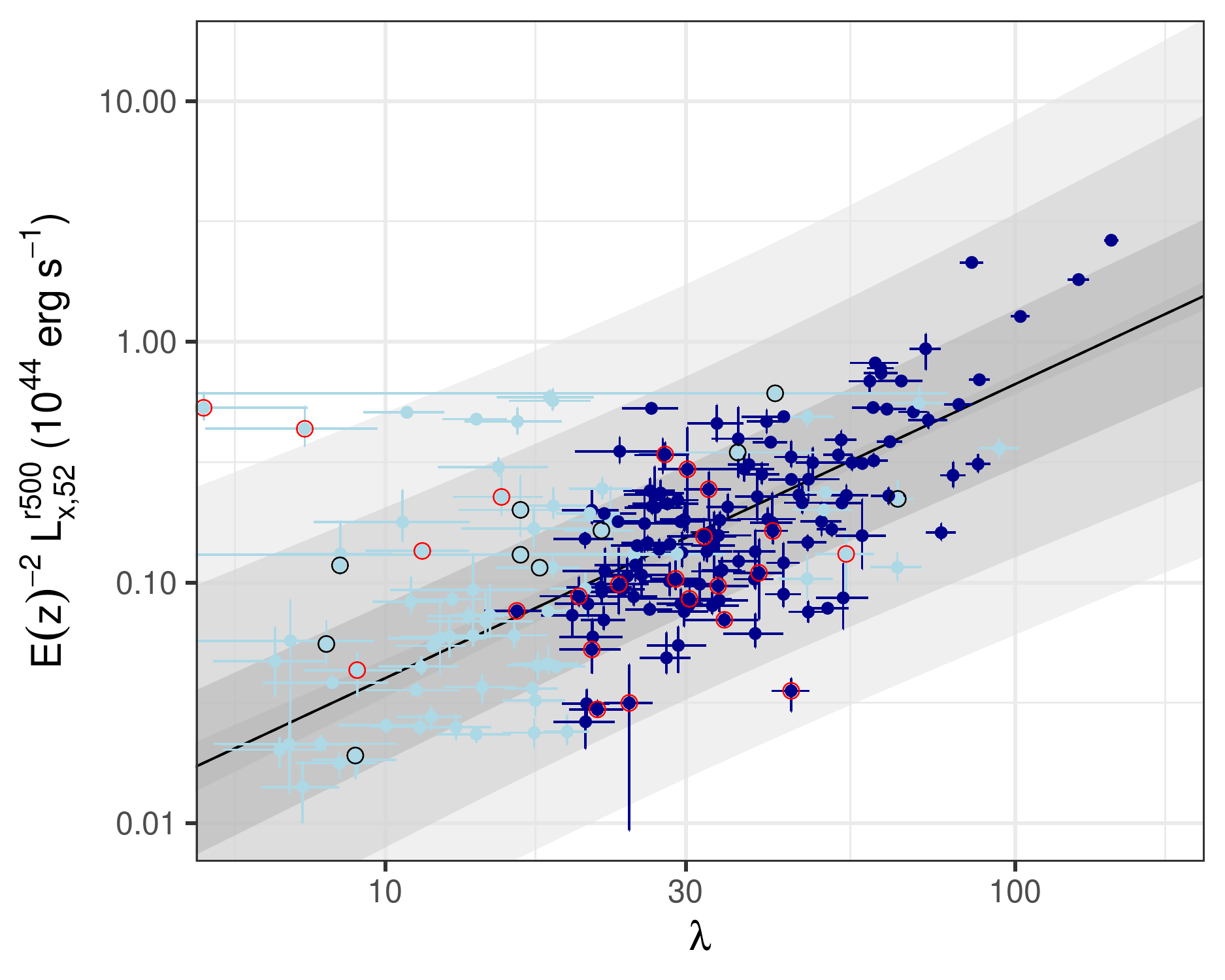}
     \includegraphics[width=0.47\textwidth]{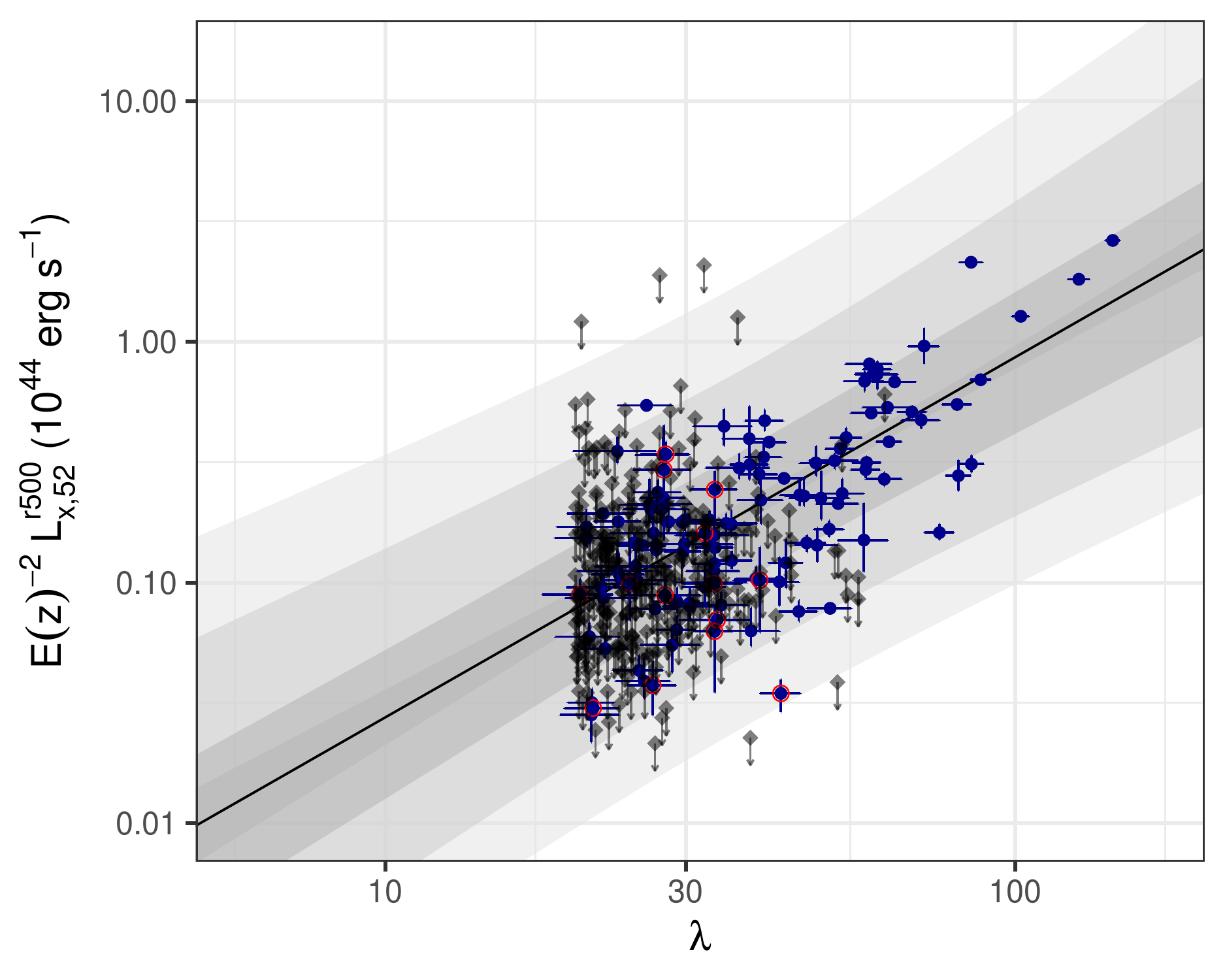}
\end{tabular}

 	\caption{ 	Luminosity-Richness ((Left): $L^{r500}_{X,52}-\lambda$) relation for the XCS$_{\rm scaling}$ sample. (Right): $L^{r500}_{X,52}-\lambda$ relation for the RM$_{\rm scaling}$ sample, including 291 upper limit luminosities (see Section \ref{sec:UL} for details) for undetected clusters (given by the black points with downward arrows for the $L^{r500}_{X,52}$ limit).  See Figure~\ref{fig:TR_scaling_relation} for further caption details.}
    \label{fig:LR_scaling_relations}  
    \end{figure*}

For the $L_{X}$-$\lambda$ relation, we fit the data using a power law relation expressed as:
\begin{align}
{\rm log}\left(\frac{L_{X,52}^{r500}}{E(z)^{\gamma_{L\lambda}} L_{0}}\right) &=  {\rm log}(A_{L\lambda}) + B_{L\lambda}{\rm log}\left(\frac{\lambda_{\rm RM}}{\lambda_{0}}\right) \pm \sigma_{L|\lambda},
\label{equ:l-lambda}
\end{align}
where $A_{L\lambda}$ denotes the normalisation, $B_{L\lambda}$ represents the slope and $\sigma_{L|\lambda}$ denotes the intrinsic scatter (once again the values are given in natural log space). $\gamma_{LT}$ is set equal to 2 as per the self-similar expectation. The values of L$_{0}$ and $\lambda_{0}$ were set as 0.7$\times$10$^{44}$~erg~s$^{-1}$ and 60 respectively.

For the $T_{X}$-$\lambda$ relation, we fit the data using a power law relation expressed as:
\begin{align}
{\rm log}\left(\frac{T_{X}^{r500}}{T_{0}}\right) &= {\rm log}(A_{T\lambda}) + B_{T\lambda}{\rm log}\left(\frac{\lambda_{\rm RM}}{\lambda_{0}}\right) \pm \sigma_{T|\lambda},
\label{equ:t-lambda}
\end{align}
where $A_{T\lambda}$ denotes the normalisation, $B_{T\lambda}$ represents the slope and $\sigma_{T|\lambda}$ and denotes the intrinsic scatter (once again the values are given in natural log space). $\gamma_{LT}$ is set equal to 2/3 as per the self-similar expectation. For a wide range of redshifts, the quantity $E(z)^{-\frac{2}{3}}T$ should be a closer reflection of halo mass, M, than temperature alone \citep{2019MNRAS.490.3341F}. The values of T$_{0}$ and $\lambda_{0}$ were set as 2.5~keV and 60 respectively.

\subsection{Fitted scaling relations}
\label{sec:fitted_scaling_relations}

\subsubsection{The L\textsubscript{x}-T\textsubscript{x} relation}

The $L_{X}$-$T_{X}$ relation for the XCS\textsubscript{scaling} sample is shown in Figure \ref{fig:LT_scaling_relations} (top), given for $L^{r500}_{X,bol}$.  We also consider the changes in the scaling relation as a function of richness using three different richness bins: 5<$\lambda$<20, 20<$\lambda$<40 and $\lambda$>40 (Figure \ref{fig:LT_scaling_relations} bottom left, middle, right respectively).  The best fit for each sample is represented by the solid black line and the grey channels represent 1$\sigma$, 2$\sigma$ and 3$\sigma$ residual scatter.  The best fit line for the optically selected sample (RM\textsubscript{scaling}) is shown by the dashed blue line. Those clusters that are excluded from the fit (see section \ref{sec:LT_xray_sample}) are circled in red.  We have tested the fit inclusive of these clusters and, as they are not significant outliers, their exclusion does not make a significant difference to the fit parameters.  Dark blue points are X-ray clusters with a counterpart in the RM\textsubscript{cut} sample. Light blue points are X-ray clusters without a counterpart in RM\textsubscript{cut} sample.  Best fit parameters for each sub-sample are given in Table \ref{Table:Scaling_Relations_Table}.
To allow easier comparison to other works, Table \ref{Table:Scaling_Relations_Table} also includes fitted values for the 0.5-2.0~keV energy band (i.e., $L^{r500}_{X,52}$).  The slope of the $L^{r500}_{X,bol}$-$T_{X}$ relation for XCS$_{scaling}$ is consistent with previous studies \citep[e.g., ][]{2009A&A...498..361P,2011A&A...535A.105E,2020ApJ...892..102L,2022A&A...661A...7B,2022MNRAS.516.3878G}. However, we find some tentative evidence that the slope of the $L^{r500}_{X,bol}$-$T_{X}$ relation is steeper when limiting the sample to lower richness (i.e. lower mass) clusters.  See section \ref{sec:lamba_disc} for further discussion. We note that the inclusion of the additional 88  clusters with $\lambda<20$ in the full X-ray sample (XCS\textsubscript{scaling}) compared to the optical sample (RM\textsubscript{scaling}) does not significantly alter the $L^{r500}_{X,bol}$-$T_{X}$ relation.

\begin{table*}[t]
\begin{center}

\vspace{1mm}
\begin{tabular}{lccccc}
\hline
\hline
 Relation & Normalisation & Slope & Residual scatter & Cluster Count & Fitted Points\\
(sample) & & & \\

\hline
$L^{r500}_{X,bol}-T^{r500}_{X}$ & $A_{LbT}$ & $B_{LbT}$ & $\sigma_{LbT}$ \\

XCS$_{scaling}$ & 0.94$\pm$0.05 & 2.79$\pm$0.11 & 0.43$\pm$0.02&208&184\\
XCS$_{scaling\_5-20}$ & 1.15$\pm$0.22 & 3.27$\pm$0.31 & 0.49$\pm$0.04&62&55\\
XCS$_{scaling\_20-40}$ & 0.92$\pm$0.1 & 2.27$\pm$0.5 & 0.42$\pm$0.04&86&72\\
XCS$_{scaling\_40+}$ & 0.97$\pm$0.11 & 2.61$\pm$0.26 & 0.38$\pm$0.03&60&57\\
RM$_{scaling}$ & 0.89$\pm$0.06 & 2.59$\pm$0.17 & 0.41$\pm$0.02&120&106\\
\hline
$L^{r500}_{X,52}-T^{r500}_{X}$ & $A_{LbT}$ & $B_{LbT}$ & $\sigma_{LbT}$ \\
XCS$_{scaling}$ & 0.36$\pm$0.02 & 2.41$\pm$0.12 & 0.46$\pm$0.02 & 208 & 184 \\
XCS$_{scaling\_5-20}$ & 0.44$\pm$0.08 & 2.9$\pm$0.31 & 0.51$\pm$0.04&62&55\\
XCS$_{scaling\_20-40}$ & 0.34$\pm$0.04 & 1.71$\pm$0.49 & 0.48$\pm$0.03&86&72\\
XCS$_{scaling\_40+}$ & 0.38$\pm$0.04 & 2.16$\pm$0.25 & 0.4$\pm$0.03&60&57\\
RM$_{scaling}$ & 0.36$\pm$0.02 & 2.19$\pm$0.17 & 0.44$\pm$0.02&120&106\\

\hline

$L^{r500}_{X}-\lambda_{\rm RM}$ & $A_{L\lambda}$ & $B_{L\lambda}$ &
$\sigma_{L\lambda}$ \\
X-ray selected & 1.31$\pm$0.12 & 1.37$\pm$0.1 & 0.74$\pm$0.02 &208&174\\
\hline
$L^{r500}_{X,52}-\lambda_{\rm RM}$ & $A_{L52\lambda}$ & $B_{L52\lambda}$ &
$\sigma_{L52\lambda}$ \\
X-ray selected & 0.51$\pm$0.04 & 1.22$\pm$0.09 & 0.67$\pm$0.02 &208&174\\
Optically selected (inc UL) & 0.57$\pm$0.05 & 1.5$\pm$0.14 & 0.58$\pm$0.02&404&383 \\
Optically selected (exc UL) & 0.56$\pm$0.05 & 1.49$\pm$0.14 & 0.58$\pm$0.02&120&106 \\
\hline
$T^{r500}_{X}-\lambda_{\rm RM}$ & $A_{T\lambda}$ & $B_{T\lambda}$ &
$\sigma_{T\lambda}$ \\
X-ray selected & 1.07$\pm$0.04 & 0.48$\pm$0.04 & 0.26$\pm$0.01 &208&174\\
Optically selected & 1.13$\pm$0.05 & 0.62$\pm$0.07 & 0.23$\pm$0.01&120&106 \\
\hline
\end{tabular}
\end{center}
\caption[]{\small Best-fit parameters for $L_{X}$-$T_{X}$, $L_{X}$-$\lambda$ and $T_{X}$-$\lambda$ scaling relations  given by equations~\ref{equ:lt},~\ref{equ:l-lambda}
and~\ref{equ:t-lambda} respectively (see Sect.~\ref{sec:fitting}).  For each relation, parameters are given for the X-ray Selected ($T_{X,err}<50\%$ and 0.1$\le$z$\le$0.9)
cluster sample. $\gamma$ is set to 1 for all the $L_{X}-T_{X}$ relations, 7/3 for the bolometric $L_{X}$-$\lambda$ relation, 2 for the $L_{X,52}$-$\lambda$ relation and 2/3 for the $T_{X}$-$\lambda$ relation, all as per self-similar expectations.}
\label{tab:bestfit}
\label{Table:Scaling_Relations_Table}
\end{table*}

\subsubsection{The T\textsubscript{x}-$\lambda$ and L\textsubscript{x}-$\lambda$ relations}
\label{sec:x-ray_lambda_relations}

The X-ray selected $T_{X}-\lambda$ relation is shown in Figure~\ref{fig:TR_scaling_relation} and the  $L^{r500}_{X,52}-\lambda$ relation is shown in Figure~\ref{fig:LR_scaling_relations} (left).  For completeness, we also show the $L^{r500}_{X,52}-\lambda$ relation for the optically selected sample with upper limits for non-detections as described in section \ref{sec:UL}.  In both Figure~\ref{fig:TR_scaling_relation} and Figure~\ref{fig:LR_scaling_relations}, we again highlight clusters that have been excluded from the fit due to uncertain luminosity or temperature measurement (red circled points).  Additionally, clusters with an uncertain $\lambda$ are circled in black and excluded from the fit (see Section \ref{sec:LT_xray_sample}).  The best-fit parameters for each relation are given in Table~\ref{tab:bestfit}.

As found in other studies (e.g., G22b), the residual scatter in the $L_{X}-\lambda$ relation is more than 3 times that of the $T_{X}-\lambda$ relation although this is consistent with the differing slopes.  The slopes for both relations are statistically similar to G22b but the measured scatter of the $L_{\rm x}-\lambda$ relation in this study is somewhat smaller ($0.80\pm{0.02}$ vs. $1.07\pm{0.06}$ in G22b). It is worth noting that when considering only the serendipitous sub-sample from G22b, our scatter is remarkably similar for the $L_{\rm X}-\lambda$ relation ($0.80\pm{0.02}$ vs $0.79\pm{0.08}$).  This should be expected given our sample is selected from survey regions (i.e., they are all serendipitously detected) and shows the possibility of creating larger samples from serendipitous X-ray cluster detections in the full XCS catalogue.  We note that this is predicated on the assumption that the serendipitous population should have a selection function that is easier to model, in comparison to archival targeted samples.  Creating larger serendipitous {\em XMM} samples is of particular importance with upcoming large area surveys (e.g. the Vera C. Rubin Observatory's upcoming Legacy Survey of Space and Time) and their overlap with the {\em XMM} archive (see Sect.~\ref{sec:lsst}).  As shown in G22b, simply matching clusters detected in an incomplete archive leads to differences between the measured scaling relations between {\em XMM} targeted and serendipitously detected clusters.  If we are able to use serendipitously detected clusters, a wealth of previously unused sources becomes available for study.  

Finally, we compare the $T_{X}-\lambda$ relation found here to that found in \citet[hereafter F19]{2019MNRAS.490.3341F}.  F19 used the redMaPPer cluster catalogue constructed from one year of DES observations to probe the $T_{X}-\lambda$ relation.  The redMaPPer catalogue in F19 was matched to all available {\em XMM} data to measure X-ray properties for their clusters.  Hence, the sample contained a mix of clusters specifically targeted by {\em XMM} and those found serendipitously (similar to the analysis of G22b).  Due to the unknown selection of targeted clusters in archival data, it is an important point of comparison to our work to investigate any systematic effects introduced by the inclusion of these clusters.  We find that the slope and residual scatter of the RM$_{\rm scaling}$ sample is consistent with that of F19 (see their Table 2 for best-fit parameters of their {\em XMM} sample).  This consistency is particularly relevant as the $T_{X}-\lambda$ relation derived in the F19 paper informs the scatter prior on the stacked mass-richness relation assumed in the DESY1 cosmology analysis \citep{2020PhRvD.102b3509A}.

\section{Discussion}
\label{sec:discussion}
\subsection{Scaling relations}

\subsubsection{Richness dependence}
\label{sec:lamba_disc}

Assuming clusters demonstrate self-similarity, the  X-ray temperature and X-ray bolometric luminosity should be related with a power law of slope 2.  In this work, the observed slope between luminosity and temperature is somewhat steeper which is consistent with other studies.  See \citet[][Table 2]{2021Univ....7..139L}, for a selection of scaling relation properties from the literature. This is likely due to the over-simplified assumption that gravity is the sole heating mechanism within the clusters' physics as well as the gas fraction increasing as a function of mass \citep[e.g.,][]{2016A&A...592A..12E}. 

The literature is more divided when it comes to comparing the slope of the $L_{X}-T_{X}$ relation between clusters and groups; for a more detailed discussion, see \cite{2021Univ....7..139L}.  Previous work has shown that scaling relations can be modelled by a broken power-law, highlighting a transition between the cluster and group scale \citep[e.g.][]{2015MNRAS.451.1460K,2015A&A...573A.118L}.  However, it is noted that other works \citep[e.g.][]{2009ApJ...693.1142S,2016MNRAS.463..820Z} do not observe any inconsistencies between the slopes of low and high mass scaling relations.  Results from simulations, however, do indicate the presence of a break, or gradual change, in the slope when modelling scaling relations \citep[e.g.,][]{2014MNRAS.441.1270L,2018MNRAS.478.2618F}. Recently, \cite{2022arXiv220511528P} used 30,000 mock haloes from the TNG300 simulations covering the $M_{500,c}=(10^{12}-2\times10^{15})~M_{\odot}$ mass range, to study various scaling relations. \cite{2022arXiv220511528P} found strong evidence for a break in the modelled relations, occurring between $M_{500,c} \sim 3\times10^{13} - 2\times10^{14}$, depending on the scaling relation considered.  

To investigate a possible break in the slope of $L_{X}-T_{X}$ relation considered in the work, we have divided the full X-ray sample into bins of $\lambda<20$, $20< \lambda <40$ and $\lambda>40$ as an illustration.  Using the mass-richness relation of \cite{10.1093/mnras/sty2711}, these bins correspond to M$_{200,m}\lesssim$1.2$\times$10$^{14}$~M$_{\odot}$, 1.2$\lesssim$M$_{200,m}\lesssim$3.0$\times$10$^{14}$~M$_{\odot}$ and M$_{200,m}\gtrsim$3$\times$10$^{14}$~M$_{\odot}$.  The slope and normalisation of the $L^{r500}_{X,bol}$-$T_{X}$ relation for the $20< \lambda <40$ and $\lambda>40$ bins are statistically similar to the total sample, but it is noted that the lower richness (and by definition, lower mass) bin relation is marginally steeper with a higher normalisation.  The steeper slope at these lower richnesses may support the broken power-law model of scaling relations, however, the slope of the $\lambda<20$ clusters is only steeper than the $\lambda>40$ clusters at the $\sim$1.6$\sigma$ level.  One plausible reason for the steepening of the relation at low richnesses is the increased fraction of AGN within clusters as a function of decreasing mass \citep[e.g.,][]{2020MNRAS.498.4095N}.  It is possible that a higher fraction of the low $\lambda$ bin has increased AGN contamination, leading to an increased luminosity and hence steepening the slope of the $L_{X}-T_{X}$ relation.  

\subsubsection{Effect of low signal-to-noise clusters on scaling relations}
\label{sec:snr_cut}

\begin{figure}
    \centering
    \includegraphics[width=\linewidth]{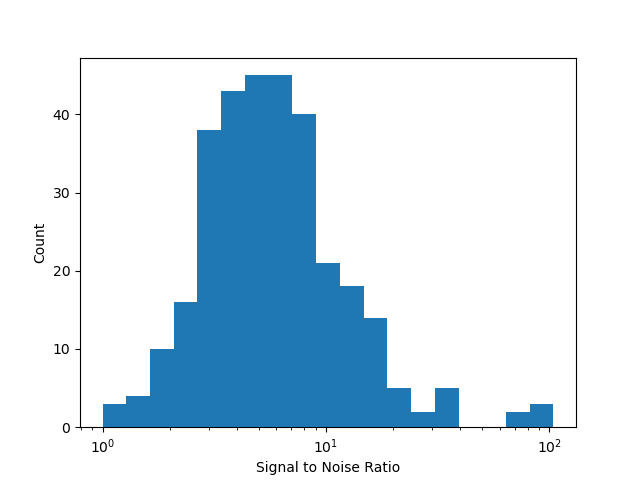}
    \caption{Signal to Noise distribution for the XCS\textsubscript{optically\_confirmed} data set. SNR is taken from the initial XAPA region for the PN camera.}
  	\label{fig:snr_dist}
\end{figure}

As per section \ref{sec:LT_xray_sample}, the only cut we make on the data is removing clusters where the temperature error bar is greater than 50\% of the central value i.e. post-processing.  However, we are aware that, due to the serendipitous nature of the X-ray detections, the X-ray sample includes low signal-to-noise observations, as shown in Figure \ref{fig:snr_dist}.  We have tested the effect of these low SNR clusters on our derived scaling relations by fitting the data (following the same fitting method as in Section~\ref{sec:fitting}), but excluding all clusters with an SNR<5. The SNR ratio used here was estimated from the PN spectra used in the {\sc XSPEC} fits (as generated in Section~\ref{sec:x-ray_analysis}) and represents an SNR within our estimate of $r_{500}$ for the cluster.  Excluding these low SNR clusters makes no statistical difference to the derived scaling relations, and so we use all data points, regardless of their SNR.

\subsection{Comparison to previous studies}
\label{sec:comp_disc}

One of the most analogous comparisons to the work presented here is that of \citet[][hereafter O22]{2022arXiv220609536O}, who constructed a photometrically selected sample of clusters from Hyper Suprime-Cam (HSC) analysed using the CAMIRA red-sequence cluster finder \cite{2014MNRAS.444..147O}.  The clusters were selected over 140~deg$^{2}$ that coincided with the eROSITA Final Equatorial-Depth Survey \citep[eFEDS][]{2022A&A...661A...1B}, and hence have complete X-ray coverage.  O22 cross-matched 41 clusters with a richness $\hat{N}_{mem}>40$ with the eFEDS cluster catalogue \citep{2022A&A...661A...2L}, finding 32 CAMIRA clusters with a match to the eFEDS catalogue, in order to probe various scaling relations.  Using these 32 clusters, O22 find a slope of the $L_{X}-T_{X}$ relation of 1.87$\pm$0.45.  Note, this result does not account for selection effects.  However, O22 do model effects of selection resulting in a steeper slope (2.08$\pm$0.46), but we compare to the uncorrected slope for consistency with the work presented here.  To provide a more robust comparison to O22, we limit the RM$_{scaling}$ scaling sample to clusters with $\lambda > 40$ and re-fit for the $L_{X}-T_{X}$ relation (following the method as in Sect.~\ref{sec:fitting}).  Using only $\lambda > 40$ RM$_{scaling}$ clusters, we find a slope of the $L_{X}-T_{X}$ relation of 2.61$\pm$0.27.  While steeper than the O22 relation, the difference is only significant at the 1.5$\sigma$ level.

Another point of comparison is the work of \citet[][hereafter G22a]{2022MNRAS.511.1227G}, who presented a comparison of optically and X-ray selected clusters over $\approx$16 deg$^{2}$ of the XXL-N region.  While the G22a optical clusters were spectroscopically selected from the Galaxy and Mass Assembly survey \citep[GAMA,][]{2011MNRAS.413..971D} group catalog \citep[version G3Cv10, constructed from the group detection routine of][]{2011MNRAS.416.2640R}, and hence different from the photometric selection used here, the comparison is still warranted.  Note the X-ray data in G22a were selected from the XXL-N survey \citep{2016A&A...592A...1P}, as used in this work.  The main result presented in G22a was an apparent increase in the scatter of the luminosity - velocity dispersion ($L-\sigma_{\rm v}$) for their X-ray selected sample compared to the optically selected sample.  

While we cannot compare a $L-\sigma_{\rm v}$ relation to the one done in G22a, the most appropriate comparison we can make is the $L_{\rm X}-\lambda$ relation between the optically and X-ray selected samples.  We find that the scatter of the XCS$_{\rm scaling}$ $L_{\rm X}-\lambda$ relation is only 10\% higher than that of the RM$_{\rm scaling}$ sample.  This is significantly less than the factor 2.7 times higher scatter of the G22a X-ray selected sample compared to their optically selected sample.  It is noted in G22a that due to the small sample size, a small number of high luminosity outliers in the X-ray selected sample were affecting the measurement of the scatter. Since our sample is significantly larger than that in G22a, we are less affected by small population outliers (but note that a larger fraction of outliers would still indeed affect our results).  As mentioned in Section~\ref{sec:fitted_scaling_relations}, removal of low SNR clusters does not significantly impact the measured relation.  We note that for all the relations studied in this work, the XCS$_{\rm scaling}$ sample presents a marginally larger scatter than the RM$_{\rm scaling}$ sample, but also note that none of the differences in scatters are significantly different.  

Our results are more in line with that of \citet[][hereafter C12]{Connelly_2012}. C12 utilised two regions of the Canadian Network for Observational Cosmology Field Galaxy Redshift Survey 2 (CNOC2) with overlapping contiguous {\em Chandra} observations to construct optically and X-ray selected samples.  Broadly, they found that the scatter of the $L_{\rm X}-\sigma_{v}$ relation were consistent for both sample of clusters, as found in the work presented here.

\subsection{Predicted X-ray cluster detections in the era of the Rubin Observatory}
\label{sec:lsst}

\begin{figure}
\centering
\includegraphics[angle=0,width=8.5cm]{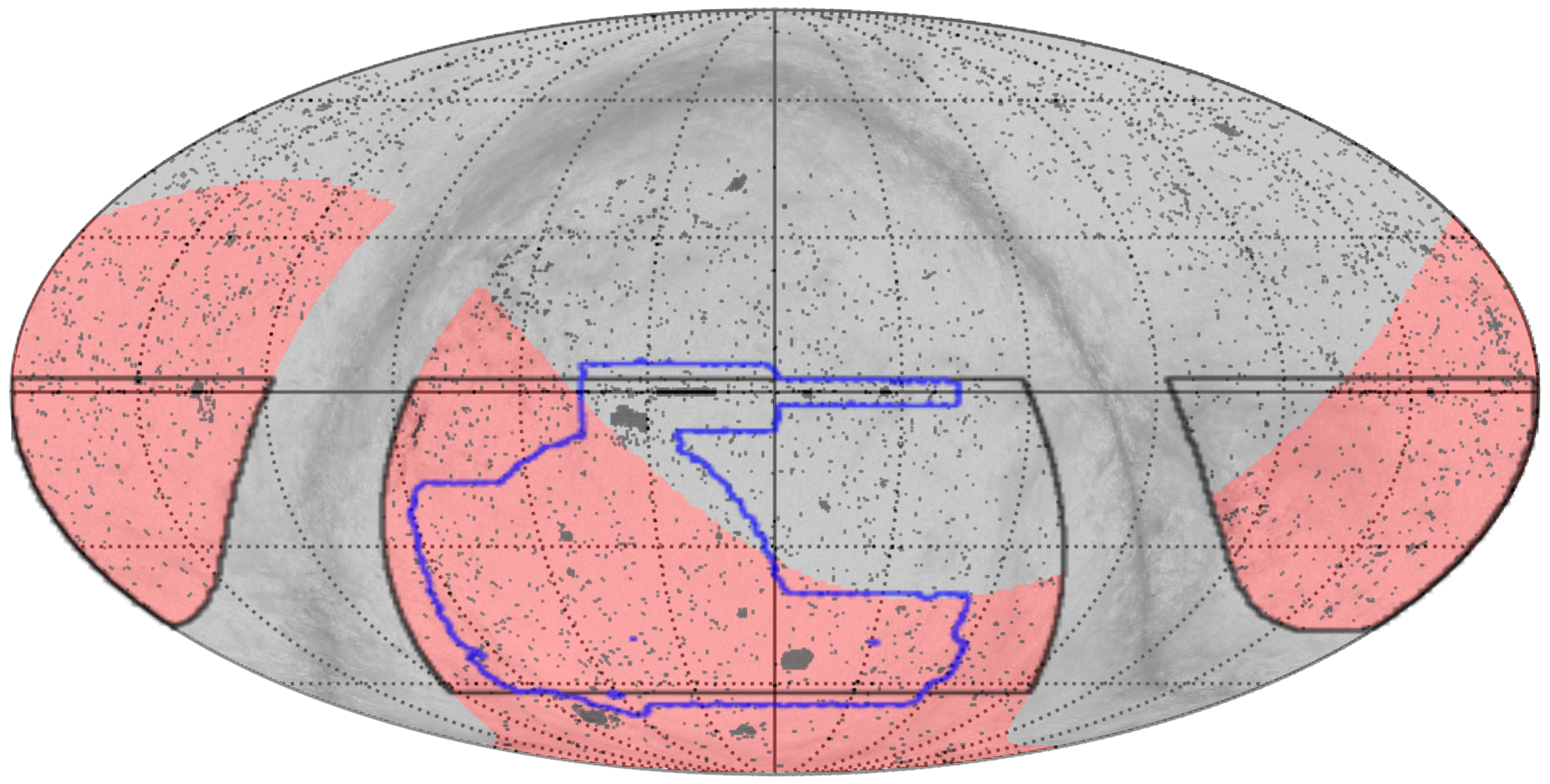}
\vspace{0.5cm}
\caption[]{{Full sky map highlighting the location and size of {\em XMM} observations given by the dark grey points.  In pink, the eRASS$^{DE}$ region (excluding the Galactic plane) is highlighted. The DES and LSST survey footprints are given by the blue and black outlines respectively.  The light-grey background map displays the polarised dust emission map from {\em Planck}}\label{fig:xcs_sky}}
\end{figure}

Given the samples constructed in this work, we can estimate the number of potential clusters in common between those detected by the upcoming Legacy Survey of Space and Time (LSST), planned to be carried out by the Rubin Observatory, and the {\em XMM} archive.  The LSST Wide Fast Deep (LSST-WFD) survey \citep {2018arXiv180901669T} will aim to cover $\approx$14,000 deg$^{2}$ of the southern sky (excluding the Galactic plane).  Currently, the {\em XMM} archive covers 504 deg$^{2}$ of the proposed footprint of the LSST-WFD (including the 57.4 deg$^{2}$ of observations used in this work).  This is shown in Figure~\ref{fig:xcs_sky}, with the sky plot highlighting the position of all currently available {\em XMM} observations (grey points).  The proposed area of the LSST-WFD is given by the black outline.  Given the incompleteness of the X-ray detections of RM clusters in this work (see Sect.~\ref{sec:completeness}), the limiting factor will be the depth of the {\em XMM} observations.  Therefore, using the current DESY3 sample as a precursor should provide a representative indication of the number of LSST-WFD clusters detected in the {\em XMM} archive.  Given the 178 cluster in the RM$_{\rm XCS}$ sample, this leads to a cluster density of 3.1 clusters per deg$^{2}$.  For the full overlap of the LSST-WFD with the {\em XMM} archive, we predict a sample of $\approx$1,500 LSST selected clusters will be detected by {\em XMM}.  Note however that $\sim$50\% of observations in {\em XMM} archive have a nominal exposure time longer than that of a typical observation used in this work, and hence the predicted number of clusters is likely a lower limit. Furthermore, these estimates are for redMaPPer\footnote{Note that redMaPPer is one of a number of cluster finders being tested for use with the LSST} with $\lambda>20$ and within 0.1$<$z$<$0.9.  

Focussing on forecasts for X-ray selected samples,  the XCS$_{\rm opt}$ sample contains 341 clusters, 298 of which have properties (i.e. redshift and richness) returned by redMaPPer.  Using these 298 clusters, we determine a source density of $\approx5.2$ clusters per deg$^{2}$.  Again expanding to the full LSST-WFD with available {\em XMM} data, we estimate there will be $\approx$2,600 X-ray selected clusters.  Due to the deeper depth of the LSST compared to DES, this is likely to be a lower limit (e.g., LSST will detect the high redshift X-ray clusters not found by RM using the current DES data, see Sect.~\ref{sec:completeness}).  Further X-ray data will come from the {\em eROSITA} all sky survey (eRASS).  Currently, only the western half (in Galactic coordinates) is due to be publicly released (by the German eROSITA Consortium, we denote this half of the sky eRASS$^{\rm DE}$).  The sky region covered by eRASS is given by the pink shaded region in Figure~\ref{fig:xcs_sky}. To estimate the number of clusters detectable by eFEDS over the LSST-WFD region, we make use of the recently released cluster catalogue from the 140 deg$^{2}$ eROSITA Final Equatorial Depth Survey \citep[eFEDS,][]{2022A&A...661A...2L}.  This sample contains 542 clusters, of which 477 are subsequently optically confirmed \citep{2022A&A...661A...4K}, using the multicomponent matched filter (MCMF) cluster confirmation tool \cite{2018MNRAS.474.3324K}.  Using the optically confirmed sample leads to an eRASS source density of $\approx$3.4 clusters deg$^{-2}$.  We estimate the overlap between the LSST-WFD and eRASS$^{\rm DE}$ to be 10,174 deg$^{2}$, leading to a potential $\approx$34,600 clusters when the final depth eRASS$^{\rm DE}$ is released.

%\section{Comparison to Giles GAMA and eRASS}
\section{Summary}
\label{sec:summary}
In this paper, we consider two samples of clusters, one selected via optical data from the Dark Energy Survey (DES), and a second using X-ray observations from {\em XMM-Newton}.  The samples are constructed from 4 survey regions observed by {\em XMM}, analysed by the {\em XMM} Cluster Survey (XCS) that overlap with the DES footprint.  We cross-match between the two samples to determine the level of overlap explaining why some optically detected clusters are not being detected in the X-ray observations.  We also explore various scaling relations, including the X-ray $L_{X}$-$T_{X}$ relation and the mass observable relations $L_{X}$-$\lambda$ and $T_{X}$-$\lambda$.
We find the following:
\begin{itemize}
    \item Across 57.4~deg\textsuperscript{2} of the four survey regions used in this work, there are 468 redMaPPer detected DES clusters within the parameter space $\lambda>20$ and $0.1<z<0.9$, of which 178 have a visually confirmed X-ray counterpart.  By comparison, there are 341 X-ray extended sources detected by XCS, with a visually identified optical red galaxy overabundance coincident with the X-ray source.
    \item From the samples derived, we find that the redMaPPer sample is $\simeq$38\% matched in terms of X-ray detections.  However, as a function of $\lambda$, X-ray completeness is $\approx$95\% above  $\lambda>60$ and entirely complete above $\lambda>70$, although the number of clusters in these subsamples is small.
    \item Based upon the constructed X-ray sample, the redMaPPer catalogue is fully matched to the XCS catalogue for $\lambda>20$ and $0.1<z<0.9$, i.e., all X-ray clusters within this range are recovered in the optical cluster catalogue.
    \item We found that 46\% of redMaPPer clusters undetected by our X-ray data can be explained by the insufficient exposure times of our current observations (based on estimating the X-ray luminosity from their richness and our measured $L_{X}-\lambda$ relation).  For the other 54\%, we found that the X-ray exposure times of current observations become insufficient if we reduce the clusters' luminosities within the measured scatter of the $L_{X}-\lambda$ relation.   
    \item The $L_{X}$-$T_{X}$ scaling relation for the overall X-ray and optical samples are consistent with each other, and also with serendipitous X-ray cluster samples in the literature.
    \item Creating sub-samples based upon $\lambda$ cuts, we find that the slope of the $L_{X}$-$T_{X}$ relation is somewhat steeper for lower richness clusters with $\lambda<20$ compared to $\lambda>40$ although only significant to the $1.6\sigma$ level.
    \item We have shown that the $L_{X}$-$\lambda$ relation slope and residual scatter is consistent with that for the relation derived from serendipitously detected clusters found in \cite{2022MNRAS.516.3878G}. 
    \item Additionally, the $T_{X}$-$\lambda$ slope and residual scatter is consistent with the results of \cite{2019MNRAS.490.3341F}.
    \item We have shown that the inclusion of low SNR X-ray clusters does not affect the scaling relation fit.  However, the binning technique used for low SNR clusters may have an effect on the extraction of X-ray properties from {\sc xspec}.  Further work will be undertaken on higher quality X-ray data, and degraded down to lower signal-to-noise, to determine how the binning effects the measurement of cluster properties.
    \item We find that the scatter in each of the scaling relations considered is consistent between the optically and X-ray selected cluster samples.  This is in tension with the previous results of \cite{2022MNRAS.511.1227G}, using clusters from a smaller area than the surveys used in this work, albeit with different optical selection methods.
    \item Finally, we estimate that there will be $\approx$1,500 {\em XMM} detected clusters from those detected by the upcoming Legacy Survey of Space and Time (LSST) and its overlap with the {\em XMM} archive.
\end{itemize}
%%%%%%%%%%%%%%%%%%%%%%%%%%%%%%%%%%%%%%%%%%%%%%%%%%
\section*{Data Availability}

The X-ray sample used for the catalogue crossmatch can be found at: \url{https://users.sussex.ac.uk/~xcs-data/XCS_tests_on_DESY3/xray_sample.csv}

The optical sample used for the catalogue crossmatch can be found at: \url{https://users.sussex.ac.uk/~xcs-data/XCS_tests_on_DESY3/optical_sample.csv}

The data used for the X-ray scaling relations can be found at: \url{https://users.sussex.ac.uk/~xcs-data/XCS_tests_on_DESY3/xray_sample_scaling_data.csv}

The data used for the optical scaling relations can be found at: \url{https://users.sussex.ac.uk/~xcs-data/XCS_tests_on_DESY3/optical_sample_scaling_data.csv}

\section*{Affiliations}
$^{1}$Department of Physics and Astronomy, University of Sussex, Brighton BN1 9QH, UK\\
$^{2}$Astrophysics Research Centre, University of KwaZulu-Natal, Westville Campus, Durban 4041, SA \\
$^{3}$School of Mathematics, Statistics, and Computer Science, University of KwaZulu-Natal, Westville Campus, Durban 4041, SA \\
$^{4}$Kavli Institute for Particle Astrophysics \& Cosmology, P. O. Box 2450, Stanford University, Stanford, CA 94305, USA \\
$^{5}$Department of Statistics and Data Sciences, The University of Texas at Austin, Austin, TX 78712, USA \\
$^{6}$AIM, CEA, CNRS, Université Paris-Saclay, Université Paris Diderot, Sorbonne Paris Cité, F-91191 Gif-sur-Yvette, France \\
$^{7}$Santa Cruz Institute for Particle Physics, University of California, Santa Cruz, 1156 High St, Santa Cruz, CA 95064, USA \\
$^{8}$Faculty of Physics, Ludwig-Maximilians-Universität, Scheinerstr. 1, 81679, Munich, Germany \\
$^{9}$Astrophysics Research Institute, Liverpool John Moores University, Liverpool Science Park, 146 Brownlow Hill, Liverpool L3 5RF, UK \\
$^{10}$Instituto de Astrof\'isica e Ci\^{e}ncias do Espa\c co, Universidade do Porto, CAUP, Rua das Estrelas, 4150-762 Porto, Portugal \\
$^{11}$Departamento de F\'isica e Astronomia, Faculdade de Ci\^{e}ncias, Universidade do Porto, Rua do Campo Alegre, 687, 4169-007 Porto, Portugal \\ 
$^{12}$Institute for Astronomy, University of Edinburgh, Royal Observatory, Blackford Hill, Edinburgh EH9 3HJ, UK \\
$^{13}$Astronomy Department, University of Michigan, Ann Arbor, MI 48109, USA \\
$^{14}$ Department of Physics and Astronomy, Uppsala University, SE-751 20 Uppsala, Sweden \\
$^{15}$Department of Physics, Lancaster University, Lancaster LA1 4YB, UK \\
$^{16}$ Department of Physics, University of Michigan, Ann Arbor, MI 48109, USA\\
$^{17}$ Institute of Cosmology and Gravitation, University of Portsmouth, Portsmouth, PO1 3FX, UK\\
$^{18}$ Sorbonne Universit\'es, UPMC Univ Paris 06, UMR 7095, Institut d'Astrophysique de Paris, F-75014, Paris, France\\
$^{19}$ University Observatory, Faculty of Physics, Ludwig-Maximilians-Universit\"at, Scheinerstr. 1, 81679 Munich, Germany\\
$^{20}$ Department of Physics \& Astronomy, University College London, Gower Street, London, WC1E 6BT, UK\\
$^{21}$ Center for Astrophysical Surveys, National Center for Supercomputing Applications, 1205 West Clark St., Urbana, IL 61801, USA\\
$^{22}$ Department of Astronomy, University of Illinois at Urbana-Champaign, 1002 W. Green Street, Urbana, IL 61801, USA\\
$^{23}$ Institut de F\'{\i}sica d'Altes Energies (IFAE), The Barcelona Institute of Science and Technology, Campus UAB, 08193 Bellaterra (Barcelona) Spain\\
$^{24}$ Astronomy Unit, Department of Physics, University of Trieste, via Tiepolo 11, I-34131 Trieste, Italy\\
$^{25}$ INAF-Osservatorio Astronomico di Trieste, via G. B. Tiepolo 11, I-34143 Trieste, Italy\\
$^{26}$ Institute for Fundamental Physics of the Universe, Via Beirut 2, 34014 Trieste, Italy\\
$^{27}$ Laborat\'orio Interinstitucional de e-Astronomia - LIneA, Rua Gal. Jos\'e Cristino 77, Rio de Janeiro, RJ - 20921-400, Brazil\\
$^{28}$ Hamburger Sternwarte, Universit\"{a}t Hamburg, Gojenbergsweg 112, 21029 Hamburg, Germany\\
$^{29}$ Centro de Investigaciones Energ\'eticas, Medioambientales y Tecnol\'ogicas (CIEMAT), Madrid, Spain\\
$^{30}$ Department of Physics, IIT Hyderabad, Kandi, Telangana 502285, India\\
$^{31}$ Fermi National Accelerator Laboratory, P. O. Box 500, Batavia, IL 60510, USA\\
$^{32}$ Jet Propulsion Laboratory, California Institute of Technology, 4800 Oak Grove Dr., Pasadena, CA 91109, USA\\
$^{33}$ Institute of Theoretical Astrophysics, University of Oslo. P.O. Box 1029 Blindern, NO-0315 Oslo, Norway\\
$^{34}$ Kavli Institute for Cosmological Physics, University of Chicago, Chicago, IL 60637, USA\\
$^{35}$ Department of Astronomy, University of Michigan, Ann Arbor, MI 48109, USA\\
$^{36}$ School of Mathematics and Physics, University of Queensland,  Brisbane, QLD 4072, Australia\\
$^{37}$ Center for Cosmology and Astro-Particle Physics, The Ohio State University, Columbus, OH 43210, USA\\
$^{38}$ Department of Physics, The Ohio State University, Columbus, OH 43210, USA\\
$^{39}$ Center for Astrophysics $\vert$ Harvard \& Smithsonian, 60 Garden Street, Cambridge, MA 02138, USA\\
$^{40}$ Australian Astronomical Optics, Macquarie University, North Ryde, NSW 2113, Australia\\
$^{41}$ Lowell Observatory, 1400 Mars Hill Rd, Flagstaff, AZ 86001, USA\\
$^{42}$ Departamento de F\'isica Matem\'atica, Instituto de F\'isica, Universidade de S\~ao Paulo, CP 66318, S\~ao Paulo, SP, 05314-970, Brazil\\
$^{43}$ George P. and Cynthia Woods Mitchell Institute for Fundamental Physics and Astronomy, and Department of Physics and Astronomy, Texas A\&M University, College Station, TX 77843,  USA\\
$^{44}$ Instituci\'o Catalana de Recerca i Estudis Avan\c{c}ats, E-08010 Barcelona, Spain\\
$^{45}$ Observat\'orio Nacional, Rua Gal. Jos\'e Cristino 77, Rio de Janeiro, RJ - 20921-400, Brazil\\
$^{46}$ Department of Physics, University of Genova and INFN, Via Dodecaneso 33, 16146, Genova, Italy\\
$^{47}$ School of Physics and Astronomy, University of Southampton,  Southampton, SO17 1BJ, UK\\
$^{48}$ Computer Science and Mathematics Division, Oak Ridge National Laboratory, Oak Ridge, TN 37831\\
$^{49}$ Lawrence Berkeley National Laboratory, 1 Cyclotron Road, Berkeley, CA 94720, USA\\
$^{50}$ Max Planck Institute for Extraterrestrial Physics, Giessenbachstrasse, 85748 Garching, Germany\\
$^{51}$ Universit\"ats-Sternwarte, Fakult\"at f\"ur Physik, Ludwig-Maximilians Universit\"at M\"unchen, Scheinerstr. 1, 81679 M\"unchen, Germany\\
$^{52}$ SLAC National Accelerator Laboratory, Menlo Park, CA 94025, USA
$^{53}$ Instituto de Fisica Teorica UAM/CSIC, Universidad Autonoma de Madrid, 28049 Madrid, Spain
\section*{Acknowledgements}
PG, KR, RW, DT and EU recognises support from the UK Science and Technology Facilities Council via grants ST/P000525/1 and ST/T000473/1 (PG, KR), ST/P006760/1 (RW, DT) and ST/T506461/1 (EU).

This paper has gone through internal review by the DES collaboration.  Funding for the DES Projects has been provided by the U.S. Department of Energy, the U.S. National Science Foundation, the Ministry of Science and Education of Spain,
the Science and Technology Facilities Council of the United Kingdom, the Higher Education Funding Council for England, the National Center for Supercomputing
Applications at the University of Illinois at Urbana-Champaign, the Kavli Institute of Cosmological Physics at the University of Chicago,
the Center for Cosmology and Astro-Particle Physics at the Ohio State University,
the Mitchell Institute for Fundamental Physics and Astronomy at Texas A\&M University, Financiadora de Estudos e Projetos,
Funda{\c c}{\~a}o Carlos Chagas Filho de Amparo {\`a} Pesquisa do Estado do Rio de Janeiro, Conselho Nacional de Desenvolvimento Cient{\'i}fico e Tecnol{\'o}gico and
the Minist{\'e}rio da Ci{\^e}ncia, Tecnologia e Inova{\c c}{\~a}o, the Deutsche Forschungsgemeinschaft and the Collaborating Institutions in the Dark Energy Survey.

The Collaborating Institutions are Argonne National Laboratory, the University of California at Santa Cruz, the University of Cambridge, Centro de Investigaciones Energ{\'e}ticas,
Medioambientales y Tecnol{\'o}gicas-Madrid, the University of Chicago, University College London, the DES-Brazil Consortium, the University of Edinburgh,
the Eidgen{\"o}ssische Technische Hochschule (ETH) Z{\"u}rich,
Fermi National Accelerator Laboratory, the University of Illinois at Urbana-Champaign, the Institut de Ci{\`e}ncies de l'Espai (IEEC/CSIC),
the Institut de F{\'i}sica d'Altes Energies, Lawrence Berkeley National Laboratory, the Ludwig-Maximilians Universit{\"a}t M{\"u}nchen and the associated Excellence Cluster Universe,
the University of Michigan, NSF's NOIRLab, the University of Nottingham, The Ohio State University, the University of Pennsylvania, the University of Portsmouth,
SLAC National Accelerator Laboratory, Stanford University, the University of Sussex, Texas A\&M University, and the OzDES Membership Consortium.

Based in part on observations at Cerro Tololo Inter-American Observatory at NSF's NOIRLab (NOIRLab Prop. ID 2012B-0001; PI: J. Frieman), which is managed by the Association of Universities for Research in Astronomy (AURA) under a cooperative agreement with the National Science Foundation.

The DES data management system is supported by the National Science Foundation under Grant Numbers AST-1138766 and AST-1536171.
The DES participants from Spanish institutions are partially supported by MICINN under grants ESP2017-89838, PGC2018-094773, PGC2018-102021, SEV-2016-0588, SEV-2016-0597, and MDM-2015-0509, some of which include ERDF funds from the European Union. IFAE is partially funded by the CERCA program of the Generalitat de Catalunya.
Research leading to these results has received funding from the European Research
Council under the European Union's Seventh Framework Program (FP7/2007-2013) including ERC grant agreements 240672, 291329, and 306478.
We  acknowledge support from the Brazilian Instituto Nacional de Ci\^encia
e Tecnologia (INCT) do e-Universo (CNPq grant 465376/2014-2).



\bibliographystyle{mnras}
\bibliography{4cont} % if your bibtex file is called example.bib

\begin{thebibliography}{}
\makeatletter
\relax
\def\mn@urlcharsother{\let\do\@makeother \do\$\do\&\do\#\do\^\do\_\do\%\do\~}
\def\mn@doi{\begingroup\mn@urlcharsother \@ifnextchar [ {\mn@doi@}
  {\mn@doi@[]}}
\def\mn@doi@[#1]#2{\def\@tempa{#1}\ifx\@tempa\@empty \href
  {http://dx.doi.org/#2} {doi:#2}\else \href {http://dx.doi.org/#2} {#1}\fi
  \endgroup}
\def\mn@eprint#1#2{\mn@eprint@#1:#2::\@nil}
\def\mn@eprint@arXiv#1{\href {http://arxiv.org/abs/#1} {{\tt arXiv:#1}}}
\def\mn@eprint@dblp#1{\href {http://dblp.uni-trier.de/rec/bibtex/#1.xml}
  {dblp:#1}}
\def\mn@eprint@#1:#2:#3:#4\@nil{\def\@tempa {#1}\def\@tempb {#2}\def\@tempc
  {#3}\ifx \@tempc \@empty \let \@tempc \@tempb \let \@tempb \@tempa \fi \ifx
  \@tempb \@empty \def\@tempb {arXiv}\fi \@ifundefined
  {mn@eprint@\@tempb}{\@tempb:\@tempc}{\expandafter \expandafter \csname
  mn@eprint@\@tempb\endcsname \expandafter{\@tempc}}}

\bibitem[\protect\citeauthoryear{Abbott et~al.,}{Abbott
  et~al.}{2018}]{PhysRevD.98.043526}
Abbott T. M.~C.,  et~al., 2018, \mn@doi [Phys. Rev. D]
  {10.1103/PhysRevD.98.043526}, 98, 043526

\bibitem[\protect\citeauthoryear{{Abbott} et~al.,}{{Abbott}
  et~al.}{2020}]{2020PhRvD.102b3509A}
{Abbott} T.~M.~C.,  et~al., 2020, \mn@doi [\prd] {10.1103/PhysRevD.102.023509},
  \href {https://ui.adsabs.harvard.edu/abs/2020PhRvD.102b3509A} {102, 023509}

\bibitem[\protect\citeauthoryear{{Anbajagane}, {Evrard}, {Farahi}, {Barnes},
  {Dolag}, {McCarthy}, {Nelson}  \& {Pillepich}}{{Anbajagane}
  et~al.}{2020}]{2020MNRAS.495..686A}
{Anbajagane} D.,  {Evrard} A.~E.,  {Farahi} A.,  {Barnes} D.~J.,  {Dolag} K.,
  {McCarthy} I.~G.,  {Nelson} D.,   {Pillepich} A.,  2020, \mn@doi [\mnras]
  {10.1093/mnras/staa1147}, \href
  {https://ui.adsabs.harvard.edu/abs/2020MNRAS.495..686A} {495, 686}

\bibitem[\protect\citeauthoryear{{Andreon}, {Serra}, {Moretti}  \&
  {Trinchieri}}{{Andreon} et~al.}{2016}]{2016A&A...585A.147A}
{Andreon} S.,  {Serra} A.~L.,  {Moretti} A.,   {Trinchieri} G.,  2016, \mn@doi
  [\aap] {10.1051/0004-6361/201527408}, \href
  {http://adsabs.harvard.edu/abs/2016A%26A...585A.147A} {585, A147}

\bibitem[\protect\citeauthoryear{Applegate et~al.,}{Applegate
  et~al.}{2014}]{10.1093/mnras/stt2129}
Applegate D.~E.,  et~al., 2014, \mn@doi [Monthly Notices of the Royal
  Astronomical Society] {10.1093/mnras/stt2129}, 439, 48

\bibitem[\protect\citeauthoryear{{Arnaud}}{{Arnaud}}{1996}]{1996ASPC..101...17A}
{Arnaud} K.~A.,  1996, in {Jacoby} G.~H.,  {Barnes} J.,  eds,  Astronomical
  Society of the Pacific Conference Series Vol. 101, Astronomical Data Analysis
  Software and Systems V. p.~17

\bibitem[\protect\citeauthoryear{{Arnaud}, {Pointecouteau}  \&
  {Pratt}}{{Arnaud} et~al.}{2005}]{2005A&A...441..893A}
{Arnaud} M.,  {Pointecouteau} E.,   {Pratt} G.~W.,  2005, \mn@doi [\aap]
  {10.1051/0004-6361:20052856}, \href
  {http://adsabs.harvard.edu/abs/2005A%26A...441..893A} {441, 893}

\bibitem[\protect\citeauthoryear{Arviset, Guainazzi, Hernandez, Dowson, Osuna
  \& Venet}{Arviset
  et~al.}{2002}]{https://doi.org/10.48550/arxiv.astro-ph/0206412}
Arviset C.,  Guainazzi M.,  Hernandez J.,  Dowson J.,  Osuna P.,   Venet A.,
  2002, The XMM-Newton Science Archive,
  \mn@doi{10.48550/ARXIV.ASTRO-PH/0206412}, \url
  {https://arxiv.org/abs/astro-ph/0206412}

\bibitem[\protect\citeauthoryear{{Bahar} et~al.,}{{Bahar}
  et~al.}{2022}]{2022A&A...661A...7B}
{Bahar} Y.~E.,  et~al., 2022, \mn@doi [\aap] {10.1051/0004-6361/202142462},
  \href {https://ui.adsabs.harvard.edu/abs/2022A&A...661A...7B} {661, A7}

\bibitem[\protect\citeauthoryear{Bocquet et~al.,}{Bocquet
  et~al.}{2019}]{Bocquet_2019}
Bocquet S.,  et~al., 2019, \mn@doi [The Astrophysical Journal]
  {10.3847/1538-4357/ab1f10}, 878, 55

\bibitem[\protect\citeauthoryear{{Brunner} et~al.,}{{Brunner}
  et~al.}{2022}]{2022A&A...661A...1B}
{Brunner} H.,  et~al., 2022, \mn@doi [\aap] {10.1051/0004-6361/202141266},
  \href {https://ui.adsabs.harvard.edu/abs/2022A&A...661A...1B} {661, A1}

\bibitem[\protect\citeauthoryear{{Carvalho}, {Bernui}, {Benetti}, {Carvalho}
  \& {Alcaniz}}{{Carvalho} et~al.}{2016}]{2016PhRvD..93b3530C}
{Carvalho} G.~C.,  {Bernui} A.,  {Benetti} M.,  {Carvalho} J.~C.,   {Alcaniz}
  J.~S.,  2016, \mn@doi [\prd] {10.1103/PhysRevD.93.023530}, \href
  {https://ui.adsabs.harvard.edu/abs/2016PhRvD..93b3530C} {93, 023530}

\bibitem[\protect\citeauthoryear{Connelly et~al.,}{Connelly
  et~al.}{2012}]{Connelly_2012}
Connelly J.~L.,  et~al., 2012, \mn@doi [The Astrophysical Journal]
  {10.1088/0004-637x/756/2/139}, 756, 139

\bibitem[\protect\citeauthoryear{{Dark Energy Survey Collaboration}
  et~al.,}{{Dark Energy Survey Collaboration}
  et~al.}{2016}]{2016MNRAS.460.1270D}
{Dark Energy Survey Collaboration} et~al., 2016, \mn@doi [\mnras]
  {10.1093/mnras/stw641}, \href
  {https://ui.adsabs.harvard.edu/abs/2016MNRAS.460.1270D} {460, 1270}

\bibitem[\protect\citeauthoryear{{Driver} et~al.,}{{Driver}
  et~al.}{2011}]{2011MNRAS.413..971D}
{Driver} S.~P.,  et~al., 2011, \mn@doi [\mnras]
  {10.1111/j.1365-2966.2010.18188.x}, \href
  {http://adsabs.harvard.edu/abs/2011MNRAS.413..971D} {413, 971}

\bibitem[\protect\citeauthoryear{{Eckert} et~al.,}{{Eckert}
  et~al.}{2016}]{2016A&A...592A..12E}
{Eckert} D.,  et~al., 2016, \mn@doi [\aap] {10.1051/0004-6361/201527293}, \href
  {https://ui.adsabs.harvard.edu/abs/2016A&A...592A..12E} {592, A12}

\bibitem[\protect\citeauthoryear{{Eckmiller}, {Hudson}  \&
  {Reiprich}}{{Eckmiller} et~al.}{2011}]{2011A&A...535A.105E}
{Eckmiller} H.~J.,  {Hudson} D.~S.,   {Reiprich} T.~H.,  2011, \mn@doi [\aap]
  {10.1051/0004-6361/201116734}, \href
  {http://adsabs.harvard.edu/abs/2011A%26A...535A.105E} {535, A105}

\bibitem[\protect\citeauthoryear{{Farahi}, {Evrard}, {McCarthy}, {Barnes}  \&
  {Kay}}{{Farahi} et~al.}{2018}]{2018MNRAS.478.2618F}
{Farahi} A.,  {Evrard} A.~E.,  {McCarthy} I.,  {Barnes} D.~J.,   {Kay} S.~T.,
  2018, \mn@doi [\mnras] {10.1093/mnras/sty1179}, \href
  {https://ui.adsabs.harvard.edu/abs/2018MNRAS.478.2618F} {478, 2618}

\bibitem[\protect\citeauthoryear{{Farahi} et~al.,}{{Farahi}
  et~al.}{2019}]{2019MNRAS.490.3341F}
{Farahi} A.,  et~al., 2019, \mn@doi [\mnras] {10.1093/mnras/stz2689}, \href
  {https://ui.adsabs.harvard.edu/abs/2019MNRAS.490.3341F} {490, 3341}

\bibitem[\protect\citeauthoryear{{Flaugher} et~al.,}{{Flaugher}
  et~al.}{2015}]{2015AJ....150..150F}
{Flaugher} B.,  et~al., 2015, \mn@doi [\aj] {10.1088/0004-6256/150/5/150},
  \href {https://ui.adsabs.harvard.edu/abs/2015AJ....150..150F} {150, 150}

\bibitem[\protect\citeauthoryear{{Freeman}, {Kashyap}, {Rosner}  \&
  {Lamb}}{{Freeman} et~al.}{2002}]{2002ApJS..138..185F}
{Freeman} P.~E.,  {Kashyap} V.,  {Rosner} R.,   {Lamb} D.~Q.,  2002, \mn@doi
  [\apjs] {10.1086/324017}, \href
  {http://adsabs.harvard.edu/abs/2002ApJS..138..185F} {138, 185}

\bibitem[\protect\citeauthoryear{{Giles} et~al.,}{{Giles}
  et~al.}{2022a}]{2022MNRAS.511.1227G}
{Giles} P.~A.,  et~al., 2022a, \mn@doi [\mnras] {10.1093/mnras/stab3626}, \href
  {https://ui.adsabs.harvard.edu/abs/2022MNRAS.511.1227G} {511, 1227}

\bibitem[\protect\citeauthoryear{{Giles} et~al.,}{{Giles}
  et~al.}{2022b}]{2022MNRAS.516.3878G}
{Giles} P.~A.,  et~al., 2022b, \mn@doi [\mnras] {10.1093/mnras/stac2414}, \href
  {https://ui.adsabs.harvard.edu/abs/2022MNRAS.516.3878G} {516, 3878}

\bibitem[\protect\citeauthoryear{{Grandis} et~al.,}{{Grandis}
  et~al.}{2021}]{2021MNRAS.504.1253G}
{Grandis} S.,  et~al., 2021, \mn@doi [\mnras] {10.1093/mnras/stab869}, \href
  {https://ui.adsabs.harvard.edu/abs/2021MNRAS.504.1253G} {504, 1253}

\bibitem[\protect\citeauthoryear{{HI4PI Collaboration} et~al.,}{{HI4PI
  Collaboration} et~al.}{2016}]{2016A&A...594A.116H}
{HI4PI Collaboration} et~al., 2016, \mn@doi [\aap]
  {10.1051/0004-6361/201629178}, \href
  {https://ui.adsabs.harvard.edu/abs/2016A&A...594A.116H} {594, A116}

\bibitem[\protect\citeauthoryear{{Hou} et~al.,}{{Hou} et~al.}{2014}]{spt_cosmo}
{Hou} Z.,  et~al., 2014, \mn@doi [\apj] {10.1088/0004-637X/782/2/74}, \href
  {https://ui.adsabs.harvard.edu/abs/2014ApJ...782...74H} {782, 74}

\bibitem[\protect\citeauthoryear{{Kaiser}}{{Kaiser}}{1986}]{1986MNRAS.222..323K}
{Kaiser} N.,  1986, \mn@doi [\mnras] {10.1093/mnras/222.2.323}, \href
  {https://ui.adsabs.harvard.edu/abs/1986MNRAS.222..323K} {222, 323}

\bibitem[\protect\citeauthoryear{{Kettula} et~al.,}{{Kettula}
  et~al.}{2015}]{2015MNRAS.451.1460K}
{Kettula} K.,  et~al., 2015, \mn@doi [\mnras] {10.1093/mnras/stv923}, \href
  {http://adsabs.harvard.edu/abs/2015MNRAS.451.1460K} {451, 1460}

\bibitem[\protect\citeauthoryear{{Klein} et~al.,}{{Klein}
  et~al.}{2018}]{2018MNRAS.474.3324K}
{Klein} M.,  et~al., 2018, \mn@doi [\mnras] {10.1093/mnras/stx2929}, \href
  {https://ui.adsabs.harvard.edu/abs/2018MNRAS.474.3324K} {474, 3324}

\bibitem[\protect\citeauthoryear{{Klein} et~al.,}{{Klein}
  et~al.}{2022}]{2022A&A...661A...4K}
{Klein} M.,  et~al., 2022, \mn@doi [\aap] {10.1051/0004-6361/202141123}, \href
  {https://ui.adsabs.harvard.edu/abs/2022A&A...661A...4K} {661, A4}

\bibitem[\protect\citeauthoryear{{Kravtsov} \& {Borgani}}{{Kravtsov} \&
  {Borgani}}{2012}]{2012ARA&A..50..353K}
{Kravtsov} A.~V.,  {Borgani} S.,  2012, \mn@doi [\araa]
  {10.1146/annurev-astro-081811-125502}, \href
  {https://ui.adsabs.harvard.edu/abs/2012ARA&A..50..353K} {50, 353}

\bibitem[\protect\citeauthoryear{{Le Brun}, {McCarthy}, {Schaye}  \&
  {Ponman}}{{Le Brun} et~al.}{2014}]{2014MNRAS.441.1270L}
{Le Brun} A. M.~C.,  {McCarthy} I.~G.,  {Schaye} J.,   {Ponman} T.~J.,  2014,
  \mn@doi [\mnras] {10.1093/mnras/stu608}, \href
  {https://ui.adsabs.harvard.edu/abs/2014MNRAS.441.1270L} {441, 1270}

\bibitem[\protect\citeauthoryear{{Le Brun}, {McCarthy}, {Schaye}  \&
  {Ponman}}{{Le Brun} et~al.}{2017}]{2017MNRAS.466.4442L}
{Le Brun} A. M.~C.,  {McCarthy} I.~G.,  {Schaye} J.,   {Ponman} T.~J.,  2017,
  \mn@doi [\mnras] {10.1093/mnras/stw3361}, \href
  {https://ui.adsabs.harvard.edu/abs/2017MNRAS.466.4442L} {466, 4442}

\bibitem[\protect\citeauthoryear{{Liu} et~al.,}{{Liu}
  et~al.}{2022}]{2022A&A...661A...2L}
{Liu} A.,  et~al., 2022, \mn@doi [\aap] {10.1051/0004-6361/202141120}, \href
  {https://ui.adsabs.harvard.edu/abs/2022A&A...661A...2L} {661, A2}

\bibitem[\protect\citeauthoryear{{Lloyd-Davies} et~al.,}{{Lloyd-Davies}
  et~al.}{2011}]{2011MNRAS.418...14L}
{Lloyd-Davies} E.~J.,  et~al., 2011, \mn@doi [\mnras]
  {10.1111/j.1365-2966.2011.19117.x}, \href
  {http://adsabs.harvard.edu/abs/2011MNRAS.418...14L} {418, 14}

\bibitem[\protect\citeauthoryear{{Lovisari}, {Reiprich}  \&
  {Schellenberger}}{{Lovisari} et~al.}{2015}]{2015A&A...573A.118L}
{Lovisari} L.,  {Reiprich} T.~H.,   {Schellenberger} G.,  2015, \mn@doi [\aap]
  {10.1051/0004-6361/201423954}, \href
  {http://adsabs.harvard.edu/abs/2015A%26A...573A.118L} {573, A118}

\bibitem[\protect\citeauthoryear{{Lovisari} et~al.,}{{Lovisari}
  et~al.}{2020}]{2020ApJ...892..102L}
{Lovisari} L.,  et~al., 2020, \mn@doi [\apj] {10.3847/1538-4357/ab7997}, \href
  {https://ui.adsabs.harvard.edu/abs/2020ApJ...892..102L} {892, 102}

\bibitem[\protect\citeauthoryear{{Lovisari}, {Ettori}, {Gaspari}  \&
  {Giles}}{{Lovisari} et~al.}{2021}]{2021Univ....7..139L}
{Lovisari} L.,  {Ettori} S.,  {Gaspari} M.,   {Giles} P.~A.,  2021, \mn@doi
  [Universe] {10.3390/universe7050139}, \href
  {https://ui.adsabs.harvard.edu/abs/2021Univ....7..139L} {7, 139}

\bibitem[\protect\citeauthoryear{Mauduit et~al.,}{Mauduit
  et~al.}{2012}]{Mauduit_2012}
Mauduit J.-C.,  et~al., 2012, \mn@doi [Publications of the Astronomical Society
  of the Pacific] {10.1086/666945}, 124, 714

\bibitem[\protect\citeauthoryear{McClintock et~al.,}{McClintock
  et~al.}{2018}]{10.1093/mnras/sty2711}
McClintock T.,  et~al., 2018, \mn@doi [Monthly Notices of the Royal
  Astronomical Society] {10.1093/mnras/sty2711}, 482, 1352

\bibitem[\protect\citeauthoryear{{Noordeh} et~al.,}{{Noordeh}
  et~al.}{2020}]{2020MNRAS.498.4095N}
{Noordeh} E.,  et~al., 2020, \mn@doi [\mnras] {10.1093/mnras/staa2682}, \href
  {https://ui.adsabs.harvard.edu/abs/2020MNRAS.498.4095N} {498, 4095}

\bibitem[\protect\citeauthoryear{{Oguri}}{{Oguri}}{2014}]{2014MNRAS.444..147O}
{Oguri} M.,  2014, \mn@doi [\mnras] {10.1093/mnras/stu1446}, \href
  {https://ui.adsabs.harvard.edu/abs/2014MNRAS.444..147O} {444, 147}

\bibitem[\protect\citeauthoryear{{Ota} et~al.,}{{Ota}
  et~al.}{2022}]{2022arXiv220609536O}
{Ota} N.,  et~al., 2022, arXiv e-prints, \href
  {https://ui.adsabs.harvard.edu/abs/2022arXiv220609536O} {p. arXiv:2206.09536}

\bibitem[\protect\citeauthoryear{{Pierre} et~al.,}{{Pierre}
  et~al.}{2016}]{2016A&A...592A...1P}
{Pierre} M.,  et~al., 2016, \mn@doi [\aap] {10.1051/0004-6361/201526766}, \href
  {http://adsabs.harvard.edu/abs/2016A%26A...592A...1P} {592, A1}

\bibitem[\protect\citeauthoryear{{Planck Collaboration} et~al.,}{{Planck
  Collaboration} et~al.}{2016}]{2016A&A...594A..24P}
{Planck Collaboration} et~al., 2016, \mn@doi [\aap]
  {10.1051/0004-6361/201525833}, \href
  {https://ui.adsabs.harvard.edu/abs/2016A&A...594A..24P} {594, A24}

\bibitem[\protect\citeauthoryear{{Pop} et~al.,}{{Pop}
  et~al.}{2022}]{2022arXiv220511528P}
{Pop} A.-R.,  et~al., 2022, arXiv e-prints, \href
  {https://ui.adsabs.harvard.edu/abs/2022arXiv220511528P} {p. arXiv:2205.11528}

\bibitem[\protect\citeauthoryear{{Pratt}, {Croston}, {Arnaud}  \&
  {B{\"o}hringer}}{{Pratt} et~al.}{2009}]{2009A&A...498..361P}
{Pratt} G.~W.,  {Croston} J.~H.,  {Arnaud} M.,   {B{\"o}hringer} H.,  2009,
  \mn@doi [\aap] {10.1051/0004-6361/200810994}, \href
  {http://adsabs.harvard.edu/abs/2009A%26A...498..361P} {498, 361}

\bibitem[\protect\citeauthoryear{Predehl et~al.,}{Predehl
  et~al.}{2021}]{Predehl_2021}
Predehl P.,  et~al., 2021, \mn@doi [Astronomy \& Astrophysics]
  {10.1051/0004-6361/202039313}, 647, A1

\bibitem[\protect\citeauthoryear{{Robotham} et~al.,}{{Robotham}
  et~al.}{2011}]{2011MNRAS.416.2640R}
{Robotham} A.~S.~G.,  et~al., 2011, \mn@doi [\mnras]
  {10.1111/j.1365-2966.2011.19217.x}, \href
  {http://adsabs.harvard.edu/abs/2011MNRAS.416.2640R} {416, 2640}

\bibitem[\protect\citeauthoryear{{Romer}, {Viana}, {Liddle}  \& {Mann}}{{Romer}
  et~al.}{1999}]{1999astro.ph.11499R}
{Romer} A.~K.,  {Viana} P.~T.~P.,  {Liddle} A.~R.,   {Mann} R.~G.,  1999, ArXiv
  Astrophysics e-prints, \href
  {http://adsabs.harvard.edu/abs/1999astro.ph.11499R} {}

\bibitem[\protect\citeauthoryear{Rykoff et~al.,}{Rykoff
  et~al.}{2014}]{Rykoff_2014}
Rykoff E.~S.,  et~al., 2014, \mn@doi [The Astrophysical Journal]
  {10.1088/0004-637x/785/2/104}, 785, 104

\bibitem[\protect\citeauthoryear{{Rykoff} et~al.,}{{Rykoff}
  et~al.}{2016}]{2016ApJS..224....1R}
{Rykoff} E.~S.,  et~al., 2016, \mn@doi [\apjs] {10.3847/0067-0049/224/1/1},
  \href {http://adsabs.harvard.edu/abs/2016ApJS..224....1R} {224, 1}

\bibitem[\protect\citeauthoryear{{Sereno}}{{Sereno}}{2016}]{2016ascl.soft02006S}
{Sereno} M.,  2016, {LIRA: LInear Regression in Astronomy}, Astrophysics Source
  Code Library (\mn@eprint {ascl} {1602.006})

\bibitem[\protect\citeauthoryear{{Sevilla-Noarbe} et~al.,}{{Sevilla-Noarbe}
  et~al.}{2021}]{2021ApJS..254...24S}
{Sevilla-Noarbe} I.,  et~al., 2021, \mn@doi [\apjs] {10.3847/1538-4365/abeb66},
  \href {https://ui.adsabs.harvard.edu/abs/2021ApJS..254...24S} {254, 24}

\bibitem[\protect\citeauthoryear{{Smith}, {Brickhouse}, {Liedahl}  \&
  {Raymond}}{{Smith} et~al.}{2001}]{2001ApJ...556L..91S}
{Smith} R.~K.,  {Brickhouse} N.~S.,  {Liedahl} D.~A.,   {Raymond} J.~C.,  2001,
  \mn@doi [\apjl] {10.1086/322992}, \href
  {http://adsabs.harvard.edu/abs/2001ApJ...556L..91S} {556, L91}

\bibitem[\protect\citeauthoryear{{Sun}, {Voit}, {Donahue}, {Jones}, {Forman}
  \& {Vikhlinin}}{{Sun} et~al.}{2009}]{2009ApJ...693.1142S}
{Sun} M.,  {Voit} G.~M.,  {Donahue} M.,  {Jones} C.,  {Forman} W.,
  {Vikhlinin} A.,  2009, \mn@doi [\apj] {10.1088/0004-637X/693/2/1142}, \href
  {http://adsabs.harvard.edu/abs/2009ApJ...693.1142S} {693, 1142}

\bibitem[\protect\citeauthoryear{{The LSST Dark Energy Science Collaboration}
  et~al.,}{{The LSST Dark Energy Science Collaboration}
  et~al.}{2018}]{2018arXiv180901669T}
{The LSST Dark Energy Science Collaboration} et~al., 2018, arXiv e-prints,
  \href {https://ui.adsabs.harvard.edu/abs/2018arXiv180901669T} {p.
  arXiv:1809.01669}

\bibitem[\protect\citeauthoryear{{Vikhlinin} et~al.,}{{Vikhlinin}
  et~al.}{2009}]{2009ApJ...692.1060V}
{Vikhlinin} A.,  et~al., 2009, \mn@doi [\apj] {10.1088/0004-637X/692/2/1060},
  \href {http://adsabs.harvard.edu/abs/2009ApJ...692.1060V} {692, 1060}

\bibitem[\protect\citeauthoryear{{Wilms}, {Allen}  \& {McCray}}{{Wilms}
  et~al.}{2000}]{2000ApJ...542..914W}
{Wilms} J.,  {Allen} A.,   {McCray} R.,  2000, \mn@doi [\apj] {10.1086/317016},
  \href {https://ui.adsabs.harvard.edu/abs/2000ApJ...542..914W} {542, 914}

\bibitem[\protect\citeauthoryear{{Wu} et~al.,}{{Wu}
  et~al.}{2022}]{2022MNRAS.515.4471W}
{Wu} H.-Y.,  et~al., 2022, \mn@doi [\mnras] {10.1093/mnras/stac2048}, \href
  {https://ui.adsabs.harvard.edu/abs/2022MNRAS.515.4471W} {515, 4471}

\bibitem[\protect\citeauthoryear{{Zhang} et~al.,}{{Zhang}
  et~al.}{2019}]{2019MNRAS.487.2578Z}
{Zhang} Y.,  et~al., 2019, \mn@doi [\mnras] {10.1093/mnras/stz1361}, \href
  {https://ui.adsabs.harvard.edu/abs/2019MNRAS.487.2578Z} {487, 2578}

\bibitem[\protect\citeauthoryear{{Zou}, {Maughan}, {Giles}, {Vikhlinin},
  {Pacaud}, {Burenin}  \& {Hornstrup}}{{Zou}
  et~al.}{2016}]{2016MNRAS.463..820Z}
{Zou} S.,  {Maughan} B.~J.,  {Giles} P.~A.,  {Vikhlinin} A.,  {Pacaud} F.,
  {Burenin} R.,   {Hornstrup} A.,  2016, \mn@doi [\mnras]
  {10.1093/mnras/stw1992}, \href
  {http://adsabs.harvard.edu/abs/2016MNRAS.463..820Z} {463, 820}

\makeatother
\end{thebibliography}

%\begin{thebibliography}{99}
%\end{thebibliography}

%%%%%%%%%%%%%%%%%%%%%%%%%%%%%%%%%%%%%%%%%%%%%%%%%%

%%%%%%%%%%%%%%%%% APPENDICES %%%%%%%%%%%%%%%%%%%%%
\appendix

\section{Image Examples}
\label{app:Appendix_A}
\begin{figure*}

\centering
\textbf{Eyeballing redMaPPer clusters for X-ray counterparts}
\begin{tabular}{ccc}
\includegraphics[width=0.33\textwidth]{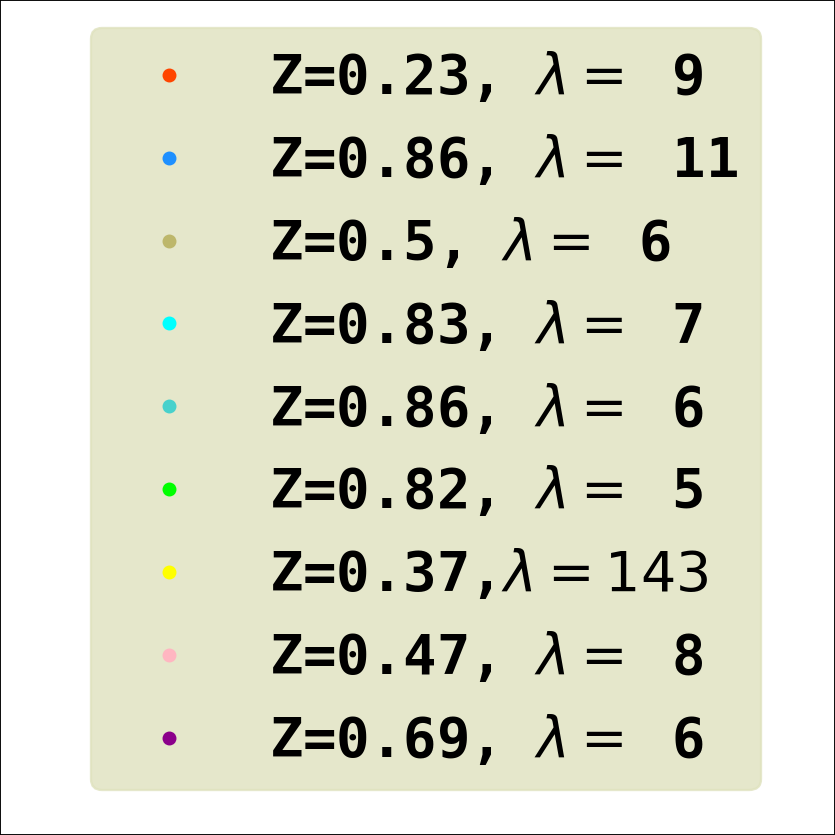} &
     \includegraphics[width=0.33\textwidth]{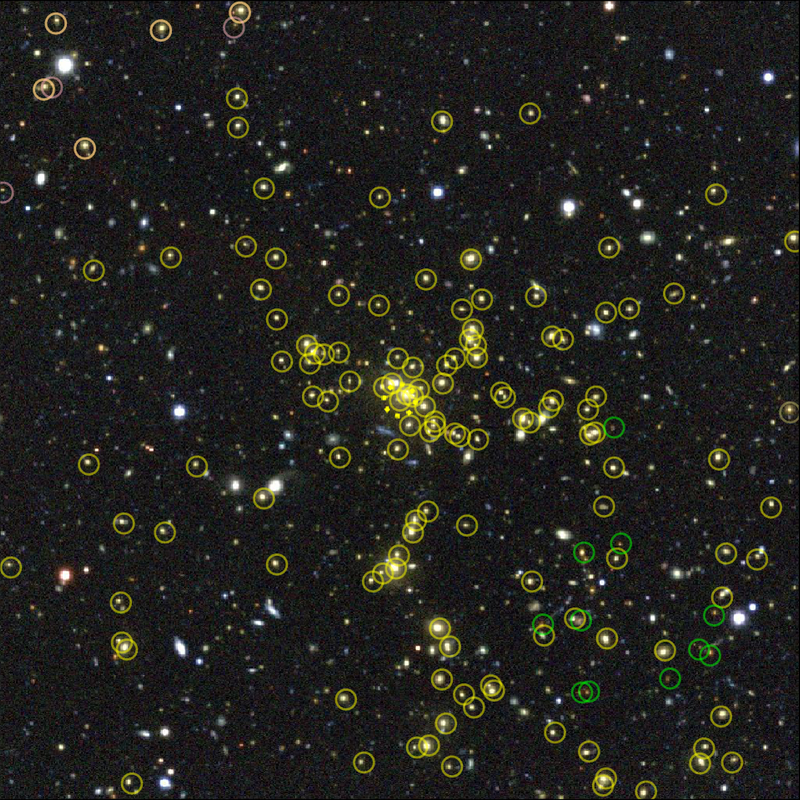} &
     
     \includegraphics[width=0.33\textwidth]{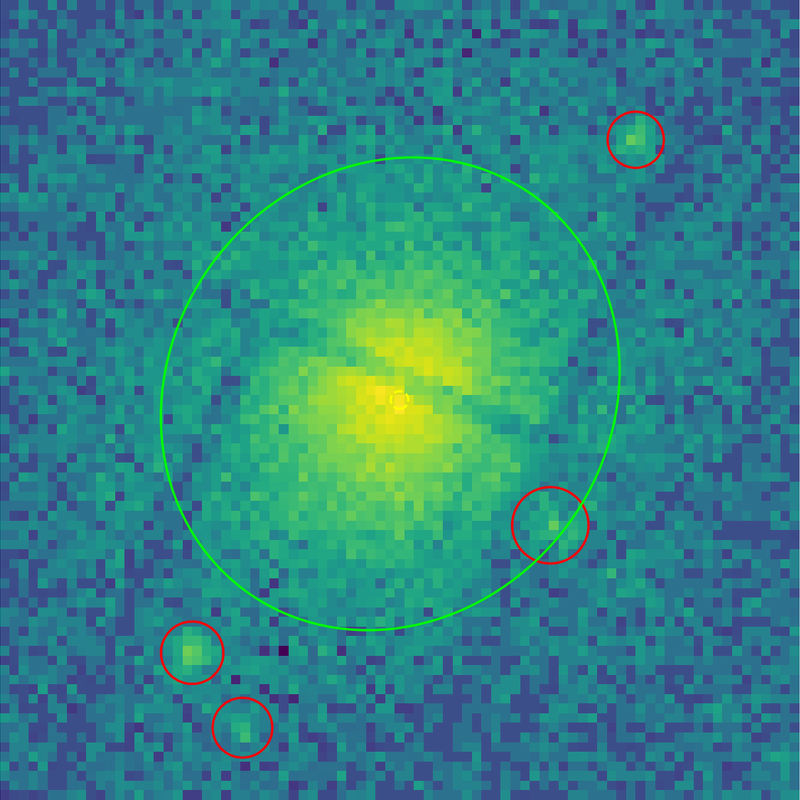}
\\

\includegraphics[width=0.33\textwidth]{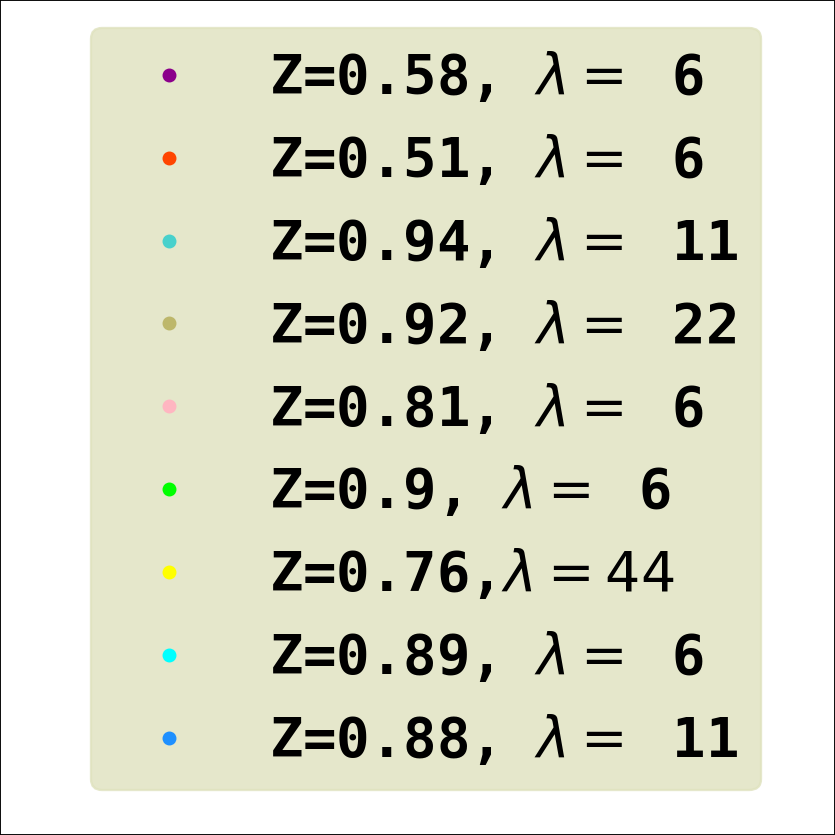} &
     \includegraphics[width=0.33\textwidth]{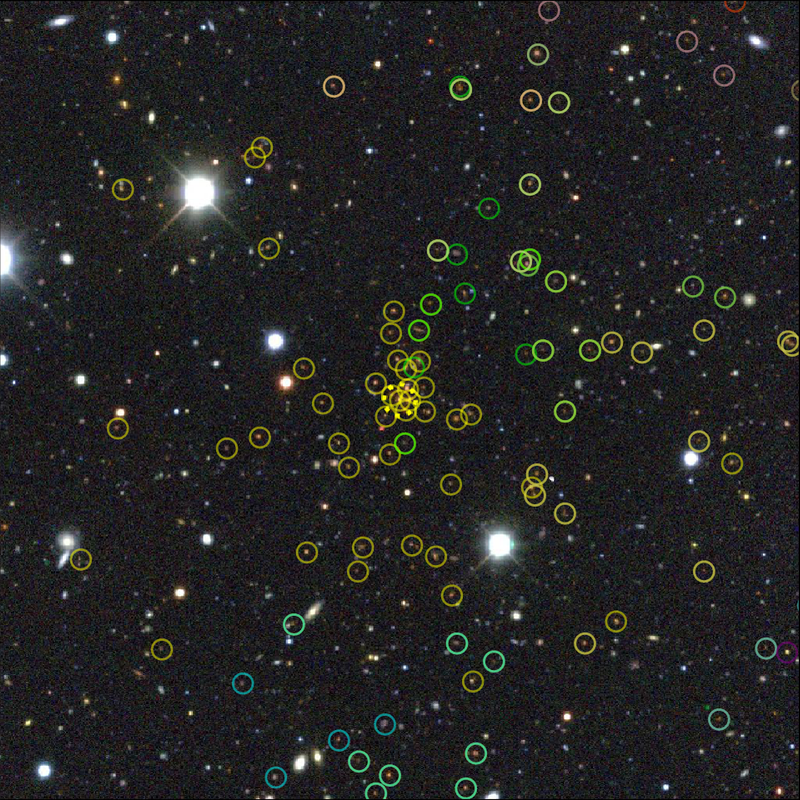} &
     
     \includegraphics[width=0.33\textwidth]{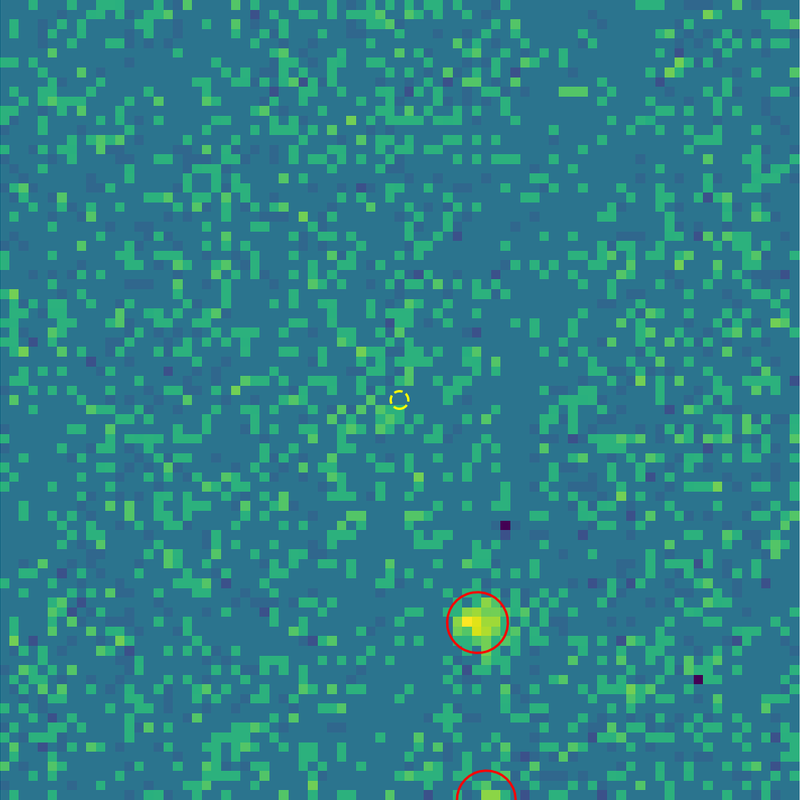} 
\\
\includegraphics[width=0.33\textwidth]{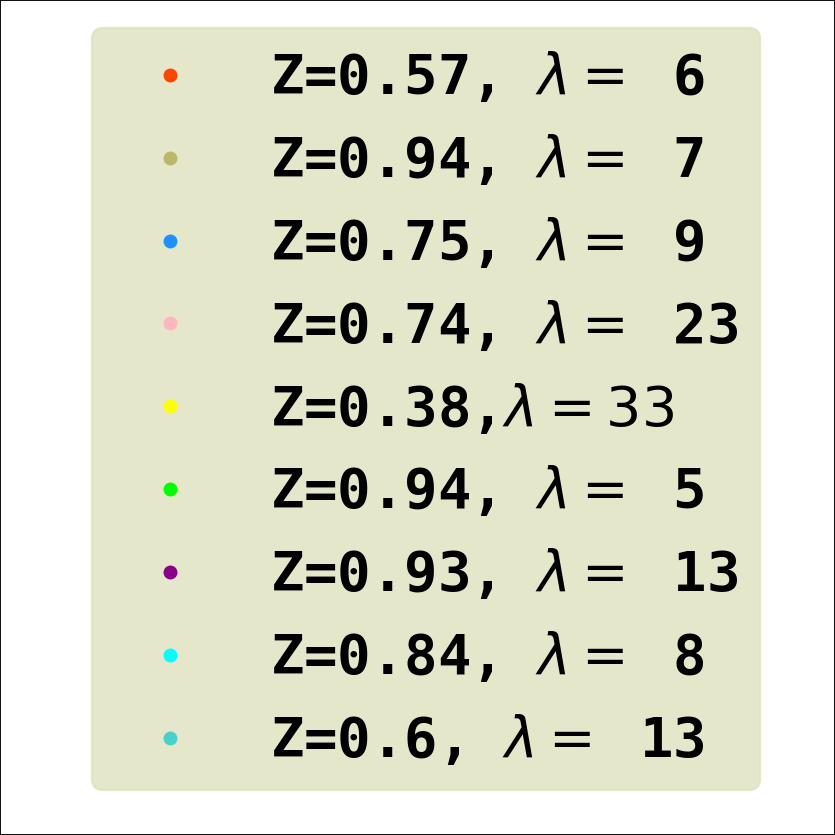} &
     \includegraphics[width=0.33\textwidth]{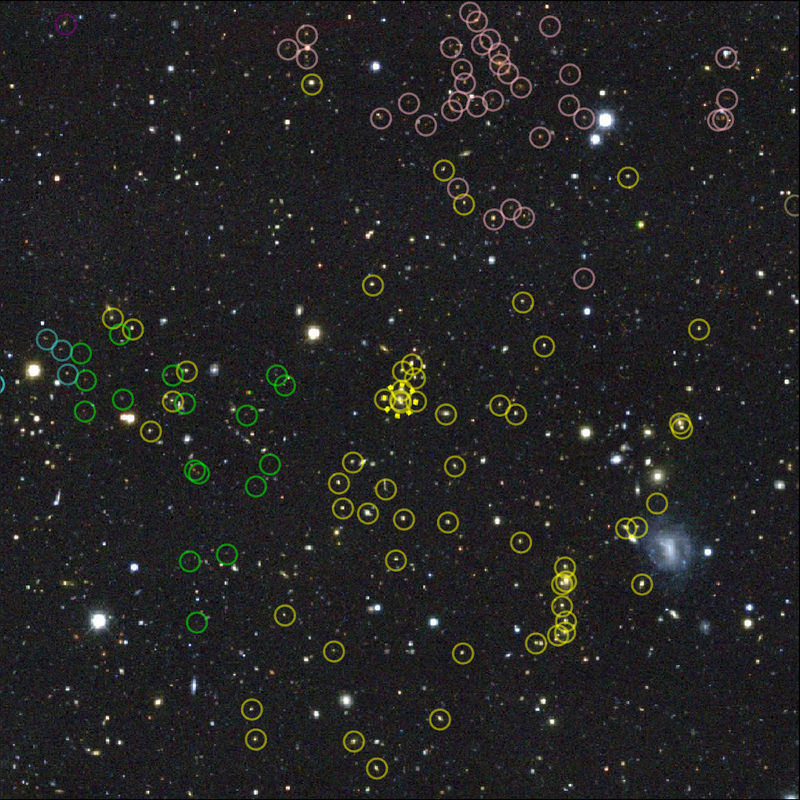} &
     
     \includegraphics[width=0.33\textwidth]{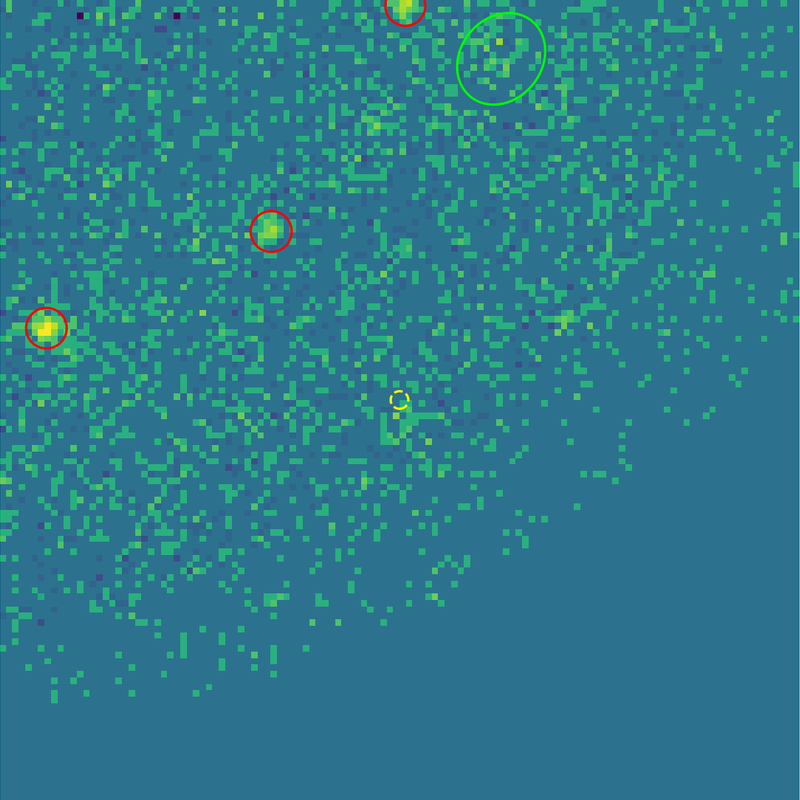}
     \\
\end{tabular}
	\caption{3 examples of associating DES optical clusters with XCS X-ray clusters.  All images are 6$\times$6 arcmins.	Left: Legend showing redshift and richnesses for DES clusters in image colour coded by members.  Middle: DES image with members of potential clusters coloured together.  Right: XCS processed image showing extended and/or point like X-ray sources. 
	\\
	\textbf{Top row:} The XMM image shows an extended associated X-ray source (green ellipse) overlaying the optical cluster.  4 point sources (red circles) are also shown.\\
	\textbf{Middle row:} The XMM image shows a point-like source (red circle) but no extended emission detected in the region.  \\
	\textbf{Bottom row:} An extended source is detected in the XMM image but it is associated with the z=0.74,$\lambda$=23 DES cluster and not the central DES cluster at z=0.38,$\lambda$=33. The lack of detection of the central cluster is likely due to the reduced efficiency of the detector towards the edge of the chip.  }
	\label{fig:app_DES_eyeball}
\end{figure*}

\begin{figure*}

\centering
\textbf{Eyeballing X-ray clusters for redMaPPer counterparts}\par\medskip
\begin{tabular}{ccc}
     \includegraphics[width=0.33\textwidth]{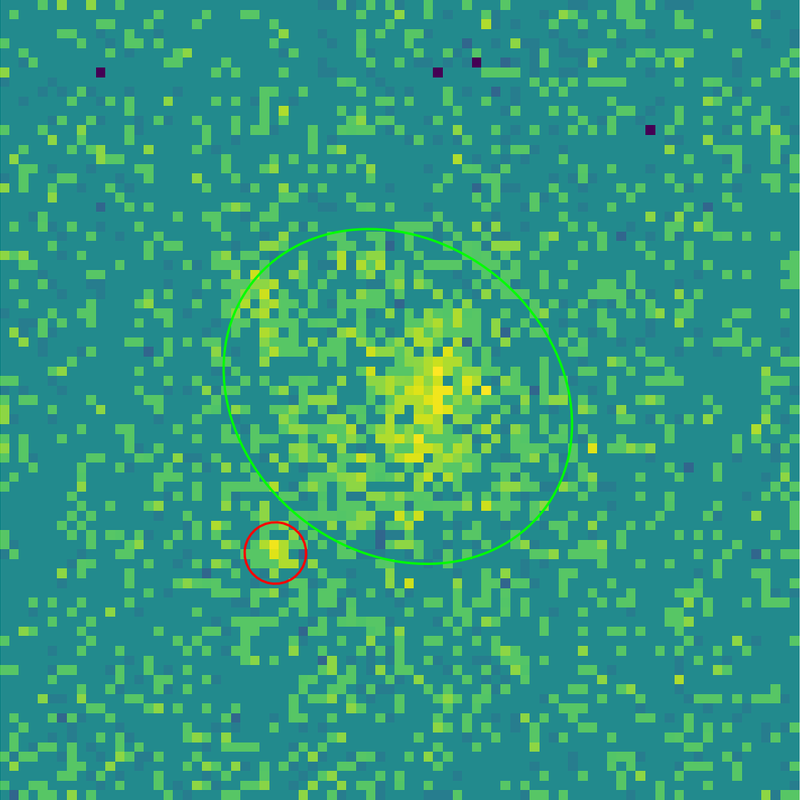} &
\includegraphics[width=0.33\textwidth]{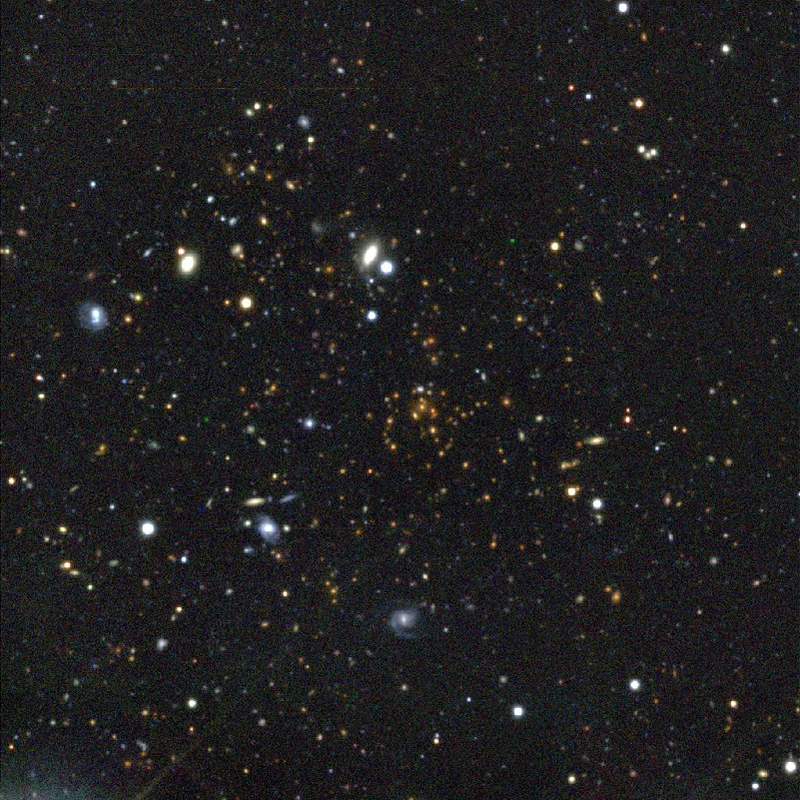} &

     \includegraphics[width=0.33\textwidth]{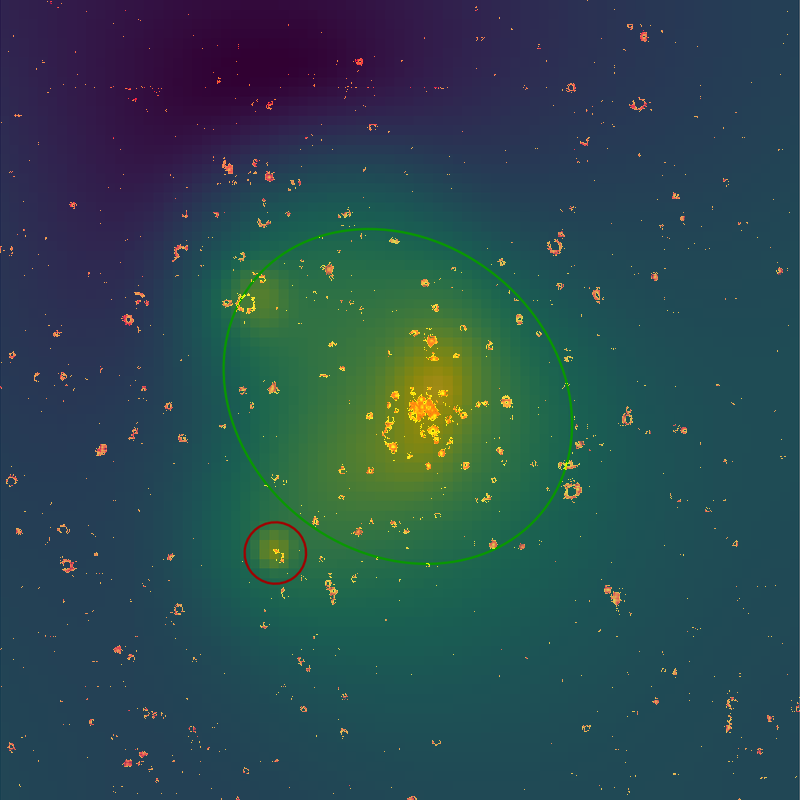}
\\
     \includegraphics[width=0.33\textwidth]{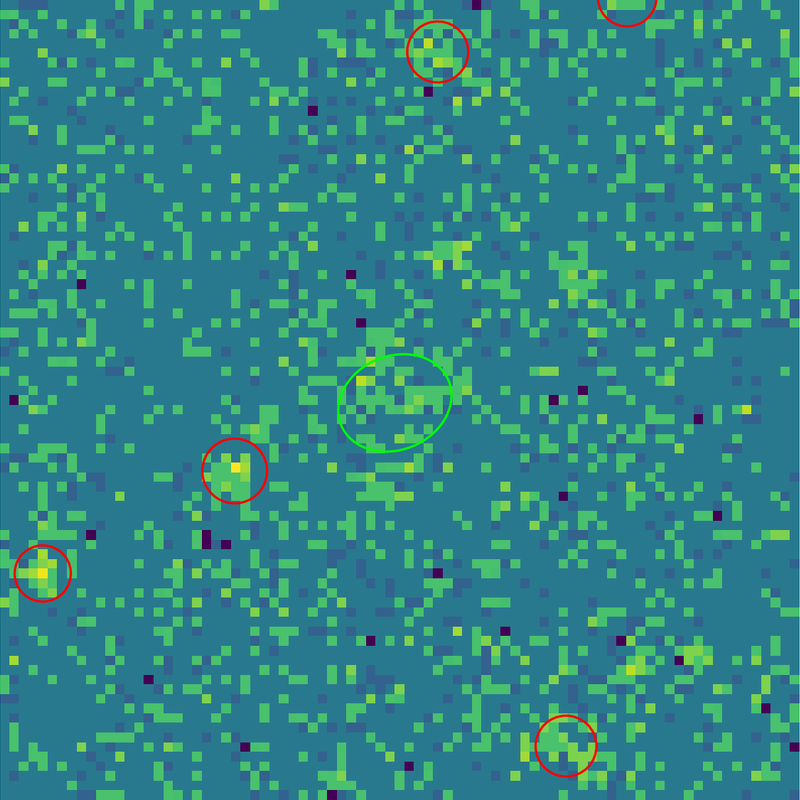} &
\includegraphics[width=0.33\textwidth]{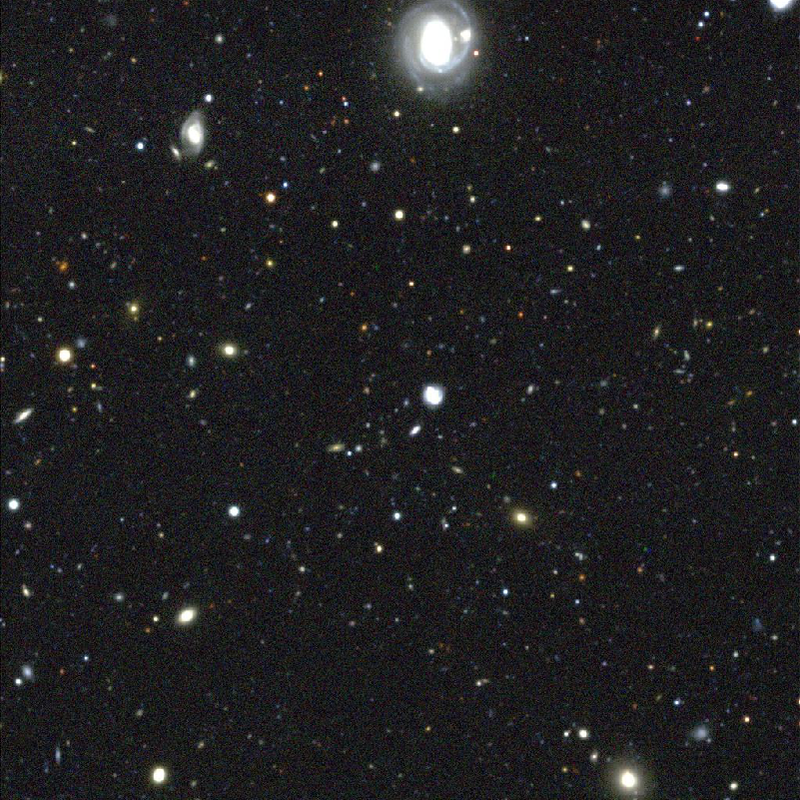} &

     \includegraphics[width=0.33\textwidth]{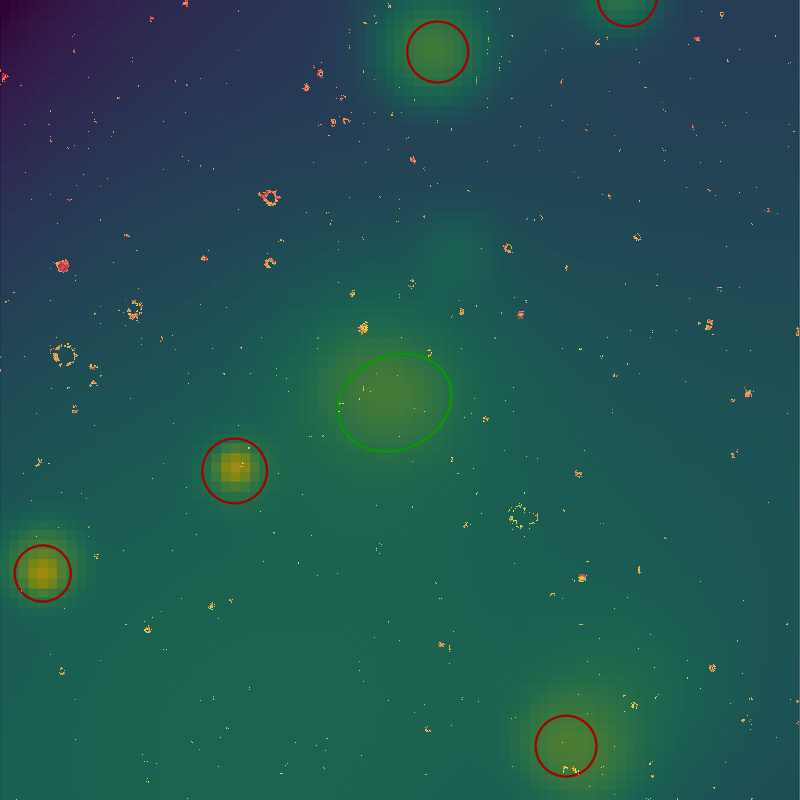}

\end{tabular}
     \caption{2 examples of using contrast enhanced DES images to confirm X-ray extended sources. All images are 6$\times$6 arcmins. Left: XMM image showing XAPA extended source detection.  Middle: DES image.  Right: XCS image showing smoothed X-ray signal and DES image enhanced red-channel to highlight red clusters.\\
     \textbf{Top row:} Clear example of red galaxy overabundance in both the pure DES image and the overlay.  This extended source is therefore optically confirmed.\\
     \textbf{Bottom row:} Example of a XAPA detection with no red galaxy over abundance showing in the DES image.  The contrast enhancements also shows no red galaxies.  This extended source is thus not optically confirmed.}
     \label{Fig:App_xray_eyeballs}
\end{figure*}

\begin{figure*}

\centering
\textbf{The 4 X-ray Detected Clusters Not Found in redMaPPer Catalogue}\par\medskip
\begin{tabular}{cc}

     \includegraphics[width=0.5\textwidth]{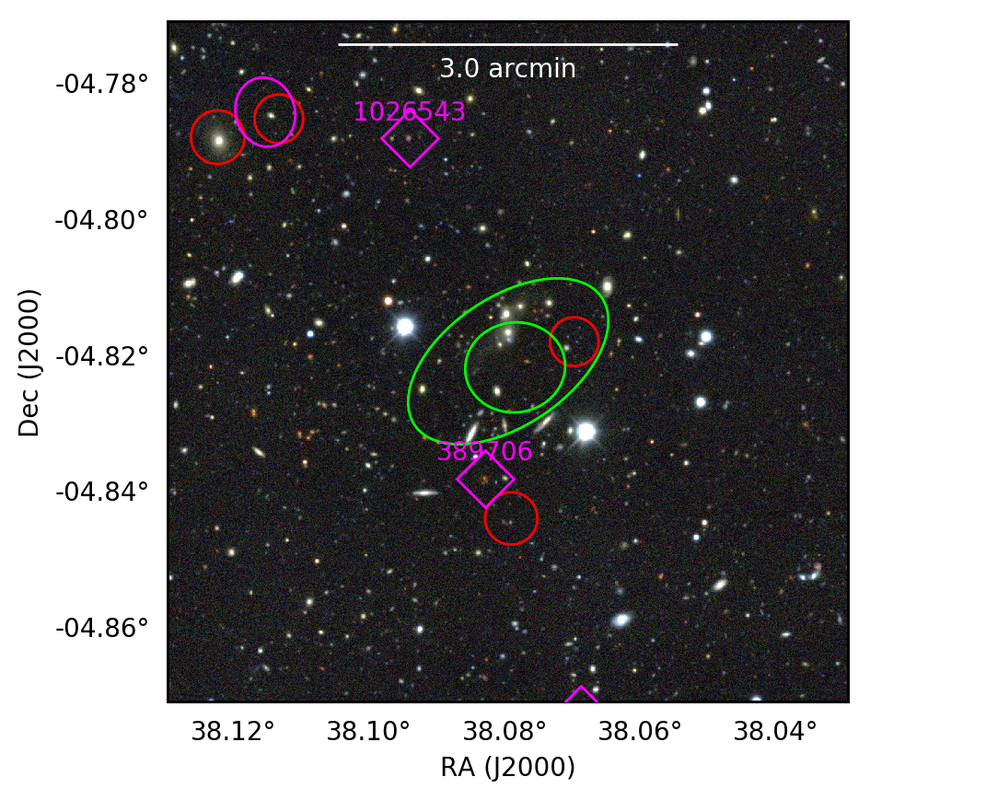} &

     \includegraphics[width=0.5\textwidth]{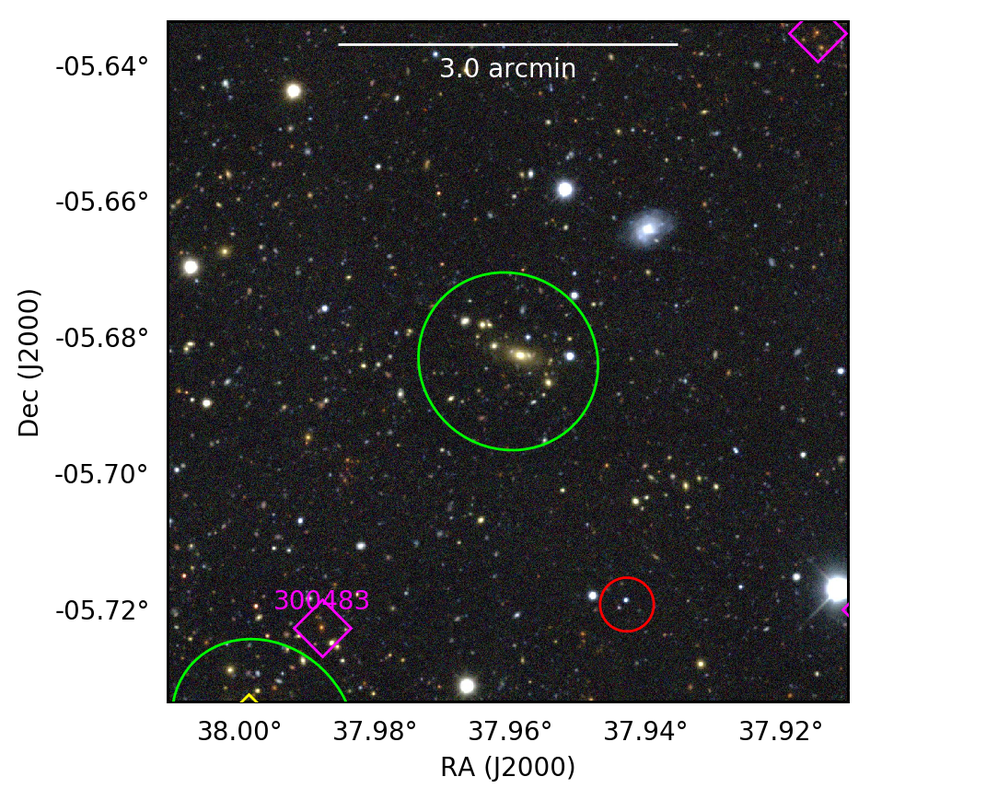}
     \\
     \includegraphics[width=0.5\textwidth]{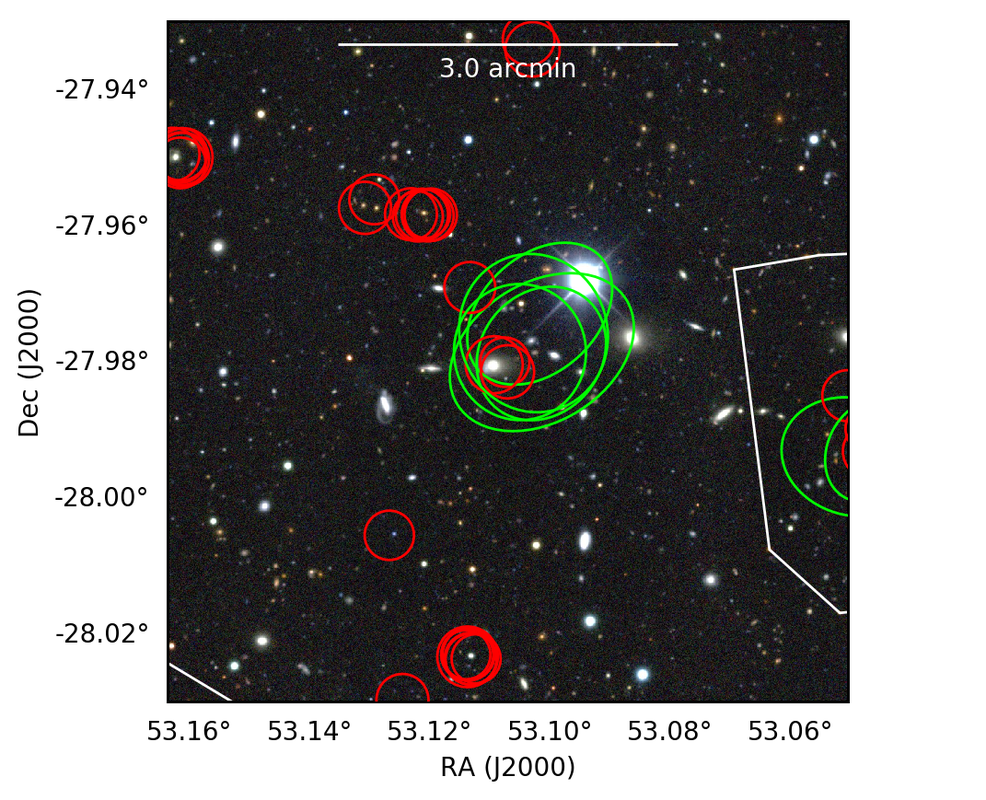} &

     \includegraphics[width=0.5\textwidth]{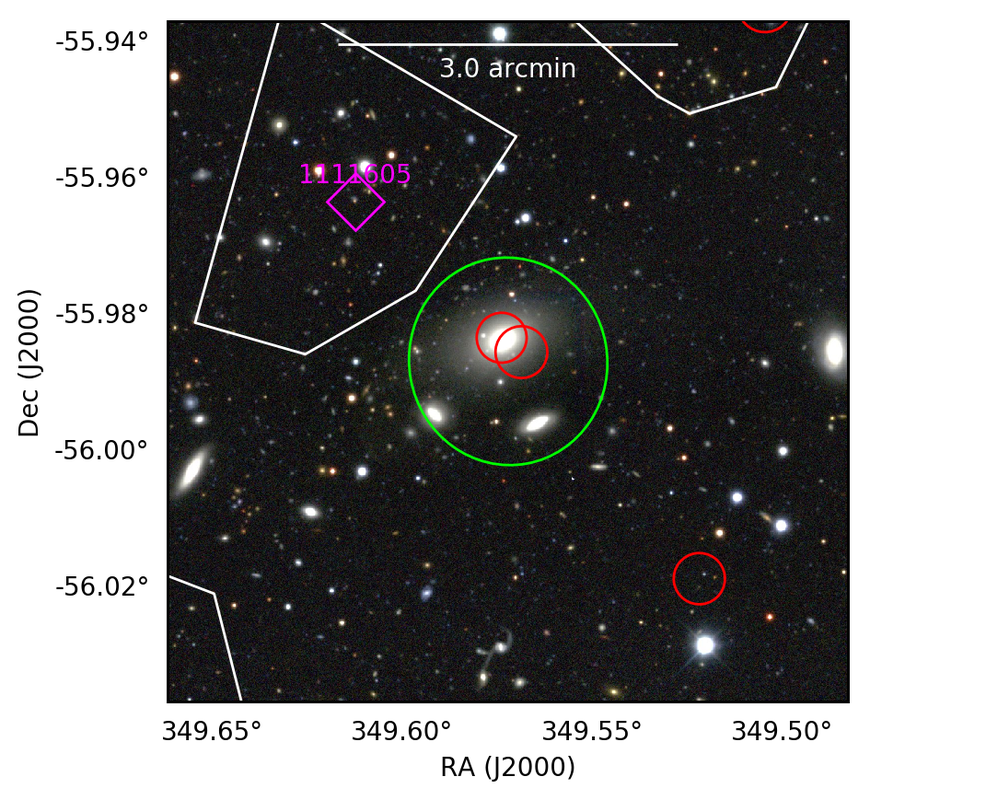}

\end{tabular}
     \caption{The four X-ray detected clusters that are not in the redMaPPer catalogue.  Green ellipses represent X-ray detected extended sources.  Purple diamonds are centred on redMaPPer clusters.\\
     \textbf{Top Left:} From redMaPPer scan - Redshift: 0.2     $\lambda: 15.26$\\
     \textbf{Top Right:} From redMaPPer scan -Redshift: 0.34    $\lambda: 9.233$\\
     \textbf{Bottom Left:} From redMaPPer scan - Redshift: 0.12     $\lambda: 5.95$\\
     \textbf{Bottom Right:} From redMaPPer scan -Redshift: 0.09    $\lambda: 13.93$
     }
\label{fig:app_missing}
\end{figure*}

Here we present some examples of various images referenced in the main paper.  Figure \ref{fig:app_DES_eyeball} shows typical images used to eyeball DES images for associated X-ray counterparts.  Figure \ref{Fig:App_xray_eyeballs} shows images used to eyeball X-ray images for optical red-galaxy overabundances.  Figure \ref{fig:app_DES_mask} is an example of a DES mask affecting detection of an optical cluster.  Figure \ref{fig:app_high_redshifts} shows the 14 high redshift clusters outside the detection limits of redMapper (and thus not in the redMaPPer catalogue) but detected in X-ray.  Figure \ref{fig:app_missing} shows the four X-ray detected clusters that are within the redMaPPer parameter space but are not found in the catalogue.

We note that visual inspection can be prone to human variation.  We mitigate this by doing classification sessions in pairs involving discussion.  When agreement is not found, a third expert is consulted.

\onecolumn

\begin{figure*}

    \centering
    \includegraphics[scale=0.5]{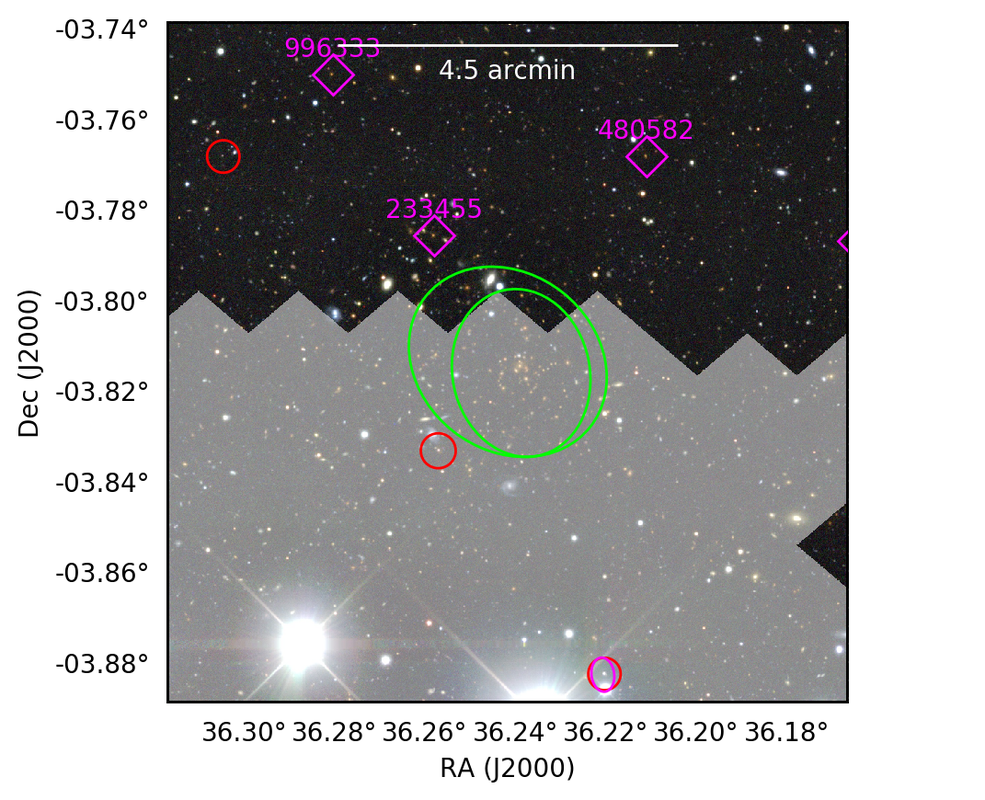}
    \caption{An example of a DES masked region that covers an obvious cluster.  This is the same cluster as Figure \ref{Fig:App_xray_eyeballs} top.  Again, green ellipses represent X-ray cluster detections whilst purple diamonds represent redMaPPer cluster centroids and their MEM\_MATCH\_ID.  The grey overlay represents the DES Mask casue by the two bright stars at the bottom of the cutout. As can be seen, the cluster is covered by a mask and therefore does not appear in the redMaPPer catalogue despite being clear and obvious.}
\label{fig:app_DES_mask}
\end{figure*}

\begin{figure*}

\centering
\textbf{The 13 high redshift clusters detected in X-ray not in the redMaPPer catalogue}\par\medskip

\begin{tabular}{ccc}
     \includegraphics[width=0.33\textwidth]{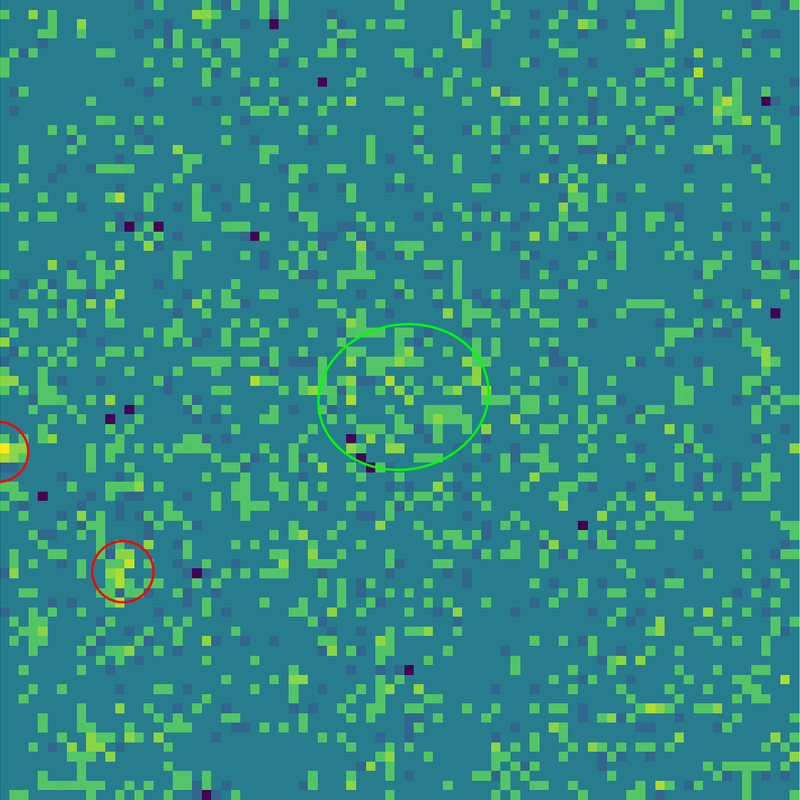} &
     \includegraphics[width=0.33\textwidth]{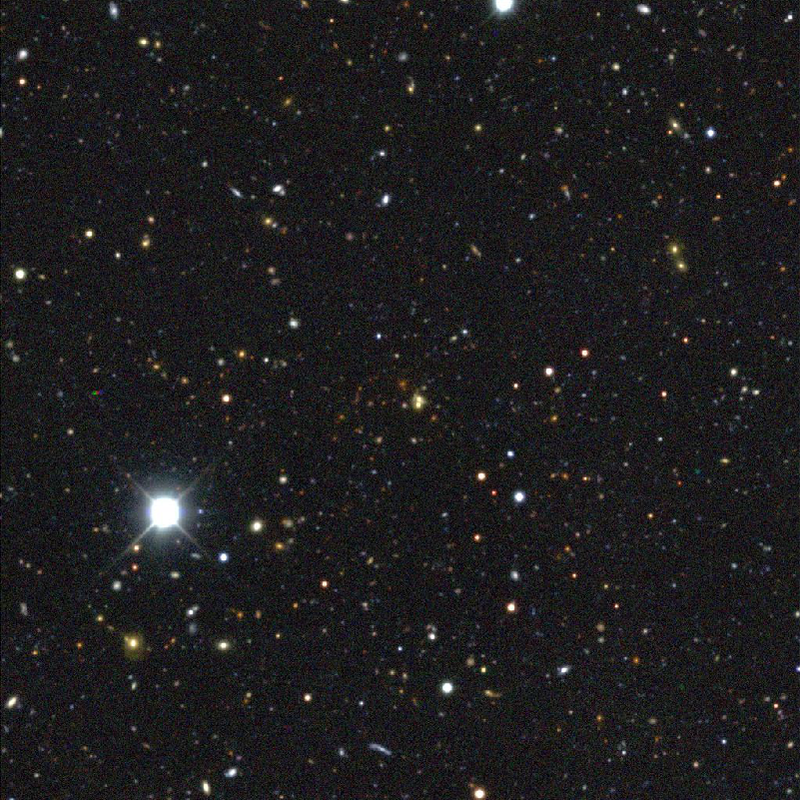} &
     \includegraphics[width=0.33\textwidth]{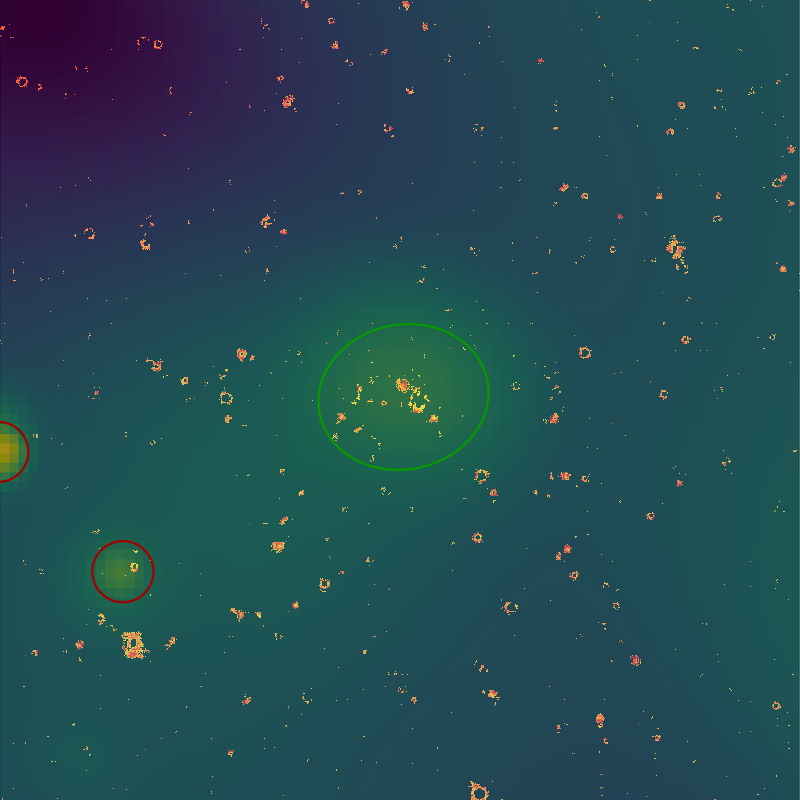}
\\
     \includegraphics[width=0.33\textwidth]{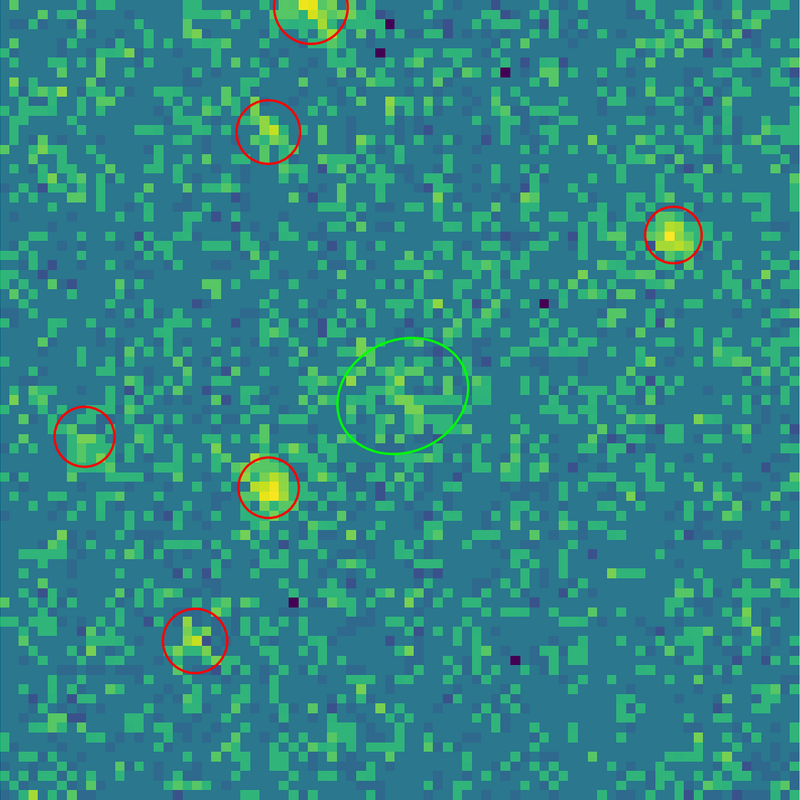} &
     \includegraphics[width=0.33\textwidth]{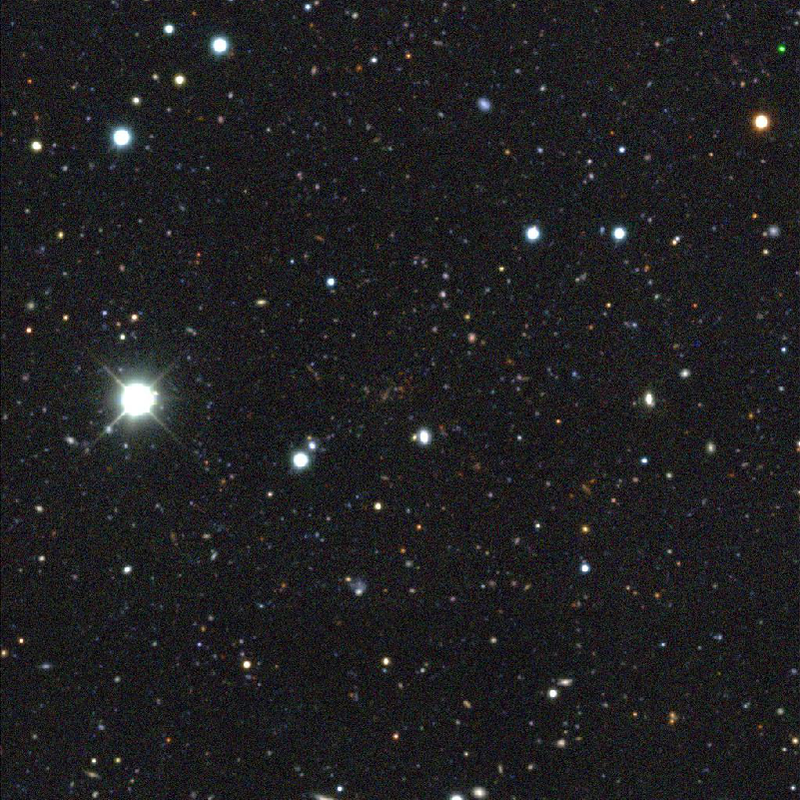} &
     \includegraphics[width=0.33\textwidth]{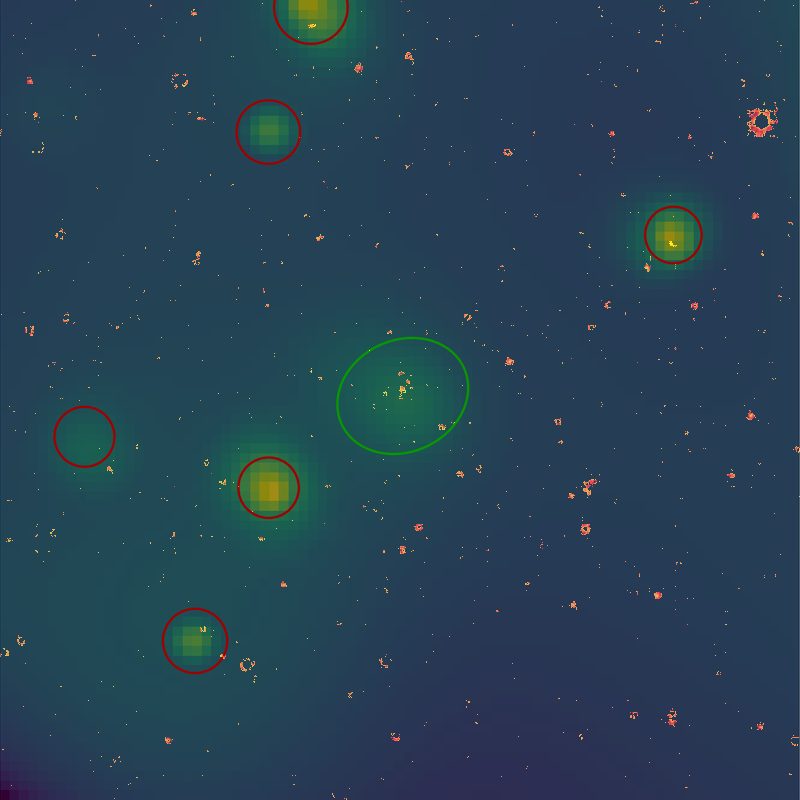}
\\
     \includegraphics[width=0.33\textwidth]{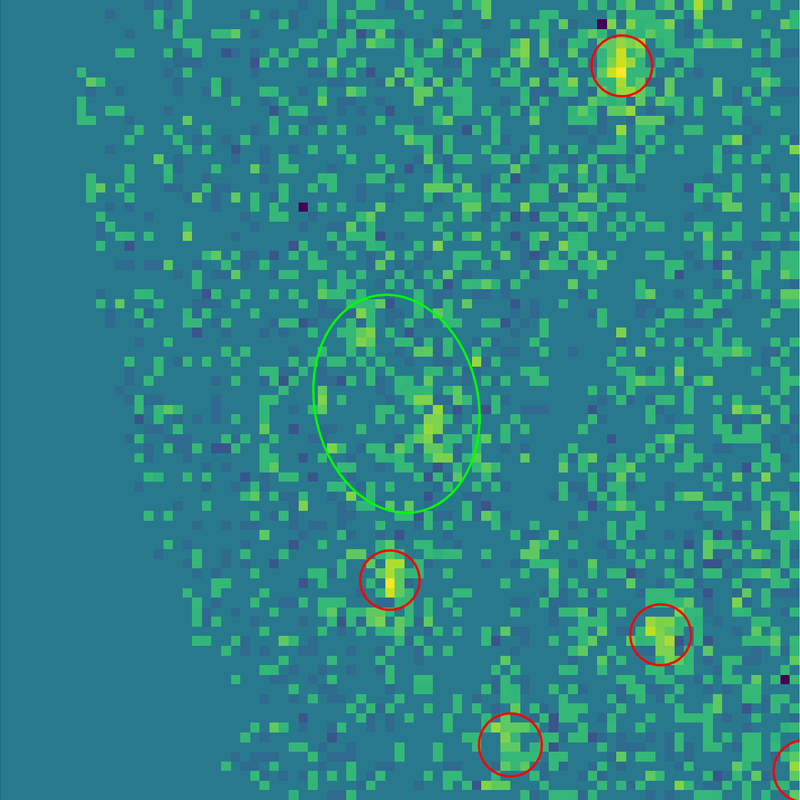} &
     \includegraphics[width=0.33\textwidth]{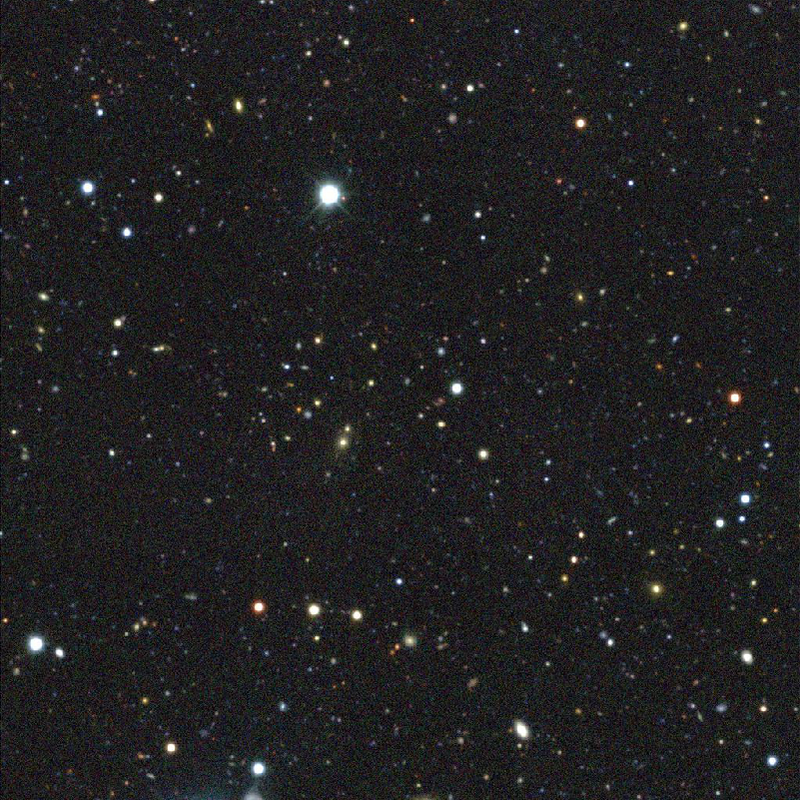} &
     \includegraphics[width=0.33\textwidth]{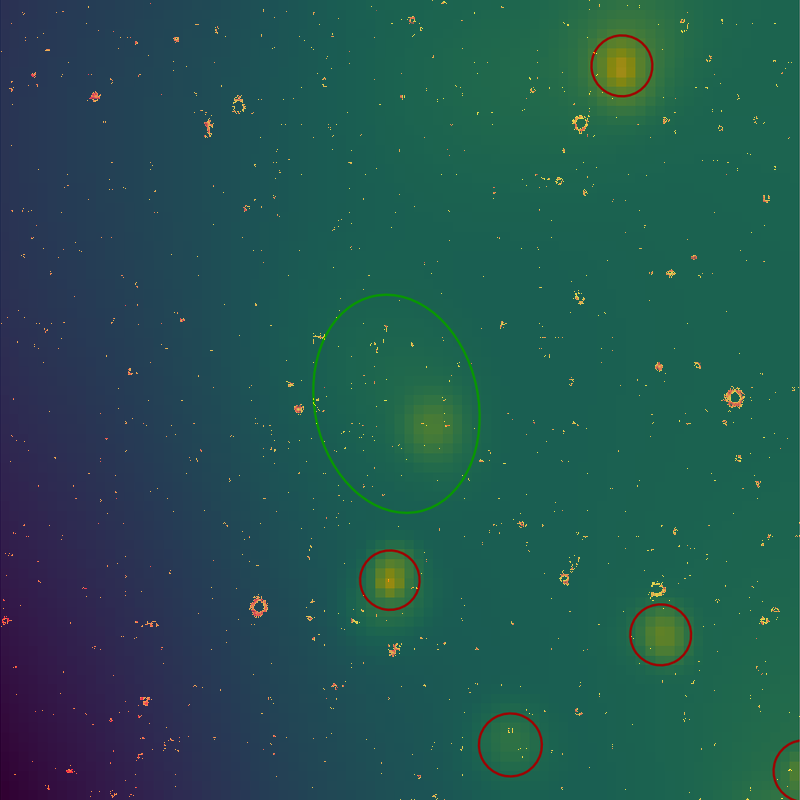}
\\
     \includegraphics[width=0.33\textwidth]{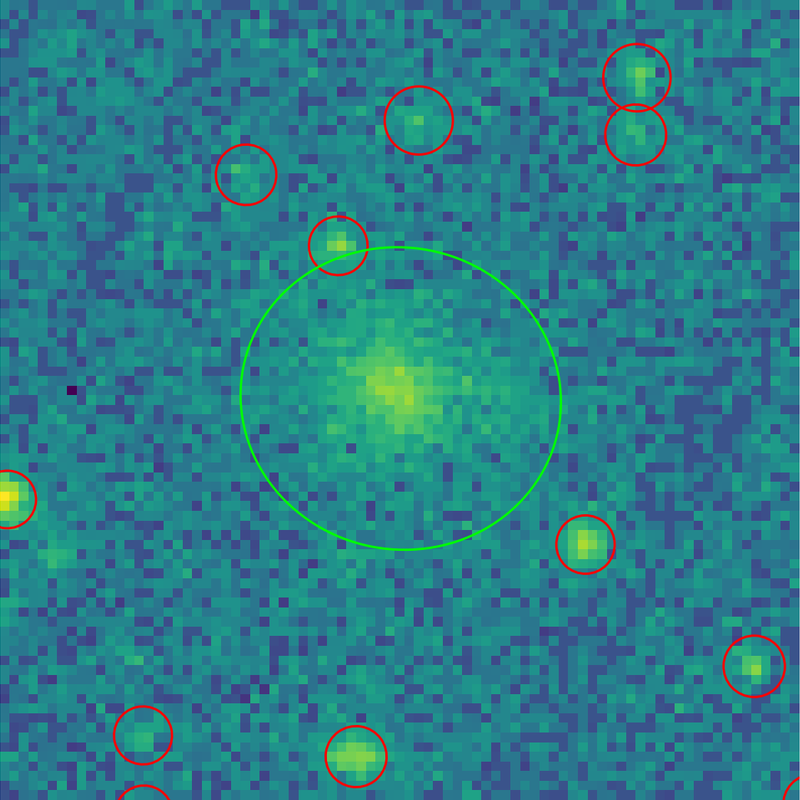} &
     \includegraphics[width=0.33\textwidth]{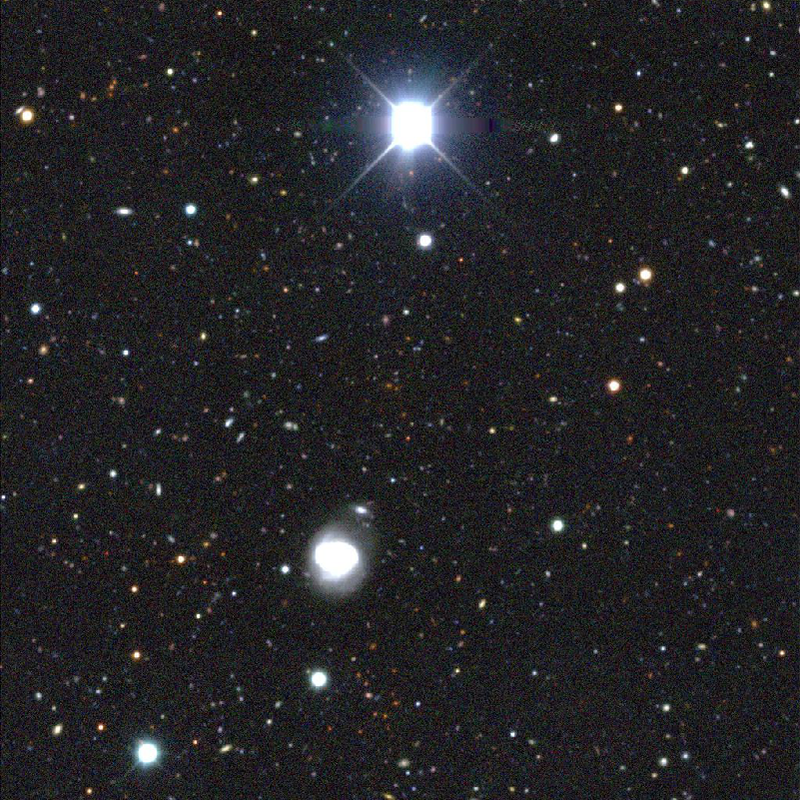} &
     \includegraphics[width=0.33\textwidth]{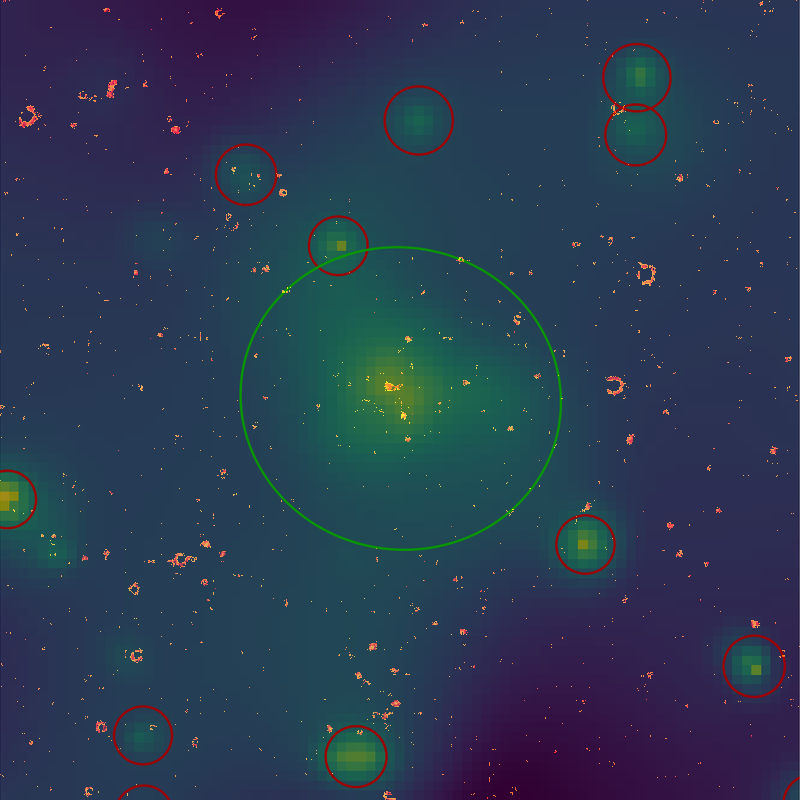}
    \end{tabular}
       \caption{cont.}
    \end{figure*}

    \begin{figure*}
\ContinuedFloat
\begin{tabular}{ccc}
     \includegraphics[width=0.33\textwidth]{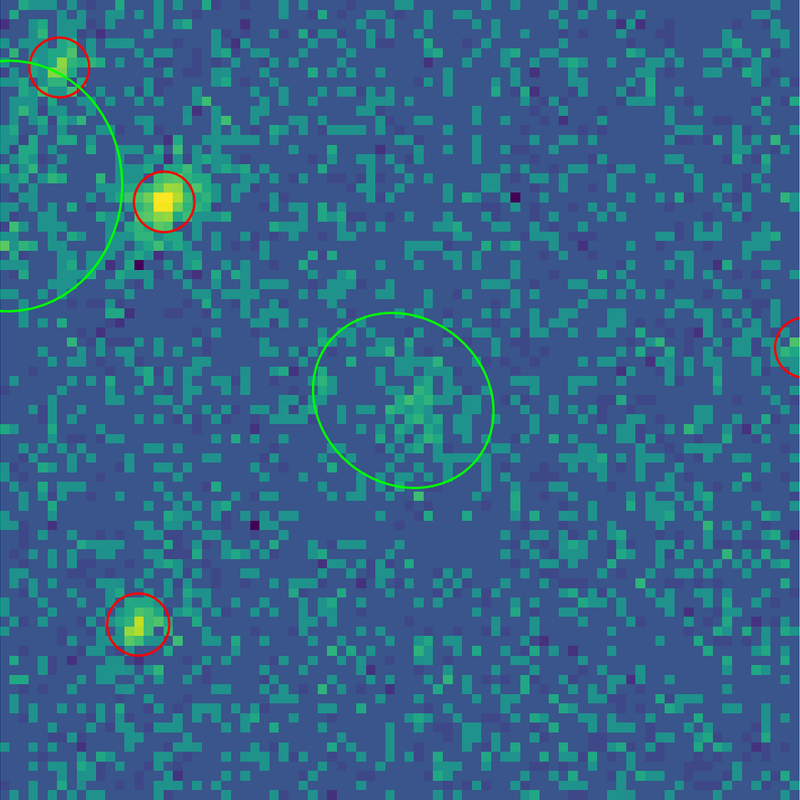} &
     \includegraphics[width=0.33\textwidth]{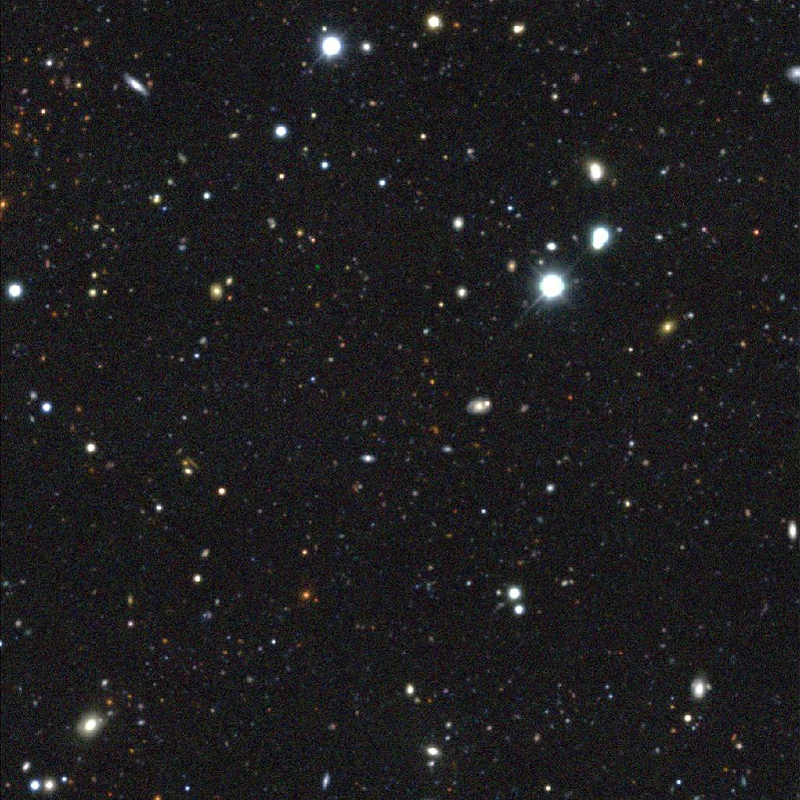} &
     \includegraphics[width=0.33\textwidth]{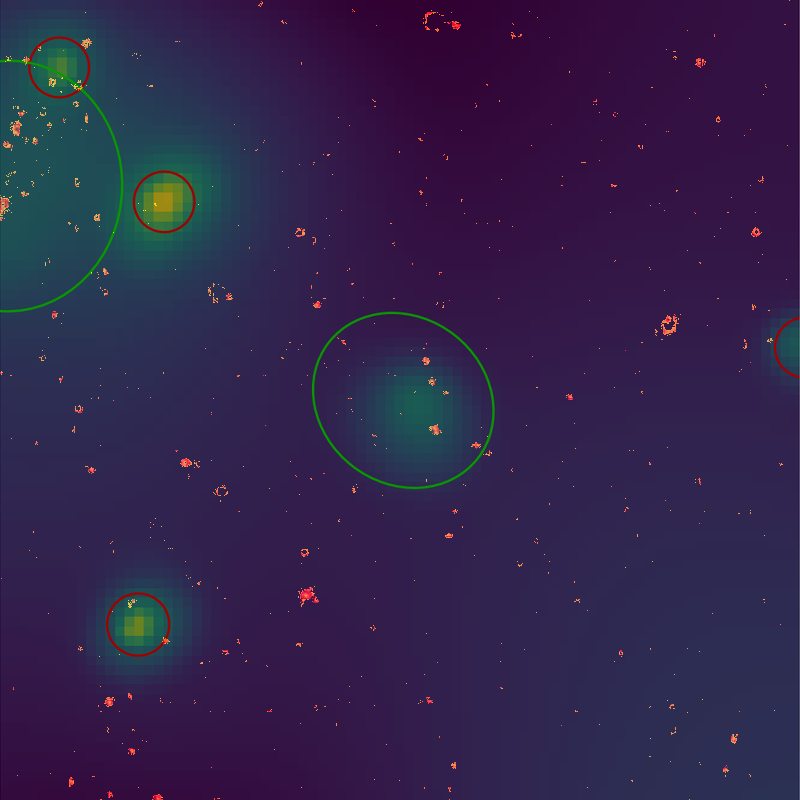}
\\
     \includegraphics[width=0.33\textwidth]{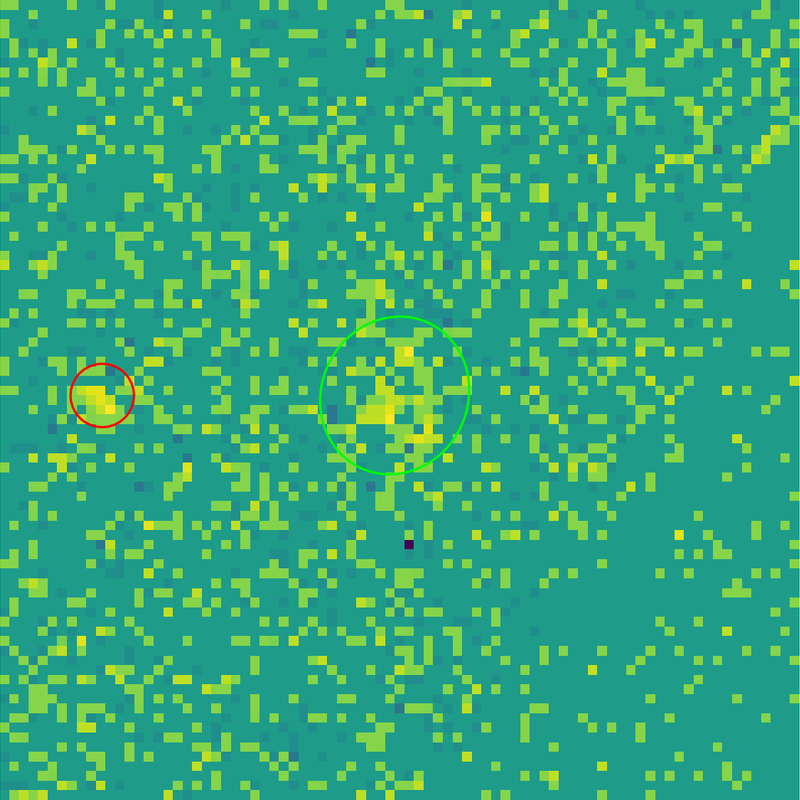} &
     \includegraphics[width=0.33\textwidth]{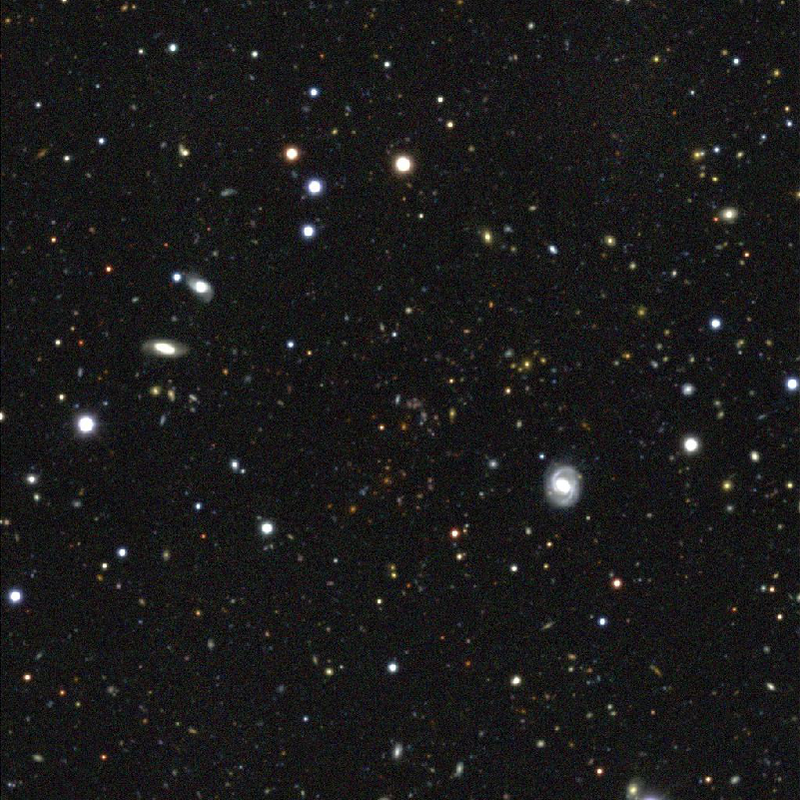} &
     \includegraphics[width=0.33\textwidth]{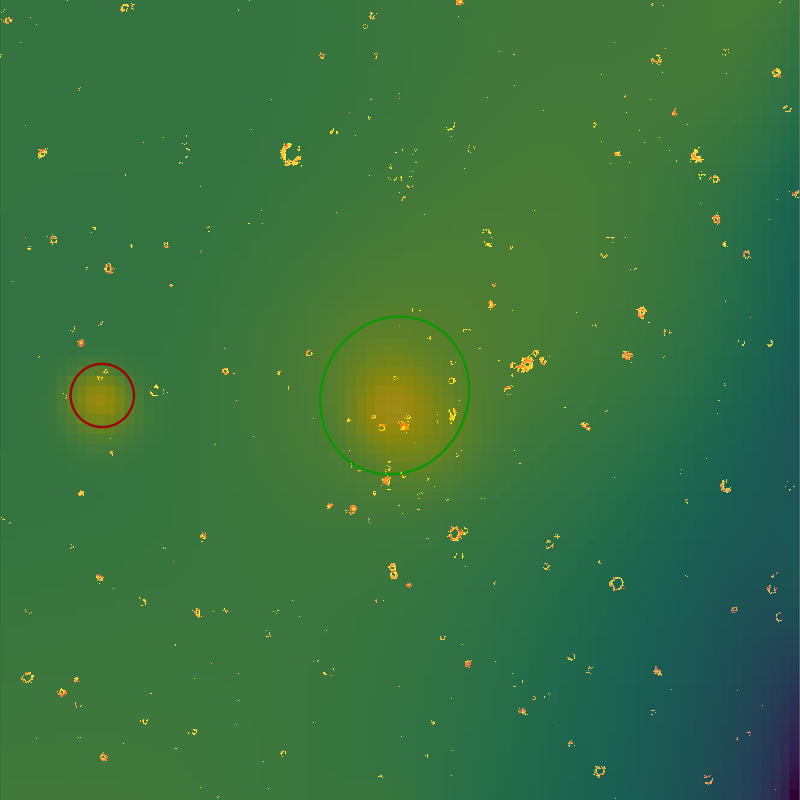}
\\
     \includegraphics[width=0.33\textwidth]{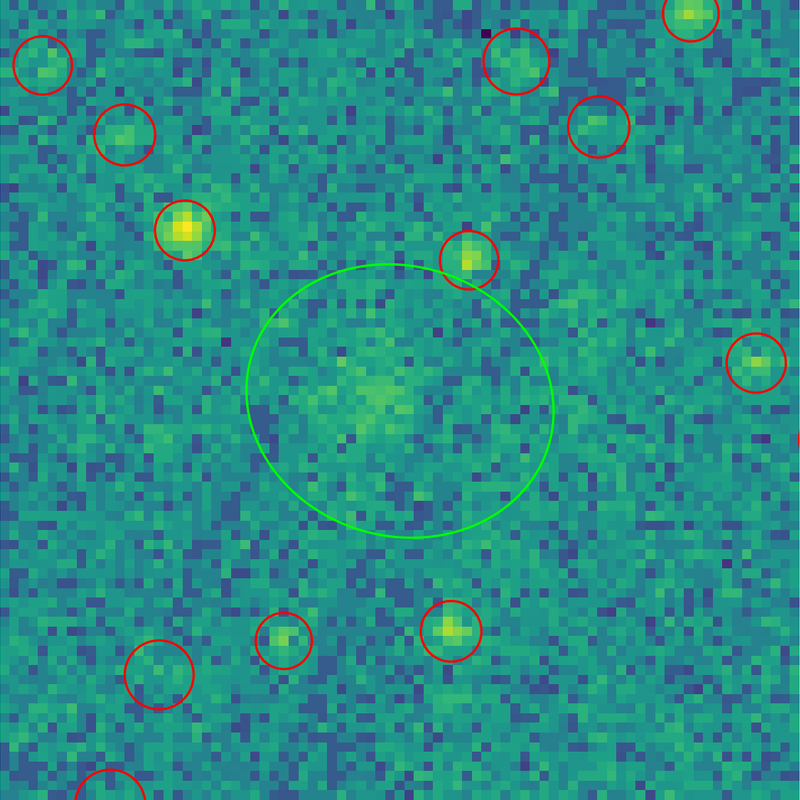} &
     \includegraphics[width=0.33\textwidth]{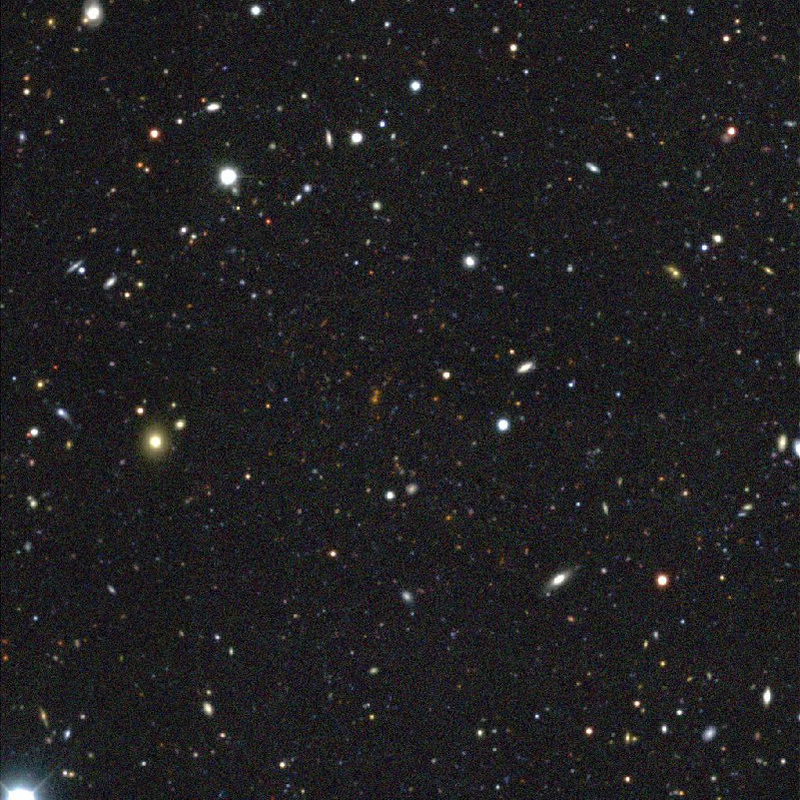} &
     \includegraphics[width=0.33\textwidth]{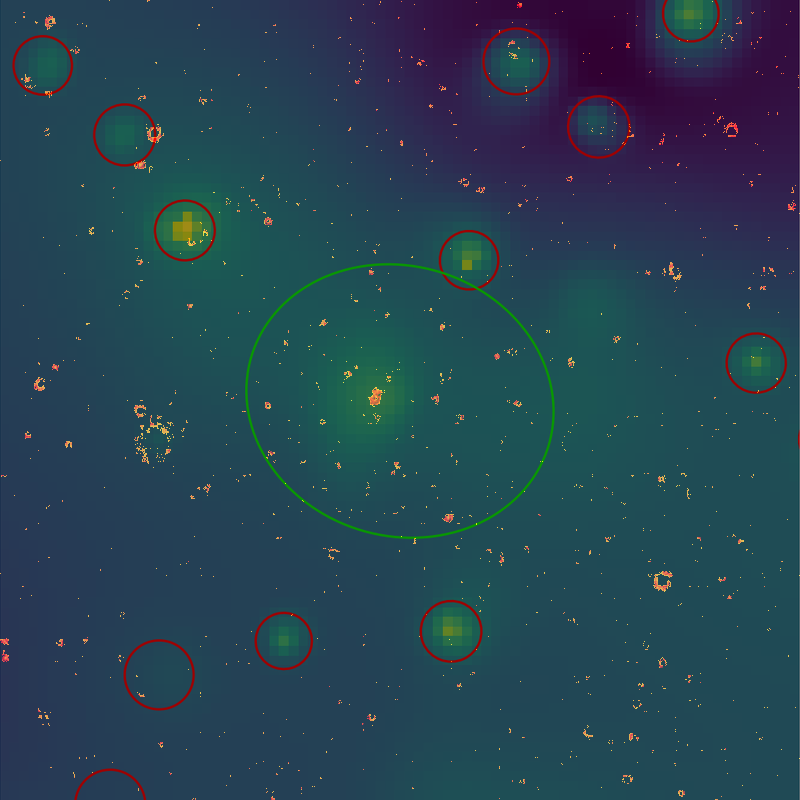}
\\
     \includegraphics[width=0.33\textwidth]{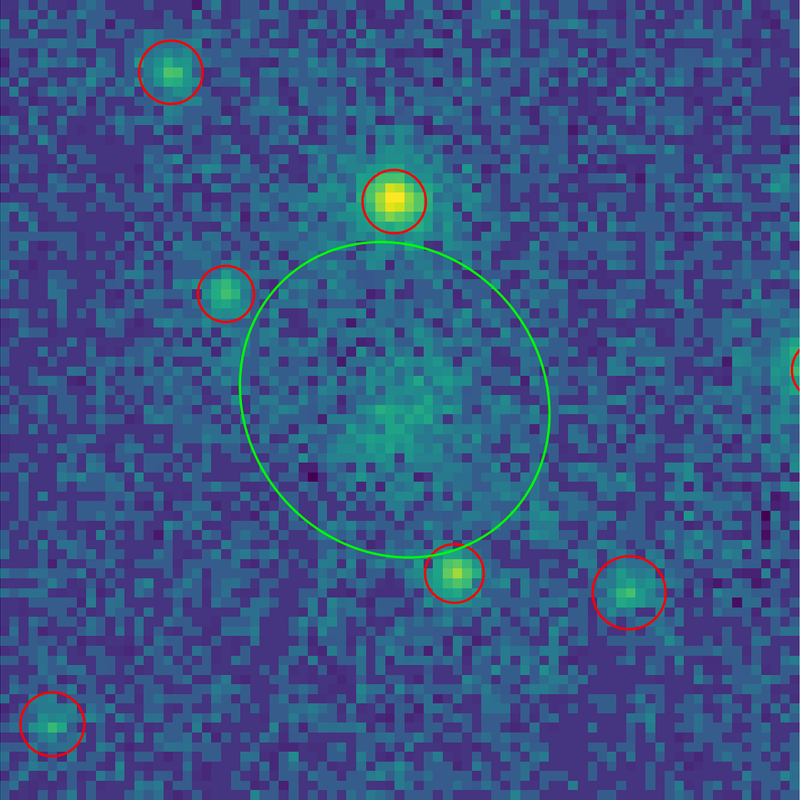} &
     \includegraphics[width=0.33\textwidth]{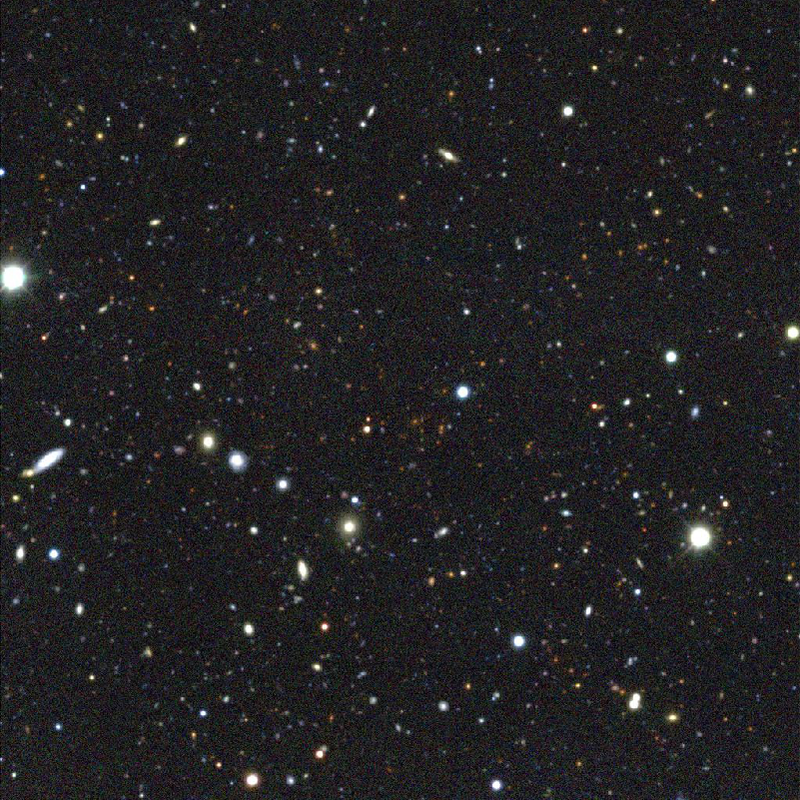} &
     \includegraphics[width=0.33\textwidth]{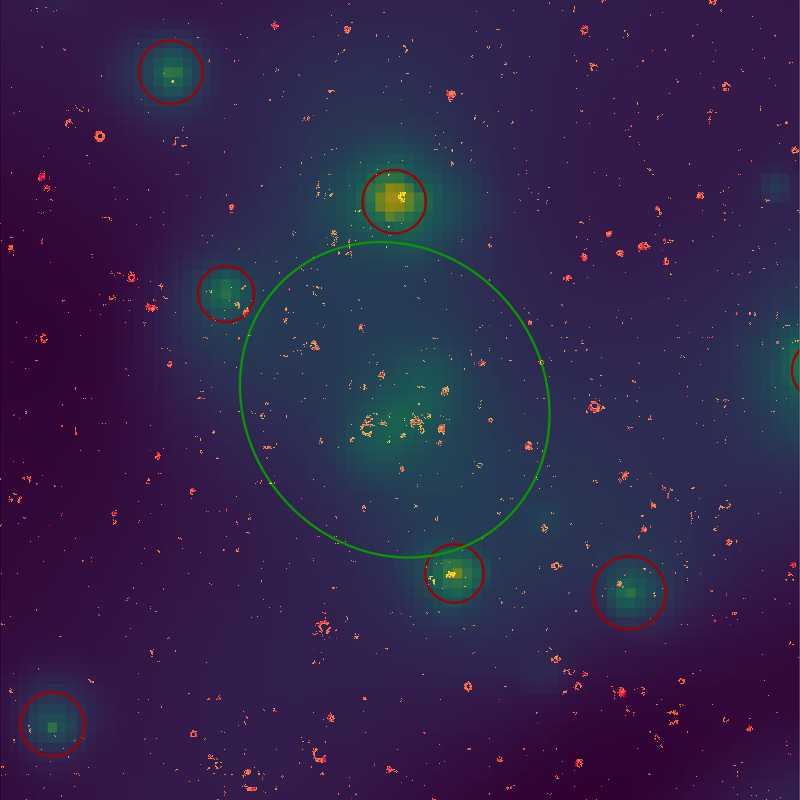}
    \end{tabular}
    \caption{cont.}
    \end{figure*}

    \begin{figure*}
\ContinuedFloat
\begin{tabular}{ccc}
     \includegraphics[width=0.33\textwidth]{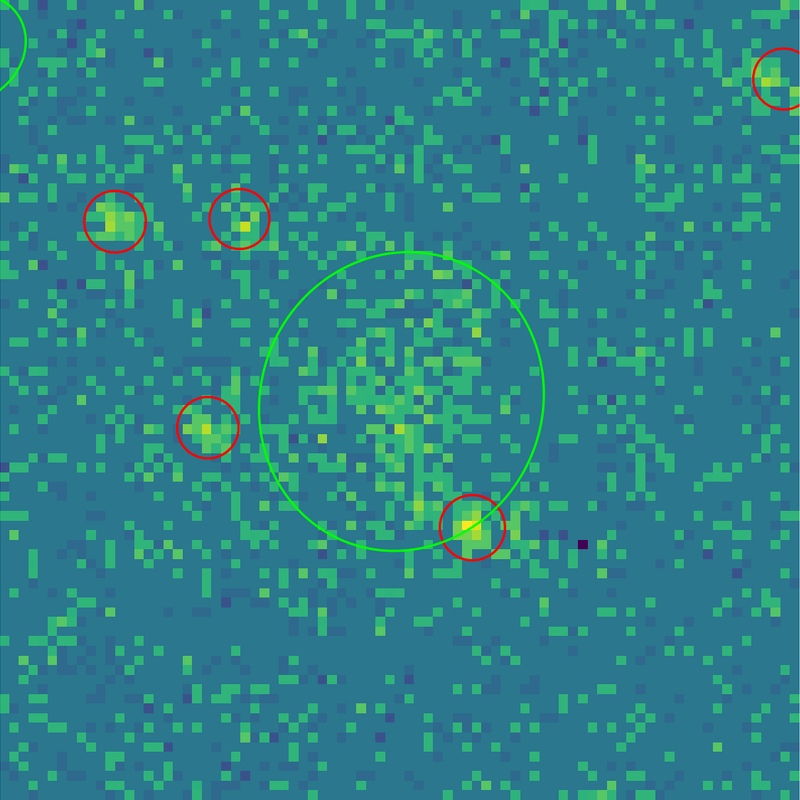} &
     \includegraphics[width=0.33\textwidth]{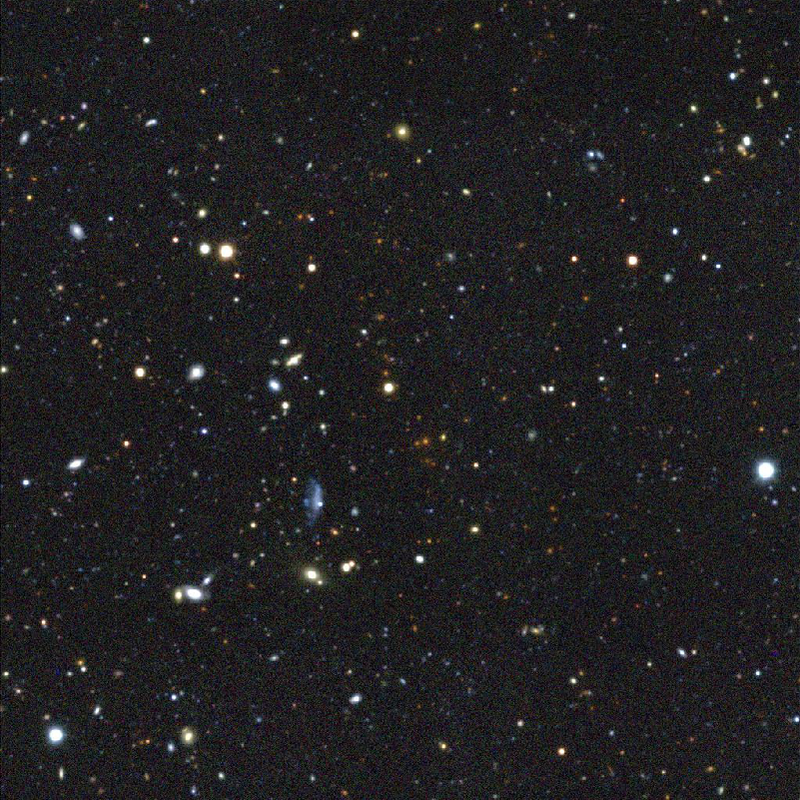} &
     \includegraphics[width=0.33\textwidth]{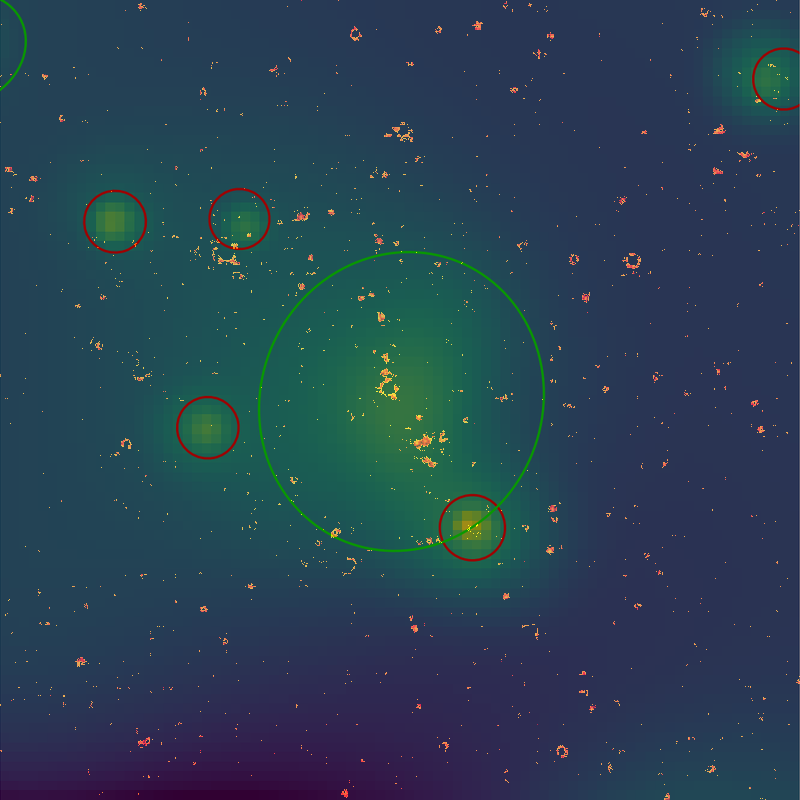}
\\
     \includegraphics[width=0.33\textwidth]{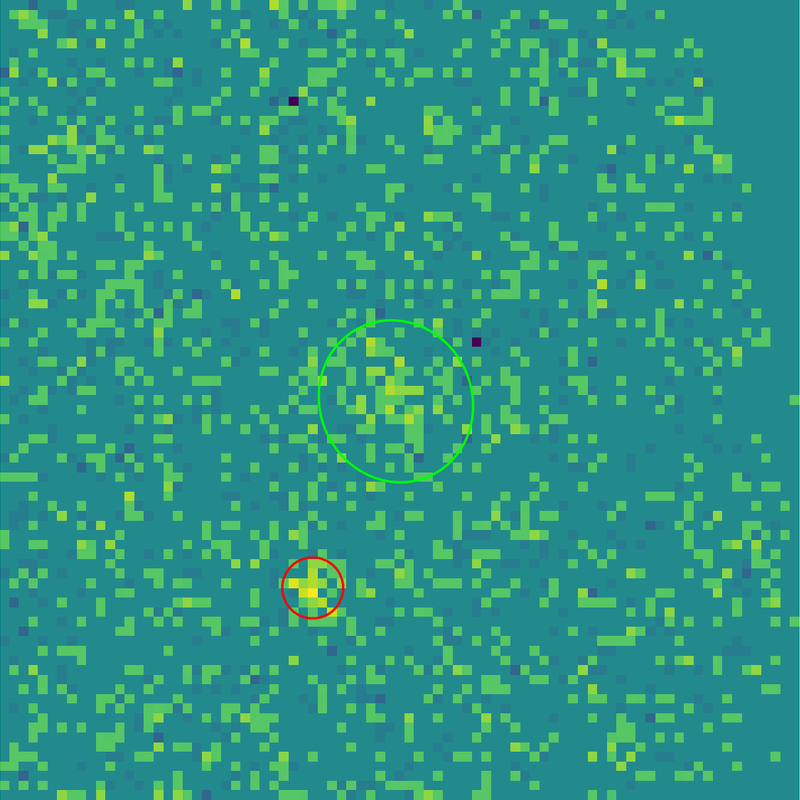} &
     \includegraphics[width=0.33\textwidth]{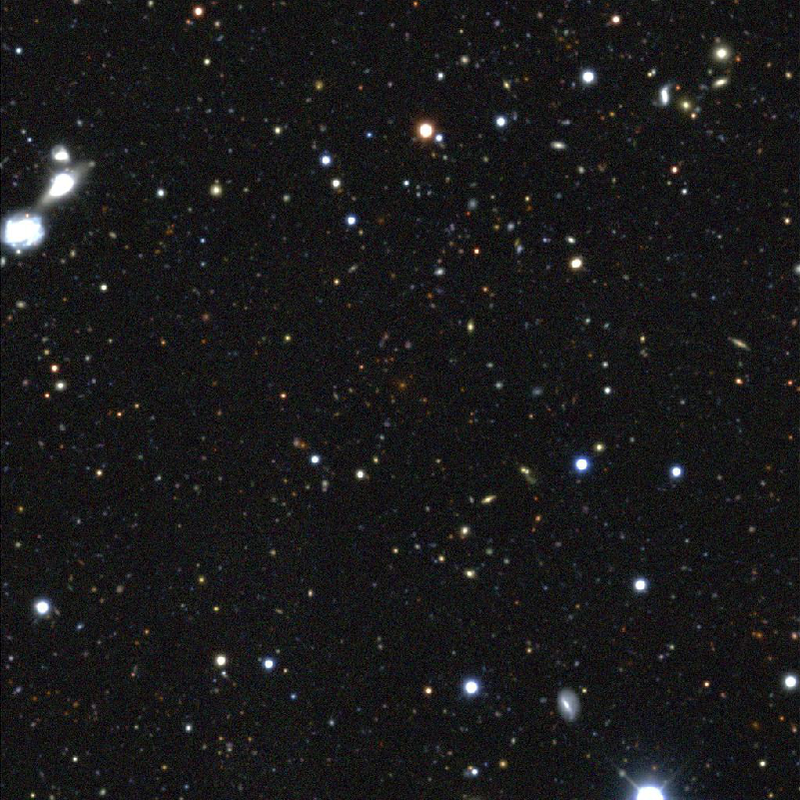} &
     \includegraphics[width=0.33\textwidth]{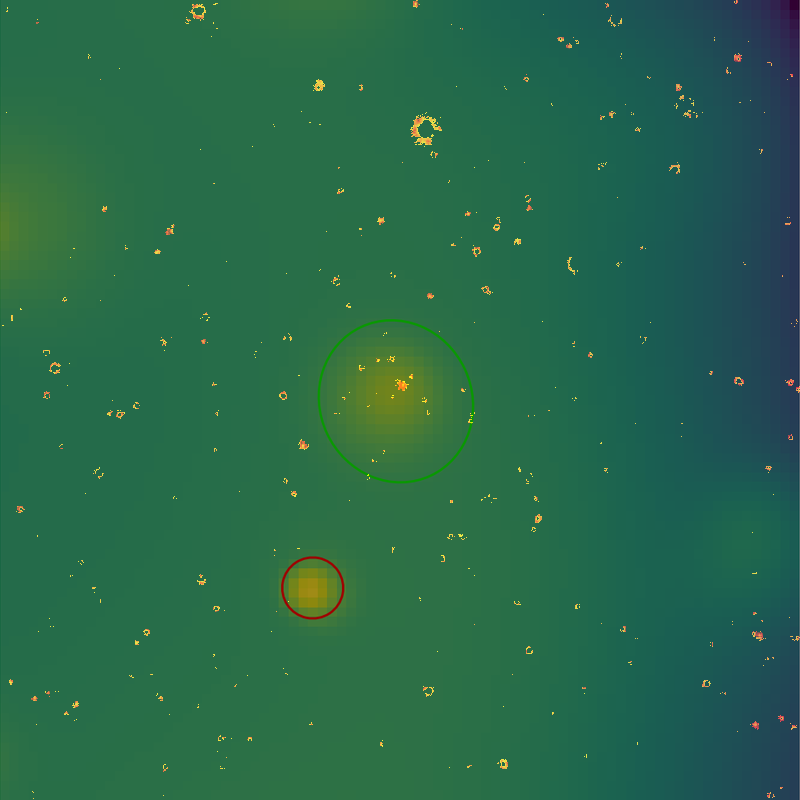}
\\
     \includegraphics[width=0.33\textwidth]{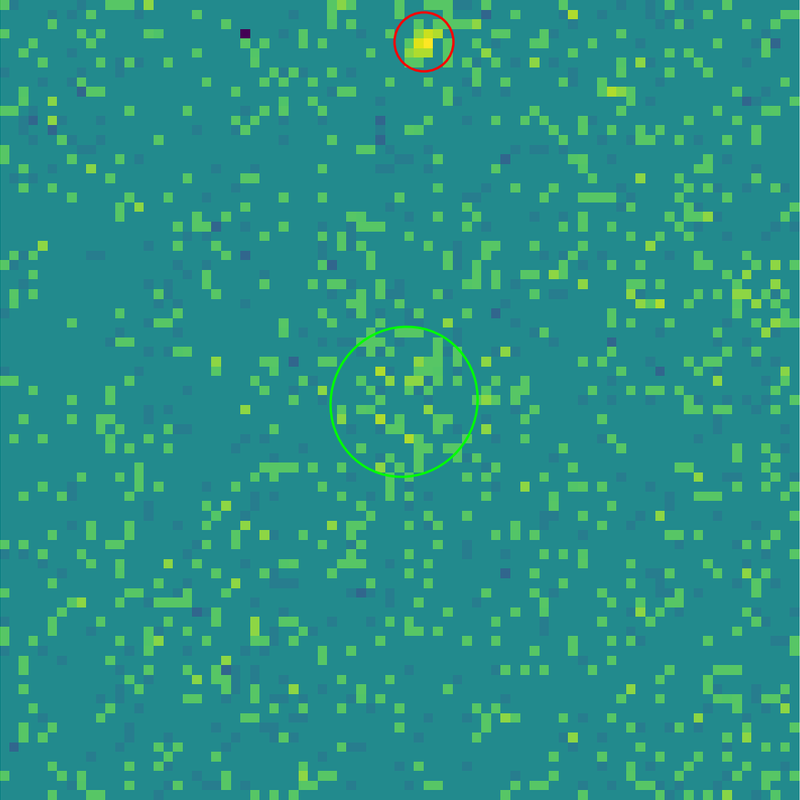} &
     \includegraphics[width=0.33\textwidth]{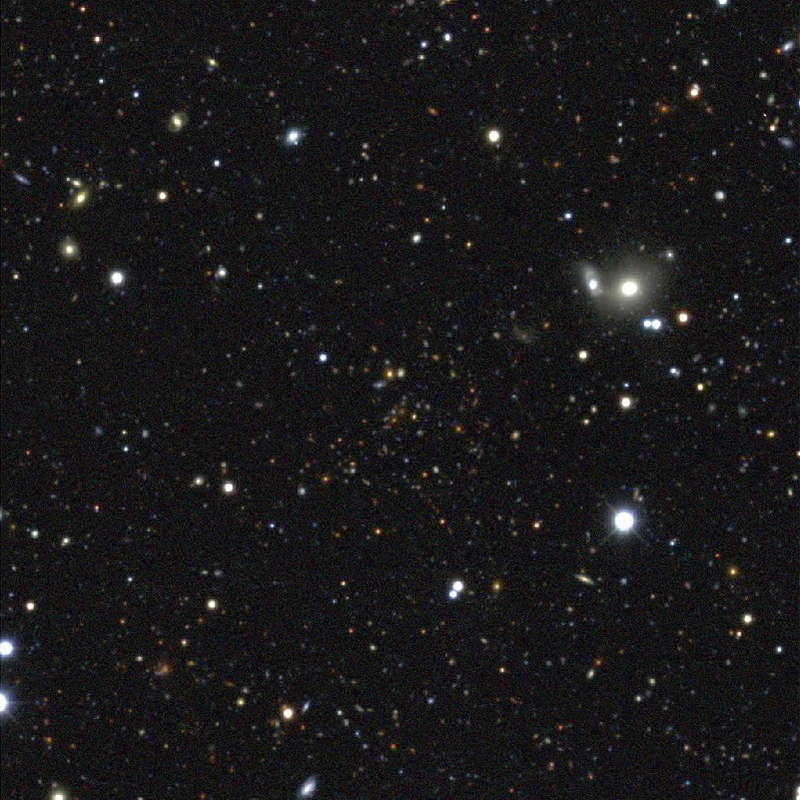} &
     \includegraphics[width=0.33\textwidth]{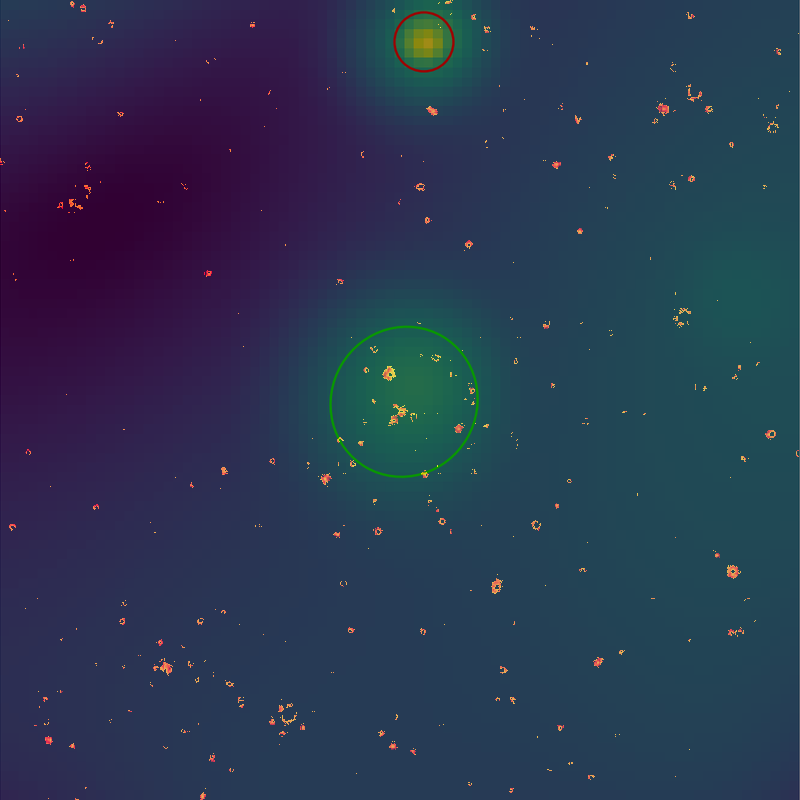}
\\
     \includegraphics[width=0.33\textwidth]{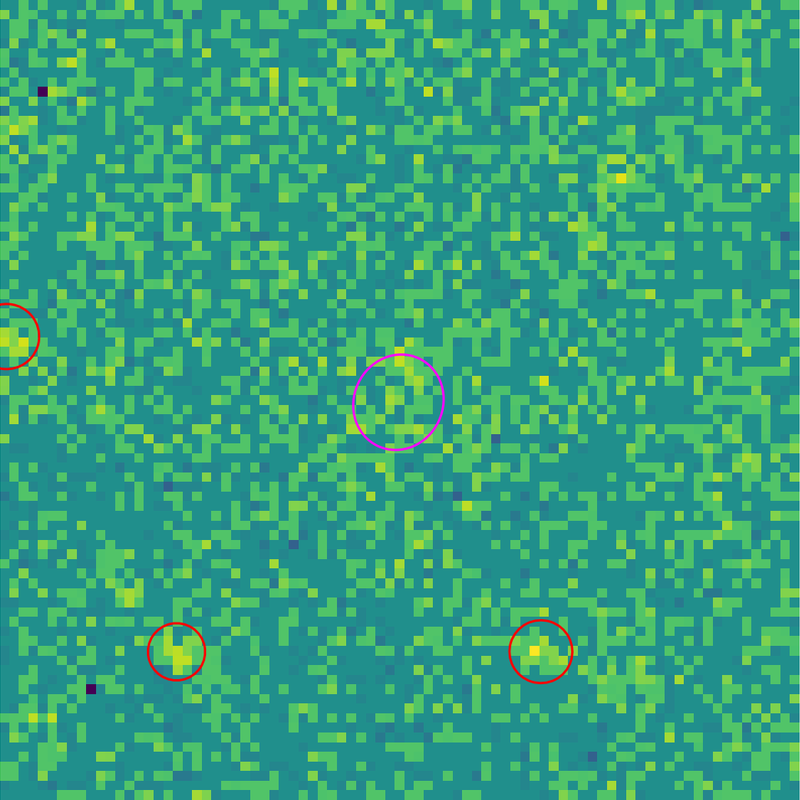} &
     \includegraphics[width=0.33\textwidth]{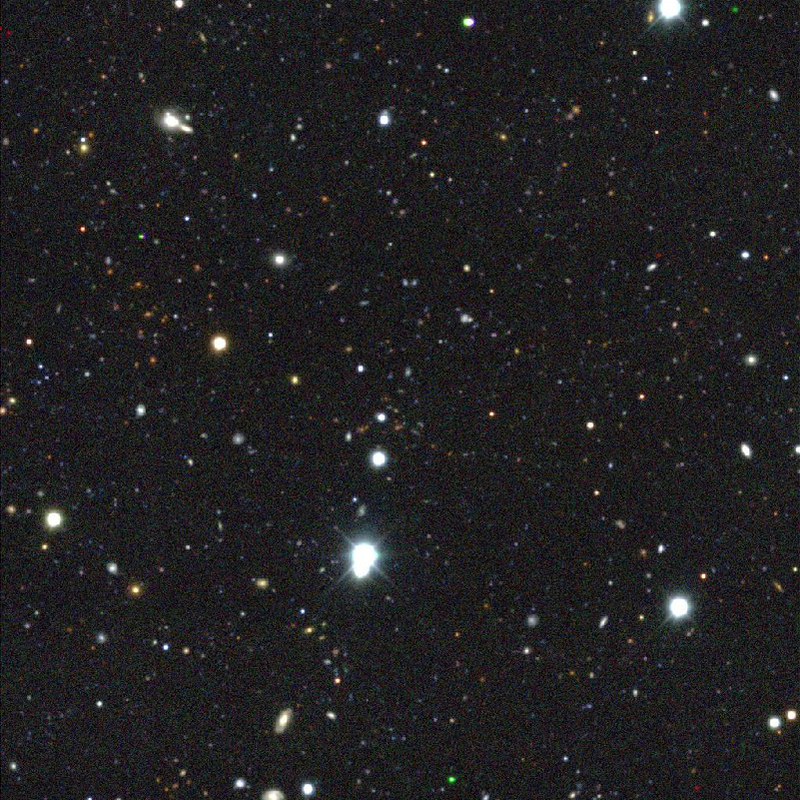} &
     \includegraphics[width=0.33\textwidth]{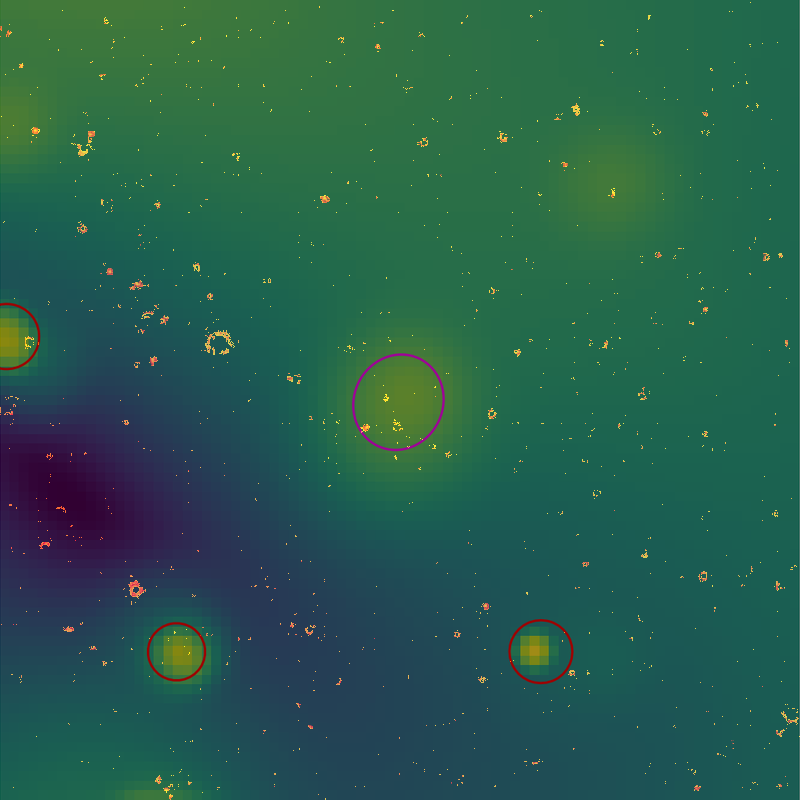}
    \end{tabular}
    \caption{cont.}
    \end{figure*}

    \begin{figure*}
\ContinuedFloat
\begin{tabular}{ccc}
     \includegraphics[width=0.33\textwidth]{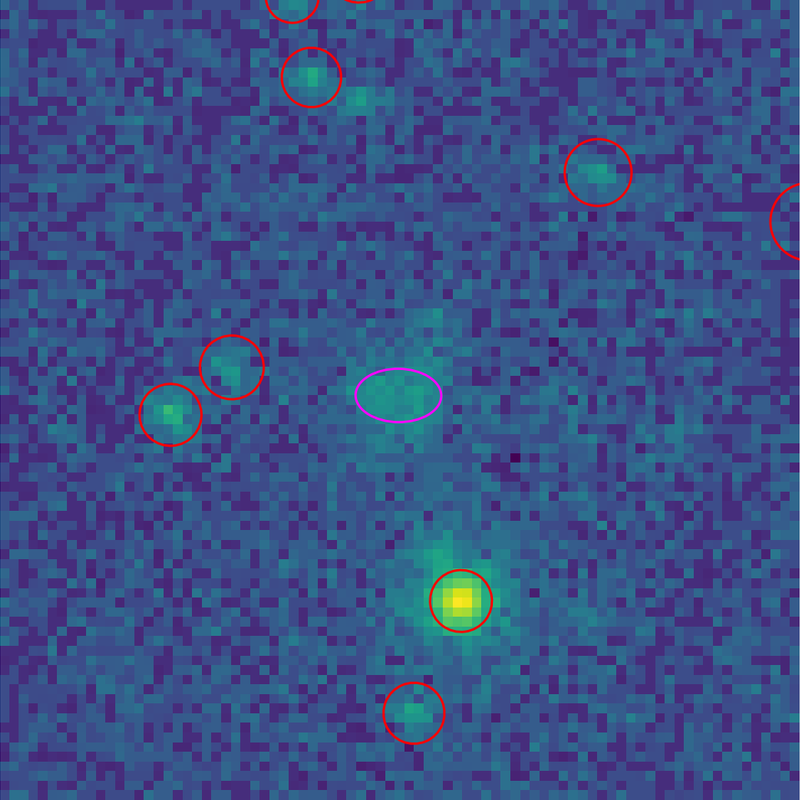} &
     \includegraphics[width=0.33\textwidth]{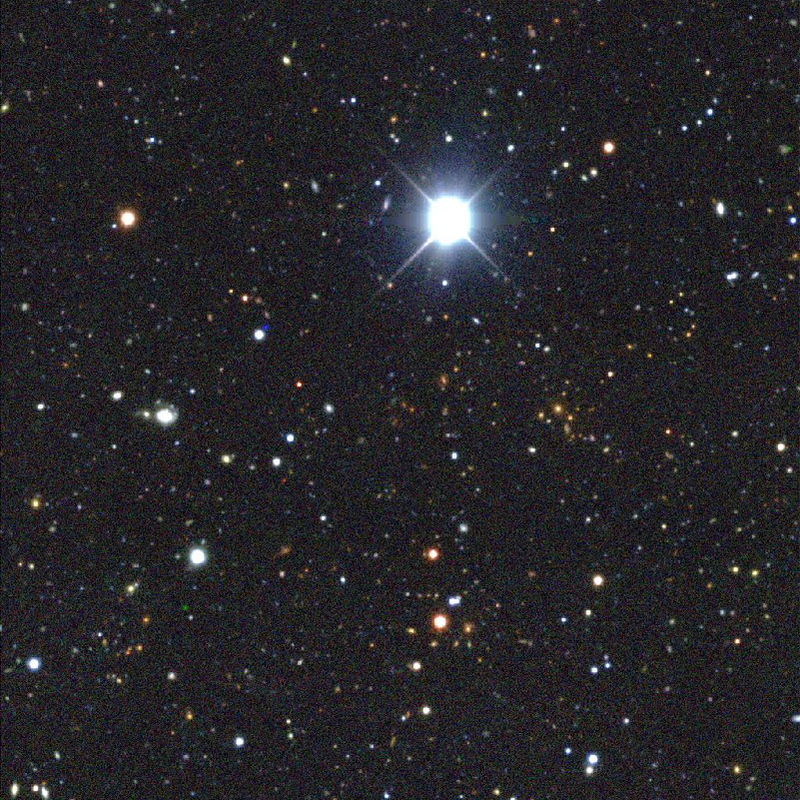} &
     \includegraphics[width=0.33\textwidth]{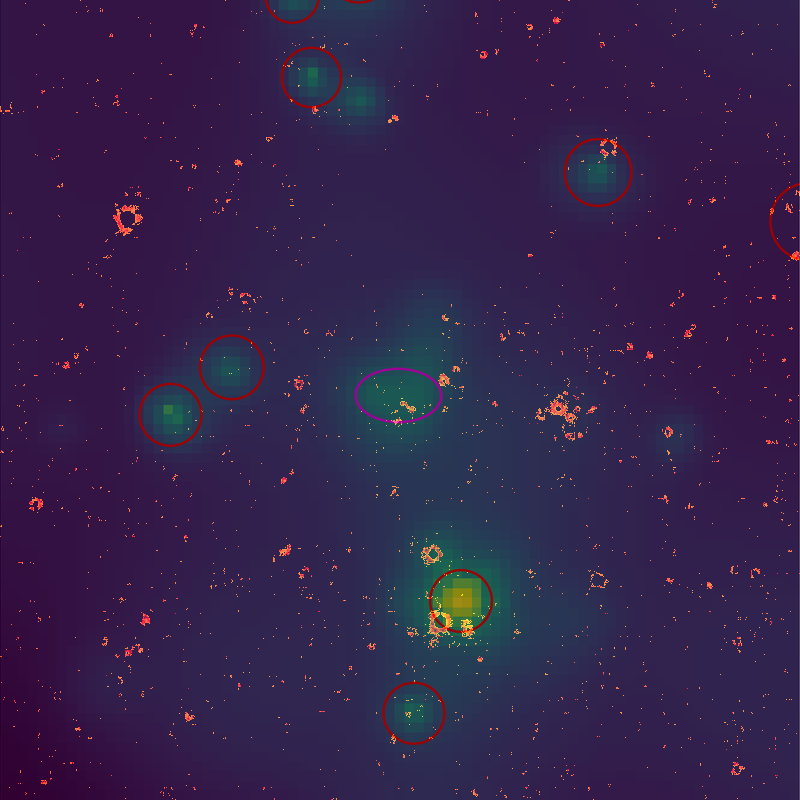}
\end{tabular}
\caption{The 13 examples of using contrast enhanced DES images to show red galaxy overabundances for high redshift X-ray extended sources that are not present in the redMaPPer catalogue. All images are 6$\times$6 arcmins. Left: XMM image showing XAPA extended source detection.  Middle: DES image.  Right: XCS image showing smoothed X-ray signal and DES image enhanced red-channel to highlight red clusters.}

\label{fig:app_high_redshifts}
\end{figure*}

%%%%%%%%%%%%%%%%%%%%%%%%%%%%%%%%%%%%%%%%%%%%%%%%%%

% Don't change these lines
\bsp	% typesetting comment
\label{lastpage}
\end{document}